\documentclass[11pt]{article}
\usepackage{old-arrows}
\usepackage[scaled=.9]{helvet}
\usepackage{charter}
\usepackage{XoohmH}
 \newmdenv[linecolor=blue!60!teal,
 linewidth=1,roundcorner=2pt,backgroundcolor=yellow!3!red!1,
 innerleftmargin=5pt,innerrightmargin=5pt,leftmargin=0pc,rightmargin=0pt,
 fontcolor=blue!60!black,
 ]{bBox}
\usepackage{cite, tabularx}

\def\SI#1{\setbox9\hbox{$\SSS#1$}\5[-1pt]{\copy9\kern-.125\wd9}{\mathscr{I}}}

\newcommand{\FF}[2][n]{F^{\sss(#1)}_{#2}}

\newcommand{\pDN}[2][]{{\pD^{\mkern-3mu\raisemath{1pt}{#1}}_{\mkern-1mu\smash{#2}}}}
\newcommand{\pDs}[2][]{\pDN[\star#1]{#2}}
\newcommand{\pFn}[2][]{{\pS^{\raisemath{1pt}{#1}}_{\mkern-2mu\smash{#2}}}}

\newcommand{\cKs}[1]{{\cK^*_{\!\smash{#1}}}}

\long\def\oMit#1{}

\def\chX{{\;\widecheck{\!\!\sX}}}
\let\fnSz=\footnotesize
\let\scSz=\scriptsize

 \SfTitles
 \allowdisplaybreaks
 \numberwithin{equation}{section}

 \omitBackreferences

\begin{document}
\displayBAstretch[.667]
\thispagestyle{empty}
\setcounter{page}{0}
\vglue5mm
\begin{center}
{\LARGE\bsf\boldmath 
  Beyond Algebraic Superstring Compactification: Part II}\\[2mm]
{\bsf Tristan H\"{u}bsch}\\
{\small\it
      Department of Physics \&\ Astronomy,
      Howard University, Washington, DC 20059, USA}\\[-1mm]
{\tt  thubsch@howard.edu}\\[5mm]
{\sf\bfseries ABSTRACT}\\[2mm]
\parbox{143mm}{\addtolength{\baselineskip}{-3pt}\parindent=2pc\noindent
The most impressively prolific exploration of superstring models (aiming for our physical reality) has been focused on worldsheet-supersymmetric gauged linear sigma models and the closely associated {\em\/complex-algebraic\/} toric geometry. Mirror duality relates this to the inherently {\em\/real symplectic\/} geometry of Calabi--Yau factors in spacetime, implying a need for a more general, heterotic framework of analysis. In turn, a closer look at possible deformations even amongst the complex-algebraic complete intersections and toric geometry models themselves indicates an a priori non-algebraic type of generalization that however perfectly aligns with requirements of mirror duality.
}
\vspace{5mm}

\begin{minipage}{.75\hsize}\small
  \baselineskip=10pt plus1pt minus 1pt
  \renewcommand{\cftbeforetoctitleskip}{\smallskipamount}
  \renewcommand{\cftaftertoctitle}{\vskip-5pt}
  \tocloftpagestyle{empty}
  \setcounter{tocdepth}{3} 
  \tableofcontents
\end{minipage}
\end{center}
\vspace{2mm}


\section{Introduction, Rationale and Summary}
\label{s:IRS}
Gauge anomaly detection and cancellation is a powerful tool in quantum field theory (QFT): 42 years ago, it opened the floodgates on superstring theory as a framework in which to construct models of our physical reality\cite{rGS-Anom, Gross:1984dd, rCHSW}. Tantamount to several anomaly cancellation conditions in the underlying worldsheet QFT of the string, Ricci-flatness of the target spacetime ensures worldsheet quantum stability\cite{rF79a, rF79b, Polchinski:1998rq, Polchinski:1998rr} as well as the self-consistency of the full, oriented loop-space reformulation\cite{rFrGaZu86, rBowRaj87, rBowRaj87a, rBowRaj87b, Oh:1987sq, rHHRR-sDiffS1, Pilch:1987eb, rBowRaj88, Bowick:1988nj, Bowick:1990wt, rBeast2}.
The initially mostly analytic analysis was soon bolstered and reframed, if not entirely replaced, by algebraic methods (see\cite{rBeast2}), especially so by the special class of worldsheet $(2,2)$-supersymmetric gauged linear sigma models (GLSMs)\cite{rPhases, rAGM01, rAGM06, rAGM04, rMP0}. Soon generalized by relaxing to $(0,2)$-supersymmetry (\mbox{{see} Refs.\cite{rD+G-DBrM,Sharpe:2024dcd}} for comprehensive reviews), 
this approach has provided us with the largest pool of \mbox{constructions\cite{Kreuzer:2000xy, wKS-CY}}, well-framed within complex-algebraic and toric geometry\cite{rGrHa, rD-TV, rO-TV, rF-TV, rGE-CCAG, rCLS-TV, rCK}, counted in terms of astounding ``{\em\/heptigoogol\/} of moles'' ($10^{723}$)\cite{Constantin:2018xkj, rBeast2}. Finding the model that matches our own world in such a mind-boggling sea of possibilities would seem like a forlorn quest, were it not for the astonishing developments in machine learning, which more and more capably and accurately extract physical data for this quest\cite{Butbaia:2024tje, Constantin:2024yxh, Berglund:2024uqv, Constantin:2024yaz, Berglund:2024psp, Rahman:2026mfy}.

The fact remains, however, that the $(2,2)$- and even $(0,2)$-supersymmetric GLSM framework has been chosen primarily for computational convenience: ultimately, worldsheet $(0,1)$-supergravity suffices to guarantee stable ground-states but it implies neither supersymmetry nor complex structure in target-spacetime physics{, nor does string theory in general require any spacetime factor or subspace to be algebraic}; these emerge in sectors with at least 
$(0,2)$-supersymmetry\cite{Sen:1986mg, Giddings:1993wn, rUDSS08, rUDSS09}. 
In particular, its a priori {\em\/complex-algebraic\/} setting frames the original approach to mirror duality\cite{rGP1, rBH, rBH-LGO+EG, Krawitz:2010FJR} (see also\cite{rSSQM3}) and dovetails with complex-algebraic toric geometry\cite{Batyrev:1993oya, rLB-MirrBH, rBatyBor3}. 
This framing was, however, neither intended nor designed to study the later-discovered (real) symplectic aspects of mirror \mbox{duality\cite{Kontsevich:1995wkA, Kontsevich:1994Mir, gross2016intrinsic}}; {this and} a recent study\cite{rBH-Fm, rBH-gB, Berglund:2022dgb, Berglund:2024zuz, Hubsch:2025sph, Hubsch:2025teh, Hubsch:2026lir} indicate a need for generalization.
{Still}, one {may} follow the general strategy of identifying/constructing the Ricci-flat manifolds of our ultimate interest in some {\em\/well-known\/} ``ambient'' spaces and by {\em\/well-known\/} methods---including the well-trodden algebraic ones---while being mindful of technical conveniences that may be relaxed.\footnote{This strategy has an illustrious history: the very first concrete solution\cite{Schwarzschild:1916uq} to Einstein's field equations made it clear that even ``empty'' (``sourceless,'' {$T_{\mu\nu}\<=0$}) observable spacetime can have a physically nontrivial geometry---that was soon aptly reconstructed by algebraic methods\cite{Kasner:1921Fin}---which makes manifest its tantalizing two-sheeted nature and physically nontrivial global geometry\cite{Einstein:1935The, rCF-BH}.}

With this goal in mind, and following\cite{Hubsch:2025teh},
this article aims to:
({\bf1})~revisit the {\em\/physics\/} (GLSM{/QFT}) {features that motivate the} links to toric geometry and identify avenues for generalization;
({\bf2})~explore in more detail the explicitly continuous showcasing {deformation} family of examples, realized as hypersurfaces in Hirzebruch scrolls, 
$\sX\<\in\FF{\sss\ora{m}}[c_1]$, which are themselves hypersurfaces in 
$\IP^n\<\times\IP^1$;\footnote{In general, the symbol $A[c_1]$ denotes the deformation family of all anticanonical, Calabi--Yau, Ricci-flat, i.e., deg-$c_1(A)$ hypersurfaces in the ambient space, $A$.}
({\bf3})~motivate extensions to {a particular subclass of} (a priori, not complex) {\em\/unitary torus manifolds\/} (UTMs)\cite{rM-MFans, Masuda:2000aa, rHM-MFs, Masuda:2006aa, rHM-EG+MF, rH-EG+MFs2, Nishimura:2006vs, Ishida:2013ab, Davis:1991uz, Ishida:2013aa, Yu:2011aa, buchstaber2014toric, Jang:2023aa} as likely candidate ambient spaces in which to {embed} mirror-models, $\chX\<\in\tP\FF{m}[c_1]$, of hypersurfaces in {\em\/non-weak-Fano}\,ambient spaces, $\sX\in\FF{m}[c_1]$.\footnote{I adopt the negation, ``non-weak-Fano''\cite{MacFadden:2025ssx} to mean ``not even weak-Fano'' over the less precise, if more descriptive, term ``limping''\cite{rBeast2}: in particular, $c_1(\FF{\!\sss\ora{m}})$ with (``taxicab''-magnitudes) $|\ora{m}|\<\geqslant3$ is positive along the fiber-$\IP^{n-1}$ but negative along the base-$\IP^1$; in the weak-Fano case, 
$|\ora{m}|\<=2$, $c_1(\FF{\!\sss\ora{m}})$ vanishes along the base.}  
This last aim necessarily involves {physics}-motivated but as yet unproven claims and conjectures, inviting more systematic and rigorous research.

The remainder of this introduction reviews the relevant key features of the worldsheet view of superstrings and how they induce target-spacetime geometry and dynamics, focusing on the $(2,2)$-supersymmetric GLSM; although widely available in the literature, it will be convenient to have these in one place for reference and reconsideration. {In particular, a key observation is that GLSM/QFT dynamics do not require all of the inherent non-negativity and convexity of complex-algebraic toric geometry\cite{rD-TV, rO-TV, rF-TV, rGE-CCAG, rCLS-TV, rCK}.}
 Section~\ref{s:PhasDefo} presents the showcasing example, where the $U(1;\IC)^2$ gauge-symmetry acting on an array of differently-charged chiral superfields already results in an intricate field space{. In its} ``geometric'' phase, this includes generalizations of Hirzebruch (rational, ruled) surface scrolls\cite{rH-Fm, rGrHa, rO-TV, rF-TV, rGE-CCAG, rCLS-TV, rKC-Fm}, {which the} worldsheet {GLSM/}QFT framework continuously connects to the ``Landau--Ginzburg orbifold'' (LGO) phase, and so also to the original formulation of the transpose-mirror construction\cite{rBH, rBH-LGO+EG, Krawitz:2010FJR}.\footnote{\label{fn:hBHK}The 1992 transposition-mirror\cite{rBH} not only exhibits $h^{p,q}(\sX)\<=h^{n-p,q}(\chX)$ but identifies the corresponding state spaces/cohomology groups as representations of ``geometric and quantum symmetries'' that govern their ring structure in this ``LGO mirror symmetry.''
Its ``classical mirror symmetry'' (routinely associated with the more geometric early formulations\cite{rGP1, Candelas:1990qd, Candelas:1990rm, Batyrev:1993oya}) was proven in 2009 via the ``cohomological LG/CY correspondence'' by Chiodo and Ruan\cite{rC+R-MirrBH, Krawitz:2010FJR, Orlov:2009Der} and in 2010 by Borisov within the vertex operator (chiral) algebra framework\cite{rLB-MirrBH}.
Substantial research has identified the transposition-mirror construction as an integral part of the inter-connected web of various ``mirror symmetry'' frameworks; for a tentative and unavoidably incomplete sampling, see also\cite{Berglund:1993ax, rCetal-2p2, Kontsevich:1995wkA, Batyrev:1996Mir, rSYZ-Mirr, Givental:1997A-m, Lian:1997Mir, rFJR-07b,  Polishchuk:1998db, Wan:2004Mir, Kontsevich:2006Aff, Gross:2012Fro, Katzarkov:2008hs, Kapustin:2009HMS, Ebeling:2012Mir, rABS-BHK3, rS-BirBHK, Clarke:2013sha, rA+P-BHK, rACG-BHK, rA+P-MM, Favero:2016vhp, rDKSS-altMM, Kelly:2015Pic, Comparin:2021sas, Filipazzi:2017hwe, Aleshkin:2019ahf, Belavin:2020xhs, Belakovskiy:2020nno, Gammage:2020wvn, Parkhomenko:2022kju, Gomez:2022xdf, Belavin:2023ldz, CuadrosValle:2023ggb, Parkhomenko:2024mxq, Cho:2024Ber, Aldi:2025mzh, Belavin:2025yjv, Aleshin:2026sbv, Aldi:2026lgf}.
Via the LGO ``phase'' of GLSM and its routinely associated toric geometry, the transposition-mirror construction is firmly tied to the toric formulation\cite{Batyrev:1993oya, rAGM01, rAGM06, rAGM04, Candelas:1994bu, Batyrev:1996Mir}---which is germane here. The central role of the ``fundamental monomial''\cite{rHY-SL2} deformation in the superpotential makes it straightforward to connect to period integrals computation and ``enumerative mirror symmetry''\cite{Batyrev:1993oya, Candelas:1994bu, Morrison:1992Mir, Berglund:1993ax, Givental:1997A-m, Lian:1997Mir, rI-QC11}, while Refs.\cite{rA+P-MM, Aldi:2025mzh, Aldi:2026lgf} use it explicitly in ``arithmetic mirror symmetry.''
In turn, as early as 1994 Candelas, de~la~Ossa and Katz had indicated\cite{Candelas:1994bu} some generalizations of the original transposition-mirror construction and its overlap with Batyrev's 1993 toric construction\cite{Batyrev:1993oya}, which the past decade's work\cite{rBH-gB, Berglund:2022dgb, Berglund:2024zuz, Hubsch:2025sph, Hubsch:2025teh, Hubsch:2026lir} and the details of the underlying physics (GLSM/QFT) show can and should be extended further.}
 Section~\ref{s:AmbiDefo} presents the explicit deformation families of Hirzebruch {scrolls} of arbitrarily high twist {and continuously connected distinct toric varieties,} to serve as ``well-known'' ambient spaces for constructing Calabi--Yau hypersurfaces.
 Turning to the construction of {transposition-}mirror models\cite{rBH-Fm, rBH-gB, Berglund:2022dgb, Berglund:2024zuz, Hubsch:2025sph, Hubsch:2025teh},  Section~\ref{s:Mirrors} explores the need for extending the complex-algebraic toric geometry framework {indicated by the Section~\ref{s:PhasDefo} observation that the mirror-dual transposed GLSM is not limited by all of the inherent non-negativity and convexity of complex-algebraic toric geometry. This indicates embedding mirror Calabi--Yau models in a particular subclass of UTMs as} a likely avenue.
 A brief summary of these features then motivates the outlook presented in the concluding Section~\ref{s:CODA}{, including a possible local surgery characterization of the relevant UTMs}.

\subsection{Worldsheet Quantum Field Theory Models and Mirror Duality}
\label{s:QFT+MD}
Worldsheet quantum field theories are all defined on a local patch of the Riemann surface swept out by the moving string. The dynamics in $(2,2)$-supersymmetric models as considered herein\,\footnote{Reducing supersymmetry $(2,2)\to(0,2)$ or even $\to(0,1)$ may then be accomplished by expanding in, say, $\vs^+,\7\vs^+$ or even in $\7\vs^-$, and treating the so-obtained expansion terms independently\cite{rUDSS08, rUDSS09}; see also\cite{rUDSS01, rUDSS02, rUDSS03, rUDSS04, rHSS, rGSS, rHP1}.}
are governed by the action functional that, in its {\em\/direct\/} form,\footnote{Also called the ``nonlinear $\s$-model'' form, this modeling is {\em\/direct\/} in that $\vev{X^\m}$ and $\vev{G_{\m\n}(X)}$ are local coordinates and metric on the target-space itself, the latter of which is most of the time unknown.}
includes the standard kinetic term
$\g^{\a\b}(\x)\,(\vd_\a X^\m)(\vd_\b X^\n)\,G_{\m\n}(X)$
for the lowest component-fields, $X^\m(\x)$, of as many chiral superfields. 
Here, $\g^{\a\b}(\x)$ is the (inverse) metric on the worldsheet, over which the partition functional of the quantum theory is integrated. In turn, $G_{\m\n}(X)$ is the target-space metric and serves as an array of coupling parameters in such worldsheet QFT models. The vacuum expectation values ({\em\/vev\/}s) of the $X^\m(\x)$ serve as local coordinates (denoted by the same symbol) on the target space itself, $\sX$. 
In turn, half of their fermionic superpartners, $\j^\m_\pm$, provide a local basis for the local tangent space, the (canonically conjugate) {complement} spanning the cotangent space.

On a {\em\/complex\/} (factor of the) target spacetime, $\sX$, this halving may be {\em\/chosen\/} {along the complex structure} (see\cite{rHSS} and references therein),
\begin{alignat}9
&&G_{\m\7\n}(X)\j_+^{\7\n}(\x) &&\eqco\j_{+\m}(\x)
  &\mapsto\vd_\m ~~&\text{and}~~ \j_+^\m(\x)&\mapsto\rd X^\m, 
 \label{e:SuSyGeom1}\\
  \text{vs.}&\quad
 &G_{\m\7\n}(X)\j_-^\m(\x) &&\eqco\j_{-\7\n}(\x)
  &\mapsto\vd_{\7\n} ~~&\text{and}~~ \j_-^{\7\n}(\x)&\mapsto\rd X^{\7\n}.
 \label{e:SuSyGeom2}
\end{alignat}
This {\em\/choice\/} maps the Fock-space elements to Dolbeault cohomology groups:
\begin{alignat}9
  h_{\7\n_1\cdots\7\n_q}^{\m_1\cdots\m_p}(X,\7X)\,
  \j_-^{\7\n_1}\cdots\j_-^{\7\n_q}\,\j_{+\m_1}\cdots\j_{+\m_p}
  \ket0 &\mapsto H^q_{\sss\7\vd}(\sX\!,\wedge^pT)
 &&=H^{n-p,q}_{\sss\7\vd}(\sX\!,\cKs{}), \label{e:n-p,q}\\
  \w_{\m_1\cdots\m_p\,\7\n_1\cdots\7\n_q}(X,\7X)\,
  \j_-^{\7\n_1}\cdots\j_-^{\7\n_q}\,\j_+^{\m_1}\cdots\j_+^{\m_p}
  \ket0 &\mapsto H^q_{\sss\7\vd}(\sX\!,\wedge^pT^*)
  &&=H^{p,q}_{\sss\7\vd}(\sX), \label{e:p,q}
\end{alignat}
where $\cKs{\sX}\coeq\wedge^nT_\sX$ is the anticanonical bundle of the {target} space, $\sX$, and the rightmost equality in~\eqref{e:n-p,q} exhibits the identity
$\wedge^pT\<{\6{\sss\rm id}{=}}\wedge^{n+(p-n)}T\<=\cKs{}{\otimes}\wedge^{n-p}T^*$.
The penultimate assignments and~\eqref{e:SuSyGeom1}--\eqref{e:SuSyGeom2} make it clear that complex conjugating only the $\j_+\iff\7\j_+$ 
swaps $\wedge^pT\iff\wedge^pT^*$, which exhibits the elementary origin {and ubiquity} of {\em\/mirror duality\/} in worldsheet $(2,2)$-supersymmetric models: 
{the Calabi--Yau space} $\chX$ is the mirror of $\sX$ if the cohomology rings\,\footnote{\label{fn:ring}This statement of mirror duality presumes the ``quantum'' (deformation of the wedge) ring structure of these cohomology groups\cite{Kontsevich:1995wkA, Kontsevich:1994Mir, rCK, rD+G-DBrM}; their dimensions must agree regardless of the ring structure.}
$H^q_{\sss\7\vd}(\sX\!,\wedge^pT)\approx 
 H^q_{\sss\7\vd}(\chX,\wedge^pT^*)$. 
Comparison of the ultimate, rightmost assignments in~\eqref{e:n-p,q} and in~\eqref{e:p,q} exhibits, in turn, that in target-spaces with a trivial (anti)canonical class, $\cKs{\sX}=\cO_\sX$, {\em\/mirror duality\/} is a symmetry of the Hodge-decomposed cohomology {\em\/ring,}\footref{fn:ring}
$H^{p,q}(\sX)\approx H^{n-p,q}(\chX)$.

The (perturbatively computed) renormalization of the worldsheet model deforms its couplings---and $G_{\m\n}(X)$ in particular, which is identified as the target space metric.
The requirement for stability under ``quantum fluctuations'' (renormalization fixed point) was found to exactly reproduce the Einstein equations in their Ricci form\cite{rF79a, rF79b},\footnote{\label{fn:gst}
This led to the introduction of the term ``geometrostasis''\cite{Braaten:1985is}, where Friedan's and some earlier work was generalized, in particular also to include geometric torsion.}
\begin{equation}
  0\overset!= \big[ R_{\m\n} - \tfrac{8\p G_{\!N}}{c^4}
  \big(T_{\m\n}{-}\tfrac1{d-2}G_{\m\n}\,G^{\r\s}T_{\r\s}\big)
  \big] + O(\a'),
 \label{e:R=T}
\end{equation}
where $d=\dim(\sX)$ and $T_{\m\n}$ is the energy-momentum density tensor for target-spacetime ``matter,'' i.e., all non-$G_{\m\n}(X)$ degrees of freedom. 
Note that the ``trace-flipping'' of the energy-momentum density tensor on the {parenthetical term in}~\eqref{e:R=T} depends on the spacetime metric $G_{\m\n}$ and its inverse.

That the {\em\/quantum\/} stability of the worldsheet {\em\/quantum\/} field theory requires the target-spacetime metric $G_{\m\n}$ to satisfy its {\em\/classical\/} equation of motion (Einstein's field equations) may be seen as a (worldsheet-to-target spacetime) layer-building generalization of Ehrenfest's theorem.
In fact, this naturally generalizes, with {\em\/quantum\/} stability of more general worldsheet {\em\/quantum\/} field theory models inducing (with technical but not substantial $\a'$-perturbative deformations) the standard, {\em\/classical\/} gauge interactions in target-spacetime. 
 Typically, it is routine to construct a target-spacetime Hamilton's action functional from which these target-spacetime {\em\/classical\/} equations of motion follow via the usual variational calculus. The standard Feynman path integral formalism then generates the corresponding fully fledged target-spacetime QFT model as a candidate for the observed Standard Model physics.
This ``layer-cake'' structure in string theory then displays the indirect nature of how string theory models our ``real-world physics'' as induced from a judiciously chosen underlying worldsheet model. In turn, it is also possible to focus on a ``sector'' within a complete model, relying on previously established connections to the rest. 

Here, we focus on target-spacetimes of the form $\IR^{1,3}\!\times\!\sX$, with
$\IR^{1,3}$ identified with the observed spacetime and
with $\sX$ henceforth denoting a compact (and $\leqslant\!10^{-32}$\,m small) spatial Ricci-flat factor, chosen at first so as to preserve an overall $N\!=\!1$ supersymmetry, which requires $\sX$ to be a compact, complex Calabi--Yau 3-fold; we revisit these choices in Section~\ref{s:Mirrors}.

\subsection{Reverse-Engineering Target-Spacetime}
\label{s:TST}
Worldsheet QFT models are built over an a priori unspecified genus-$g$ Riemann surface, $\S_g$, with a Lorentzian metric,
$\g_{\a\b}(\x)$. The Feynman path integration then sums over all
$(\S_g,\g)$-choices, as prescribed by the Deligne--Mumford ``universal curve,'' over specific boundary (periodic/anti-periodic, etc.) conditions, and it also allows the inclusion of worldsheet boundary ``sources.'' 

Aiming to dynamically determine the Calabi--Yau space, $\sX$, one must start {\em\/somewhere,} and so one starts with a well-understood, well-known and computationally convenient {\em\/ambient space,}\footnote{\label{fn:A}One seeks to work in a space where one {knows} just about everything one needs for the purposes of computing the desired characteristics and properties of the subspaces of interest; for starters, see Ref.\cite{rBeast2}.} $A$,
to provide an auxiliary scaffolding within which to construct the $\sX$ of our ultimate interest.
In this sense, and in contradistinction to the discussion in Section~\ref{s:QFT+MD}, these are {\em\/indirect\/} constructions of Calabi--Yau spaces via intermediate but convenient ambient spaces.
GLSMs\cite{rPhases, rAGM01, rAGM06, rAGM04, rMP0} provide one such large class of $(2,2)$-supersymmetric worldsheet models, the low-energy limit of which (in simple cases) corresponds one-to-one to nonlinear, constrained complex projective-space models~(\cite{Eichenherr:1978SUN, rChaSM, rMargD, rUDSS08, rUDSS09} and references therein) and so provides their vast generalization; see also\cite{Bykov:2020tao, Bykov:2021dbk}. 
GLSMs are well-specified by providing:
\begin{enumerate}[labelsep=13pt]
\item 
 A list of $n{+}r$ chiral $(2,2)$-superfields, $X_i(\x)$, the lowest 
 (bosonic) component fields of which provide (complex) coordinate fields for 
 the (complex $n$-dimensional) compact factor in the target spacetime.
\item 
 A list of $r$ twisted-chiral $(2,2)$-superfields, $\s_a$, one for
 each gauged $U(1)$ symmetry (1+1-dimensional gauge 2-vector potentials have 
 no propagating degrees of freedom but induce gauge-equivalences, and 
 they are accompanied by a complex scalar field each.)
 \item \label{i:qai}
 The worldsheet $(2,2)$-supersymmetric $U(1)^r$ gauge symmetry is
 automatically complexified, $U(1)^r\to U(1;\IC)^r=(\IC^*)^r$, and
 it acts by nonzero complex rescaling,
 $X_i\to\l{\cdot}X_i:=\prod_a\l_a^{q_{ai}}X_i$, where
 $q_{ai}=q_a(X_i)$ is the charge of $X_i$ with respect to 
 the $a^\text{th}$ $U(1)$, and where 
 $\l_a\in\IC^*$ are nonzero complex-valued chiral superfields.
\end{enumerate}
The Lagrangian super-density consists of the standard $U(1)^r$ gauge-invariant kinetic term (with the standard coupling {of} the chiral and twisted-chiral superfields) and a superpotential for $X$-exclusive (``Yukawa'') interactions.
\begin{enumerate}[resume, labelsep=13pt]
\item \label{i:S(X)}
 A choice of the superpotential, of the general form 
 $W(X)=\sum_{\a=1}^K X_0^\a\,f_\a(X)$, where we focus on the 
 $K=1$ (single hypersurface) case and omit the index $\a$. The function $f(X)$ is chosen quasi-homogeneous, $f(\l{\cdot}X)=\prod_a\l_{{a}}^{q_{af}}f(X)$, and $q_a(X_0)=q_{a0}$, so that the superpotential $X_0{\cdot}f(X)$ will be $U(1)^r$-invariant:
\begin{equation}
  q_{a0} +{\big(q_{af}=}\sum\nolimits_{i=1}^{n+r}q_{ai}{\big)}=0,\quad 
   \text{for each}~~a=1,\cdots r.
 \label{e:CY}
\end{equation}
\end{enumerate}
In the low-energy regime, the $X_0$ field indeed behaves as a Lagrange multiplier, enforcing the constraint $f(X)\!=\!0$---in the so-called ``geometric'' phases (see below), giving the first hints about the geometry of the so-described target spaces to be regarded as hypersurfaces $\{f(X)=0\}\subset A$---which provides the technical-ease bias towards ``well-known'' ambient spaces,\footref{fn:A} $A$, where all required computational information is as readily available as possible.
\begin{remk}\label{r:key}
The
({\bf1})~superpotential gauge invariance condition~\eqref{e:CY} ({\bf2})~guarantees the anomaly cancellation for each $U(1)$ gauge symmetry and is ({\bf3})~the condition for the ground-state hypersurface 
$\{f(x)=0\}\subset\big(\{X_i\}/U(1)^r\big)$ to have a vanishing 1st Chern class and so
({\bf4})~{be a Calabi--Yau hypersurface and admit a Ricci-flat K{\"a}hler metric, 
thus 
({\bf5})~exhibiting geometrostasis in its original sense\cite{Braaten:1985is},\footref{fn:gst} but also in the sense that 
({\bf6})~the coupling strengths of the {$U(1)^r$ gauge} interactions remain constant in renormalization.}
\end{remk}

After integrating out the ``auxiliary fields'' (the equations of motion of which are algebraic, non-dynamical), the effective Lagrangian density contains the potential that is the sum of the positive semi-definite terms\cite{rPhases, rAGM01, rAGM06, rAGM04, rMP0}:
\begin{equation}
 \underbrace{\sum_a\!\Big(\!\sum_iq_{ai}|X_i|^2-t_a\!\Big)^2}_{D\text{-terms}}
 ~+~
 \underbrace{|f(X)|^2
  +|X_0|^2\sum_i\Big|\frac{\vd f}{\vd X_i}\Big|^2}_{F\text{-terms}}
 ~+~
 \underbrace{\sum_{a,b}\7\s_a\s_b\sum_iq_{ai}q_{bi}|X_i|^2}_{\text{mixed terms}}.
 \label{e:U}
\end{equation}
The ground state is therefore determined by the vanishing of each summand, each of which is usually analyzed in turn and as provided in~\eqref{e:U}; see, e.g.,\cite{rBH-gB}. Here we discuss in detail the choice of $f(X)$ in a concrete showcasing example and complementing\cite{rBH-gB, Berglund:2022dgb, Berglund:2024zuz, Hubsch:2025sph, Hubsch:2025teh}, rather than attempting to specify a general and invariably rather complex meticulous algorithm.

In this framework, the geometry (metric) of the target space---spanned by the ground state degrees of freedom---emerges only through iterative computations in an interplay between so-called $D$- $F$- and mixed terms~\eqref{e:U}.
 In particular, gauge symmetry identifications result in the $X_{{i}}$-space spanning a $U(1;\IC)^r$ gauge-quotient, which induces much of its nontrivial topology and phase structure. 
Without the gauge-quotient, the minima of the potential would form a contractible cone of well-nigh trivial topology.
The manyfold variations in gauge-quotienting and choices of $f(X)$, however, makes the myriads of possible potentials~\eqref{e:U} yield an embarrassment of riches in the variety of topology and geometry of stringy spacetimes, which is regarded as both the boon and the bane of string theory---depending on whom one asks.
In a rather welcome and marvelous turn of fate, the accelerating development of various computer-intensive methods and techniques (neural networks, machine learning, artificial intelligence, etc.) over the past decade now enables computing, with fast-increasing accuracy and breadth, the metric and other characteristic quantities on Calabi--Yau manifolds,\footnote{Along
 with various aspects of curvature, the Ricci-flat metric on Calabi--Yau spaces has recently become increasingly accessible via machine learning\cite{Berglund:2022gvm}, so that analyses such as those presented herein should aim for complementary and streamlining insight. The rapidly growing literature on the subject is itself fascinating and would take us too far afield to provide an even remotely fair and functional review; suffice it here to direct the reader to the relatively recent works\cite{Berglund:2022gvm, Butbaia:2024tje, Constantin:2024yxh, Berglund:2024uqv, Constantin:2024yaz, Berglund:2024psp, Rahman:2026mfy} and references therein for starters.} which until recently had to be analyzed without this explicit knowledge.

\begin{remk}\label{r:noW}
The worldsheet supersymmetry in the 1+1-dimensional (QFT) sigma models of interest protects the superpotential from {(perturbative QFT)} renormalization. This makes it possible to completely omit it, whereupon the GLSM simply describes the {ambient} space as the gauge-quotient, $\{X_i\}/U(1)^r$, which we address in Section~\ref{s:PhasDefo}---relying on the physics/QFT aspects and by switching to the concrete showcasing sequence of examples for simplicity.
\end{remk}

\section{Gauge-Quotient Phases of the Showcasing Sequence}
\label{s:PhasDefo}
Certainly the best-known example of a convenient ambient space in which to construct subspaces of interest is the complex projective space $\IP^n$ encoded for toric geometry purposes\cite{rD-TV, rO-TV, rF-TV, rGE-CCAG, rCLS-TV, rCK} by the fan of complex multiples of the generating vectors $\nu_i$:
\begin{equation}
 \pFn{\IP^n} = 
 \big\{ \n_i=\hat{e}_i,\quad i=1,\cdots n,\quad\text{and}\quad
        \n_{n+1}\<={-}\sum\nolimits_i\hat{e}_i \big\}
 \label{e:nPnFan}
\end{equation}
This ({\bf0}-centered) fan is generated by $n{+}1$ vectors, each corresponding to a complex (homogeneous) coordinate, {$X_i\mapsto\n_i$}, 
on which the single 
$\IC^*\<\approx U(1;\IC)$ symmetry acts isotropically: $X_i\simeq\l^1 X_i$. In the GLSM, this $U(1)$ is gauged, turning the field space into the very-well-studied $U(1;\IC)\<\approx\IC^*$ gauge-quotient; see\cite{Witten:1987tv} and references therein.

Following Ref.\cite{rBH-Fm, rBH-gB, Berglund:2022dgb, Berglund:2024zuz, Hubsch:2025sph, Hubsch:2025teh}, we recall the definition of Hirzebruch $n$-fold scrolls:
\begin{defn}[Hirzebruch $n$-folds]\label{D:HnFm}
Hirzebruch $n$-folds are defined {equivalently} as:
\begin{enumerate}[labelsep=13pt, topsep=-2pt]

 \item\label{i:PrBnFm}
  the {projective bundle,}
  $\FF{m}:=\IP\big(\cO_{\IP^1}(m)\oplus\cO_{\IP^1}^{\oplus(n-1)}\big)$,

 \item\label{i:twBnFm}
  an {$m$-twisted $\IP^{n-1}$-bundle over $\IP^1$}, and

 \item\label{i:byPnFm}
  the {biprojective hypersurface,}
   $\{p_0(x,y):= x_0\,y_0^m+x_1\,y_1^m=0\}\subset\IP^n_{\!x}\times\IP^1_{\!y}$.
\end{enumerate}

  Each such $n$-fold also has a toric rendition:
  
\begin{enumerate}[resume, labelsep=13pt, topsep=-2pt]

 \item\label{i:tornFm}
  the {toric variety}, itself defined up to $\GL(n,\ZZ)$ 
  lattice transformations by the $n$-dimensional {fan} spanned by the vectors
\begin{equation}
  \pFn{\FF{m}}=
   \Big\{\nu_1\<\coeq{-}\!\sum_{i=1}^{n-1}\hat{e}_i,\quad
         \underbrace{\nu_{k+1}:=\hat{e}_k}_{k=1,\cdots n},\quad
         \nu_{n+2}\<\coeq{-}m\sum_{i=1}^{n-1}\hat{e}_i-\hat{e}_n
   \Big\}
 \label{e:nFmFan}
\end{equation}
  where $\{\nu_1,\cdots \nu_n\}$ span the fiber-$\IP^{n-1}$,
  and $\{\nu_{n+1},\nu_{n+2}\}$ the base-$\IP^1$.
\end{enumerate}
\end{defn}

The $n=2$ examples, $\FF[2]{m}$, are the well-known Hirzebruch surfaces\cite{rH-Fm}, seen to be a minor modification
(at $\nu_1$ and $\nu_{n+2}$) of $\pFn{\IP^n}$~\eqref{e:nPnFan}:
the fan $\pFn{\FF{m}}$~\eqref{e:nFmFan} is generated $n{+}2$ vectors, has as many complex (Cox) coordinates, and so encodes a $U(1;\IC)^2$-quotient, one of each $U(1;\IC)$ ``inherited'' from the fiber-$\IP^{n-1}$ and the 
base-$\IP^1$; this will be made precise in Section~\ref{s:AmbiDefo}.

This variety of equivalent definitions and widespread use and study of these varieties in diverse subfields of algebraic geometry provides for computational versatility, which makes them exceptionally appealing as ``well-known'' ambient spaces; see footnote~\ref{fn:A}.
The computational aspects of this goal benefit from the coordinate-level identification between the biprojective embedding (item~\ref{i:byPnFm}) and the toric specification (item~\ref{i:tornFm})\cite{Berglund:2022dgb, Berglund:2024zuz, Hubsch:2025sph, Hubsch:2025teh}.

The particular choice of vectors~\eqref{e:nFmFan} in fact encodes the $U(1;\IC)^2$-charges of $X_i$ by way of the relation
\begin{equation}
  \sum_{i=1}^{n+2}q_a(X_i)\,\n_i=-q_a(X_0){\n_0},\quad 
  q_{ai}\<\coeq q_a(X_i),
 \label{e:q-nu}
\end{equation}
resulting in the structure tabulated here for $n=3$:
\begin{equation}
{\small
  \begin{array}{r|@{~}r|rrr@{~}|@{~}rr@{~}|}
 \boldsymbol{n=3} & \n_0 &\nu_1 & \nu_2 & \nu_3 & \nu_4 & \nu_5\\ \toprule
  \multirow3*{\raisebox{13pt}{$\pFn{\FF[3]m}\left\{\rule{0mm}{6mm}\right.$}}
      &0 &-1 & 1 & 0 & 0 & -m \\[-2pt]
      &0 &-1 & 0 & 1 & 0 & -m \\[-2pt]
      &0 & 0 & 0 & 0 & 1 & -1 \\[-2pt] \hline
      &1 & 0 & 0 & 0 & 0 &  0 \\[-1pt] \midrule
  q_1 & -3 & 1 & 1 & 1 & 0 & 0 \\[-2pt]
  q_2 & (m{-}2) &-m & 0 & 0 & 1 & 1 \\[-2pt]
  q_3 & -2(m{+}1) &0 & m & m & 1 & 1 \\[-2pt]
  q_4 & 0 &-2(m{+}1) &(m{-}2) &(m{-}2) & 3 & 3 \\[-1pt] \bottomrule
 \makebox[0pt][r]{\textbf{Cox variables:}}
     & X_0 & X_1 & X_2 & X_3 & X_4 & X_5\\
  \end{array}
}
 \label{e:q1-4}
\end{equation}
The generators $\n_i$ are given as column 3-vectors, and the origin $\n_0$ is associated with the Lagrange-multiplier-like $X_0$. 
All $q_1,\cdots q_4$ satisfy:
({\bf a})~their defining equation~\eqref{e:q-nu},
({\bf b})~the key Calabi--Yau requirement~\eqref{e:CY} (and they are chosen so as to vanish for at least one of the $X_i$, which will be needed below),
({\bf c})~that any two of the four 6-vectors $q_a$ should be linearly independent and provide an a priori equally valid $U(1)^2$-charge basis,
({\bf d})~ that any two of these 6-vectors $(q_{a0},\cdots q_{a5})$ stacked underneath the $3{+}1$ rows of $\nu_0,\cdots\nu_5$ should form a regular $6{\times}6$ matrix,\footnote{For $\FF{m}$, the determinant of this matrix is $2n{+}(n{-}1)m^2$.} so that those six rows and six columns form six linearly independent 6-vectors,
({\bf e})~that the same holds if one omits the fourth, separated row in~\eqref{e:q1-4} and the $\n_0$-column, which corresponds to the fiber of {$\cKs{\FF{m}}$,} the anticanonical bundle of $\FF{m}$. Being a unit distance {\em\/above\/} the origin {\bf0} in the $(n{+}1)^{\text{th}}$ direction, $\n_0$ is the apex of a pyramid {\em\/over\/} the  ``base'' spanned by $\{\n_1,\cdots\n_{n+2}\}$ in a $\ZZ^n$-lattice in the real $n$-space, $(\ZZ^n{\otimes_{\sss\IR}}\IR^n)$---encoding the fiber of the anticanonical bundle over $\FF{m}$.
 Standard (complex-algebraic) toric geometry identifies $q_1,q_2$ as the {\em\/Mori vectors\/} of $\FF{m}$, $q_3$ being their non-negative linear combination\cite{rO-TV, rCLS-TV, rBKK-tvMirr}; the linear combination $q_4\<=(m{-}2)q_1+nq_2$ is also non-negative for $m\geqslant2$.

{\begin{remk}\label{r:Mori+}
For the GLSM, the $\n(X_i)$-column entries in the upper half of the rows in~\eqref{e:q1-4} specify a basis of $X_i$-charges with respect to non-gauged (global) $U(1)$-symmetries. The conditions~\eqref{e:q-nu} then specify anomaly cancellations for their mixing with the gauged $U(1)^2$ symmetry, generated by any two of the $q_a(X_i)$ in the lower half of the tabulation~\eqref{e:q1-4}.\footnote{In a 1+1-dimensional worldsheet QFT the standard Adler--Bell--Jackiv anomaly cancellation conditions stem from ``bi-angle'' Feynman diagrams (rather than triangles, as in 3+1-dimensional spacetime), whence the quadratic/mixing (rather than cubic) nature of~\eqref{e:q-nu}.}  The GLSM QFT framework has no inherent non-negativity requirement to restrict to the Mori vectors, $q_1,q_2$, and the $q_3,q_4$ choices are just as relevant in the GLSM QFT framework, already hinting at reaching beyond standard (complex-algebraic) toric geometry.
\end{remk}}

\subsection{Field Space Gauge-Orbits}
\label{s:GO}
Since the field space spanned by the $X_i$ is in fact a $U(1;\IC)^2$-gauge quotient, and since the $U(1;\IC)^2$-transformation of $(X_0;X_1,\cdots X_5)\in\IC^6$ is evidently not uniform, one must apportion ({\em\/stratify\/}) the affine field-space $\IC^6$ into regions of uniform $U(1;\IC)^2$-transformation (gauge-orbits) to specify a well-defined quotient. 

\paragraph{The $\IP^n$ template:}
For a simple example~\eqref{e:nPnFan}, the group of all complex nonzero rescalings
\begin{equation}
  \IC^2\ni(z_1,z_2) \mathrel{{\to}} \l{\cdot}(z_1,z_2)
  :=(\l z_1,\l z_2)\in\IC^2,\qquad 
  (\l\neq0)\in\IC^*\approx U(1;\IC),
 \label{e:C*C2}
\end{equation}
is {\em\/free\/} on the subset $\big((z_1,z_2)\neq(0,0)\big)\in\IC^2$, since 
$\l(z_1,z_2)=\l'(z_1,z_2)$, which implies that $\l=\l'$. The $\IC^*$-transformation~\eqref{e:C*C2} clearly fails to be {\em\/free\/} at the origin, 
${\bf E}_o{:=}(0,0)\in\IC^1$, which alone is {\em\/fixed\/} by (invariant under) the $\IC^*$-transformation~\eqref{e:C*C2}. When constructing ``$\IC^*$-gauge quotients'' one must separate these two $\IC^*$-orbits and their respective quotients (``$\sqcup$'' denotes disjoin union):
\begin{equation}
   \IC^2/\IC^* ~\leadsto~
    \big((\IC^2\ssm{\bf E}_o)/\IC^*\big) ~\sqcup~ {\bf E}_o/\IC^*
    ~=~ \IP^1 ~\sqcup~ {\bf E}_o.
 \label{e:P1+.}
\end{equation}
Although the region $(\IC^2\ssm\{(0,0)\})\subset\IC^2$ that results in the quotient $\IP^1$ includes points that are infinitesimally near the other region, $(0,0)\in\IC^2$, the resulting quotients $\IP^1$ and $(0,0)$ are well separated, and the 
$(z_1,z_2)\in\IC^2$ space has been partitioned into two $U(1;\IC)$-orbits~\eqref{e:P1+.}: one (complex) one-dimensional, the other zero-dimensional.

In general, we denote by ${\bf E}_{\cdots}$ the fixed-point locus of a $\IC^*$-transformation, such as in~\mbox{\eqref{e:C*C2}--\eqref{e:P1+.}.}

\paragraph{The {$\FF{m}$} sequence:}
In the showcase models~\eqref{e:nFmFan}, the gauge symmetries $U_a(1)$ act with charges $q_a$ given as in~\eqref{e:q1-4} on (complex) chiral superfields, and so equally on their lowest, scalar component fields, $X_0,X_1,\cdots X_{n+2}$:
\begin{small}
\begin{alignat}9
  U_1(1;\IC)\!:~
  &(\l_1^{-n}X_0,\,\l_1X_1,\,\l_1X_2,\, \dots\ \l_1X_n\,;\;
     \underline{X_{n+1}},\underline{X_{n+2}}),
   \label{e:Q1}\mkern60mu\\*
  U_2(1;\IC)\!:~
  &(\l_2^{m-2}X_0,\,\l_2^{-m}X_1,\,\underline{X_2},\, \dots\ 
     \underline{X_n}\,;\; \l_2X_{n+1},\,\l_2X_{n+2}),
 \label{e:Q2}\\
  U_3(1;\IC)\!:~
  &(\l_3^{-[(n-1)m+2]}X_0,\,\underline{X_1},\,\l_3^mX_2,\, \cdots\ 
     \l_3^mX_n\,;\; \l_3X_{n+1},\,\l_3X_{n+2}),
 \label{e:Q3}\\*
  U_4(1;\IC)\!:~
  &(\underline{X_0},\,\l_4^{-[(n-1)m+2]}X_1,\,\l_4^{m-2}X_2,\, \cdots\ 
     \l_4^{m-2}X_n\,;\; \l_4^nX_{n+1},\,\l_4^nX_{n+2}),
 \label{e:Q4}
\end{alignat}
\end{small}%
where the $U_a(1)$-neutral fields have been underlined: their nonzero {\em\/vacuum expectation values\/} (abbreviated {\em\/vev\/}s) leave that particular $U_{{a}}(1)$ unbroken; these vevs specify the different choices of a ``classical background'' and so also the broken-vs.-unbroken gauge symmetry in those ``phases.''

Extending the example~\eqref{e:C*C2}--\eqref{e:P1+.}, these (toric) actions define special subsets of the field space by separating regions that are fixed by these $\IC^*$-transformations from the complement that is being $\IC^*$-rescaled. The tabulation of charges~\eqref{e:q1-4} and the complexified (toric) actions~\eqref{e:Q1}--\eqref{e:Q4} make it evident that the field space should be regarded as the tensor product of the factors fixed by the indicated rank-1 subgroups of the $U(1;\IC)^2$ gauge group:
\begin{equation}
\begin{array}{@{}r|c|c|c|c@{}}
 \textbf{Factor} & \IC^1_0\!=\!\{X_0\} &\IC^1_1\!=\!\{X_1\}
  &\IC^{n-1}_2\!=\!\{X_2,\cdots X_n\} &\IC^2_3\!=\!\{X_{n+1},X_{n+2}\}\\*[2pt]\toprule
 \textbf{Fixed by} & U_4(1;\IC) & U_3(1;\IC) &U_2(1;\IC) & U_1(1;\IC)\\ \bottomrule
 \textbf{Excep.\ set} & {\bf E}_0=(0) & {\bf E}_1=(0)
  &{\bf E}_2=(0,\cdots 0) & {\bf E}_3=(0,0)\\
\end{array}
 \label{e:theEs}
\end{equation}
Underneath each factor are listed also the ``exceptional sets''\cite[p.\,207]{rCLS-TV}, each of which is actually fixed (as a subset of its factor, 
${\bf E}_I\subset\IC^{d_I}_I$, indicated at the top of its column in~\eqref{e:theEs}) by all of $U(1;\IC)^2$.
 Indeed, the only location in the full field space, $\IC^{n+3}_{0123}$, that is fixed (left invariant) by the full $U(1;\IC)^2$ gauge symmetry is the field-space origin:
{%
\begin{alignat}9
 {\bf E}_{0123}
 &:=\{X_0,\, X_1,\, X_2,\, \cdots\ X_n,\, X_{n+1},\, X_{n+2} = 0\}
     \subset\IC^{n+3}_{0123}. \label{e:E0123}
\intertext{Analogously,}
 {\bf E}_{12}
 &:=\{X_1,\, X_2,\, \cdots\ X_n = 0\}
     \subset\IC^{n}_{12}, \label{e:E12}\\
 {\bf E}_{23}
 &:=\{X_2,\, \cdots\ X_n,\, X_{n+1},\, X_{n+2} = 0\}
     \subset\IC^{n+1}_{23}, \label{e:E23}
\end{alignat}
and so on.}

\paragraph{Gauge-equivalence strata:}
({\bf I})~Omitting $X_0$, we see that, for example:
\begin{equation}
 \Big(\underbrace{\big((\IC^n_{12}\ssm{\bf E}_{12})/U_1(1;\IC)\big)}
      _{=\,\IP^{n-1}_\text{fiber}}\times
      \big(\IC^2_3\ssm{\bf E}_3\big)\Big)\Big/U_2(1;\IC)
  \simeq \FF{m}
 \label{e:nFmQuot}
\end{equation}
where $U_2(1;\IC)$ projectivizes $\IC^2_3\ssm{\bf E}_3\to\IP^1_\text{base}$ while also $m$-twisting\,\footnote{\label{n:mTw}{The}
 $U_2(1)$-transformation is trivial on $\IP^1_{\sss\text{base}}\ni(X_{n+1},X_{n+2})\simeq(\l\,X_{n+1},\l\,X_{n+2})$ but it is a non-trivial coordinate reparametrization on $\IP^{n-1}_{\sss\text{fiber}}\ni(X_1,X_2,\cdots X_n)\not\simeq(\l_2^{-m}\,X_1,X_2,\cdots X_n)$.}
 $\IP^{n-1}_\text{fiber}$. This then describes an $m$-twisted $\IP^{n-1}_\text{fiber}$-bundle over $\IP^1_\text{base}$, the Hirzebruch $n$-fold scroll, $\FF{m}$.

({\bf II})~Alternatively, we also have:
\begin{equation}
 \Big((\IC^1_1\ssm{\bf E}_1)\times
      \underbrace{\big((\IC^{n+1}_{23}\ssm{\bf E}_{23})/U_3(1;\IC)\big)}
      _{=\,\IP^n_{(m:\cdots:m:1:1)}}\Big)\Big/U_2(1;\IC),
 \label{e:buWCP}
\end{equation}
which describes a blowup of $\IP^n_{(m:\cdots:m:1:1)}$ along 
$(X_2,\cdots X_n;\,0,0)$ and is parametrized by $X_1$. 
This co-dimension-2 linear subspace is fixed by the discrete subgroup $\ZZ_m\subset U_3(1;\IC)$, which may be seen as follows. The
linear subset
\begin{equation}
  S=\big((X_2,\cdots X_n;0,0)/U_3(1;\IC)\big)\subset\IP^n_{(m:{\cdots}:m:1:1)}
 \label{e:sWCP}
\end{equation}
is fixed by the $\ZZ_m\subset U_3(1;\IC)$ 
\begin{equation}
  U_3(1;\IC)\supset\ZZ_m\!:~(\l_3^m\,X_2,\cdots \l_3^m\,X_n\,;\, 0,0),\quad
  \l_3^m\<=1,
\end{equation}
and it is a $\ZZ_m$-singular locus in $\IP^n_{(m:{\cdots}:m:1:1)}$. Subsequently, the $U_2(1;\IC)$-transformation separates
\begin{equation}
  \big(\IC^1_1\<=\{X_1\<\in\IC\}\big)\times S \to
  \big(\{X_1\<=0\}\times S\big) ~\sqcup~
  \big(\{X_1\<\neq0\}\times S\big)
 \label{e:F1:F1c}
\end{equation}
as the trivial and non-trivial $U_2(1;\IC)$-orbits, respectively. The former of these leaves the $X$-space $\ZZ_m$-singular at $S$ and identified as the singular weighted projective space $\IP^n_{(m:{\cdots}:m:1:1)}$; the latter provides its MPCP-desingularization\cite{Batyrev:1993oya} along $S$.

Since $U_1(1;\IC)\times U_2(1;\IC) \approx U_3(1;\IC)\times U_2(1;\IC)$, the iterated quotient~\eqref{e:buWCP} is isomorphic to~\eqref{e:nFmQuot}, thus giving the Hirzebruch $n$-fold $\FF{m}$ one more alternative description as the MPCP-desingularization of the weighted{-projective space,} $\IP^n_{\!\sss(m:{\cdots}:m:1:1)}$.

\begin{figure}[htb]
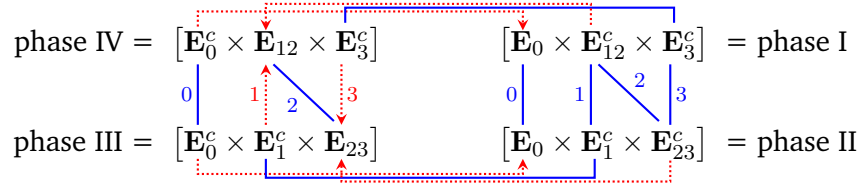

$$
 \begin{array}{c@{\qquad\qquad}c}
   \makebox[0pt][r]{phase~IV~=~~}
   \big[{\bf E}^c_0\times{\bf E}_{12}\times{\bf E}^c_3\big]
  &\big[{\bf E}_0\times{\bf E}^c_{12}\times{\bf E}^c_3\big]\makebox[0pt][l]{~~=~phase~I}\\*[8mm]
  \TikZ{\path[use as bounding box](0,0);
    \draw[thick,blue](.35,.35)tonode[left=-2pt]{$\SSS0$}++(0,.8);
    \draw[thick,blue](1.35,1.15)--node[below left=-2pt]{$\SSS2$}++(.8,-.75);
    \draw[thick,blue](4.65,.35)--node[left=-2pt]{$\SSS0$}++(0,.8);
    \draw[thick,blue](5.55,.35)--node[left=-2pt]{$\SSS1$}++(0,.8);
    \draw[thick,blue](5.65,1.15)--node[above right=-2pt]{$\SSS2$}++(.8,-.75);
    \draw[thick,blue](6.6,.35)--node[right=-2pt]{$\SSS3$}++(0,.8);
    \draw[thick,red,densely dotted,-stealth]
       (.35,1.65)to++(0,.2)to++(4.3,0)to++(0,-.25);
    \draw[thick,blue](2.3,1.7)to++(0,.2)to++(4.35,0)to++(0,-.2);
    \draw[thick,red,densely dotted, stealth-]
       (1.25,1.6)to++(0,.35)to++(4.3,0)to++(0,-.3);
    \draw[thick,red,densely dotted,-stealth](1.25,.35)
       --node[left=-2pt]{$\SSS1$}++(0,.8);
    \draw[thick,red,densely dotted,-stealth](2.25,1.15)
       --node[right=-2pt]{$\SSS3$}++(0,-.8);
    \draw[thick,red,densely dotted,-stealth]
       (.35,-.1)to++(0,-.2)to++(4.3,0)to++(0,.2);
    \draw[thick,blue](1.25,-.1)to++(0,-.25)to++(4.35,0)to++(0,.25);
    \draw[thick,red,densely dotted, stealth-]
       (2.25,-.1)to++(0,-.3)to++(4.35,0)to++(0,.3);
            }
  \makebox[0pt][r]{phase~III~=~~}
  \big[{\bf E}^c_0\times{\bf E}^c_1\times{\bf E}_{23}\big]
  &\big[{\bf E}_0\times{\bf E}^c_1\times{\bf E}^c_{23}\big]\makebox[0pt][l]{~~=~phase~II} \\[1mm]
 \end{array}\rule[-12mm]{0mm}{27mm}
$$
 \caption{A schematic diagram relating the available $U(1;\IC)^2$ gauge equivalence class choices in the field space specified in~\eqref{e:q1-4}}
 \label{f:EvsP}
\end{figure}

The above two choices, {\bf I}~\eqref{e:nFmQuot} and {\bf II}~\eqref{e:buWCP}, correspond to the two ``geometric'' phases.
 Replacing the complements $(\IC^d_{\cdots}\ssm{\bf E}_{\cdots})$ of the fixed-loci with the fixed loci ${\bf E}_{\cdots}$ themselves, and combinatorially in the various factors of~\eqref{e:nFmQuot} and~\eqref{e:buWCP}, gives the ``complementary'' strata in the field space:
\begin{alignat}9
\textbf{III}:&~~&
[F_m^{\sss(n)}]\strut_{\text{LGO}}:~
 &\Big((\IC^n_1\ssm{\bf E}_1)\times
         \big(({\bf E}_{23})/U_3(1;\IC)\big)\Big)\Big/U_2(1;\IC),\\
\textbf{IV}:&~~&
[F_m^{\sss(n)}]\strut_{\text{hyb.}}:~
 &\Big(\big[({\bf E}_{12})/U_1(1;\IC)\big] \times
      \big(\IC^2_3\ssm{\bf E}_3\big)\Big)\Big/U_2(1;\IC).
\end{alignat}
Writing ${\bf E}^c_{\cdots}:=(\IC_{\cdots}\ssm{\bf E}_{\cdots})$ for the complement of the exceptional set, the (complexified) gauge symmetries reduce the field-space to the $U(1;\IC)^2$-equivalence classes in the following four field-space regions, as shown in Figure~\ref{f:EvsP},
where the $\IC^1_0$ factor is now also included.
 The (red) dotted arrows indicate the (dimensional) collapse ${\bf E}^c_i\to{\bf E}_i$ from the complement of the (gauge-fixed) exceptional set to the exceptional set itself. The plain (blue) solid lines trace unchanged factors in the tensor product.

The diagram in Figure~\ref{f:EvsP}
presents the field space
$(X_0,X_1,\cdots X_{n+2})\in\IC^{n+3}/U(1)^2$ decomposed into four distinct regions, labeled as ``phases'' (adopting the by-now-standard nomenclature\cite{rPhases}). From~\eqref{e:theEs} and the diagram in Figure~\ref{f:EvsP},
we have:
\begin{equation}
\begin{array}[t]{@{~}r|cccc@{~}}
                 I &0 &1 &2     &3 \\ \toprule
 \dim({\bf E}_I)   &0 &0 &0     &0 \\
 \dim({\bf E}_I^c) &1 &1 &n{-}1 &2 \\
\end{array}
\quad\Rightarrow\quad
\begin{array}[t]{@{~}r|cccc@{~}}
 \text{phase} &\text{I} &\text{II} &\text{III} &\text{IV} \\ \toprule
 \dim         &n &n &0     &1 \\
\end{array}
\end{equation}
where quotienting by $U(1)^2$ reduces the dimension-count to those tabulated at right.

As in Ref.\cite{rPhases}, the variable $X_0$ is seen to provide a fiber-coordinate for a degree-$\pM{n\\2{-}m}$ line bundle over $\FF{m}$, which is set to zero in phases~I and~II, these describing the Calabi--Yau hypersurface in the base space 
$\FF{m}[c_1]$. In phase~III, the base of the bundle collapses to ${\bf E}_{23}=\{\text{pt.}\}$ and $|X_1|=\sqrt{\frac{(2{-}m)r_1{-}nr_2}{(n{-}1)m{+}2}}$, while 
$|X_0|=\sqrt{\frac{-mr_1{-}r_2}{(n{-}1)m{+}2}}$. Finally, in phase~IV the bundle collapses to ${\bf E}_{12}\times\IP^1_{\sss\text{base}}$ and 
$|X_0|=\sqrt{-r_1/n}$.
\begin{remk}\label{r:phases}
Complementing the analysis in Ref.\cite{rBH-gB}, the above-detailed $U(1)^2$ gauge-orbit separation (stratification) shows that, far from a simple 
$(X_0,X_1,\cdots X_5)\in\IC^6$, the field space has a{n} intricate structure: this rather straightforward if detailed QFT reasoning thus fully recovers the complex-algebraic toric geometry-standard Gelfand--Kapranov--Zelevinsky (GKZ) decomposition\cite{Oda:1991aa} {for $\cKs{\FF{m}}$}; see below, and compare also with Refs.\cite{rPhases, rAGM01, rAGM06, rAGM04, rMP0, rBH-gB}.
\end{remk}

\subsection{VEVs and Phases}
\label{s:VEVs}
Having stratified the field space into uniform $U(1)^2$-orbits, consider now the possible choices of ground-state vevs, which maintain the vanishing of the potential~\eqref{e:U}. In particular, with the $(q_1,q_2)$-basis in~\eqref{e:q1-4} the vanishing of the ``$D$-terms'' implies:
\begin{alignat}9
& \Big({-}n|X_0|^2{+} \sum_{i=1}^n|X_i|^2 -t_1\Big)
 \overset!=0, \label{e:VEVD1}\\
& \Big((m{-}2)|X_0|^2 {-}m |X_1|^2+\sum_{j=1}^2|X_{n+j}|^2 -t_2\Big)
 \overset!=0, \label{e:VEVD2}
\end{alignat}
where ``$\overset!=$'' denotes a required equality. Note that these equations are independent of the choice of the superpotential, and so in fact describe the ambient space $\FF{m}$ itself.

Depending on the values of the so-called ``Fayet--Iliopoulos'' couplings, $(t_1,t_2)$, the following options emerge (following\cite{rMP0, rBH-gB}), which are fully consistent with~\eqref{e:Q1}--\eqref{e:Q4} above:
\begin{enumerate}[labelsep=2pt, label=(\emph{\roman*})]
 \item
 $U_1(1)$ is preserved if $X_{n+1},X_{n+2}\neq0$ but $X_0=0=X_i$ for 
 $i=1,\ldots n$.
 Since $\mathrm{LCM}[q_2({X_{n+1},X_{n+2}})]=1$, $U_2(1)$ is broken 
 completely. 
 In this case,~\eqref{e:VEVD1} and~\eqref{e:VEVD2} imply: $t_1=0$ and $t_2\geqslant 0$;
 this is the positive $(0,1)$-direction in the $(t_1,t_2)$-plane.
 \item
 $U_2(1)$ is preserved if $X_2,\cdots X_n\neq0$ but 
 $X_0=0=X_1=X_{n+1}=X_{n+2}$.
 Since $\mathrm{LCM}[q_1({X_2,\cdots X_n})]=1$, 
 $U_1(1)$ is broken completely.
 Now~\eqref{e:VEVD2} and~\eqref{e:VEVD1} imply: $t_2=0$ and $t_1\geqslant 0$
 along the positive $(1,0)$-direction.
 \item
 $U_3(1)$ is preserved if only $X_1\<\neq0$.
 Since $q_1(X_1)\<=1$, $U_1(1)$ is broken completely, but
 $q_2(X_1)\<={-}m$ implies $U_2(1)\<\to\ZZ_m$, and
 $q_4(X_1)\<={-}(n{-}1)m{-}2$ implies $U_4(1)\to\ZZ_{(n{-}1)m{+}2}$.
 Now~\eqref{e:VEVD2} and~\eqref{e:VEVD1} imply: $m\,t_1{+}\,t_2=0$
 along the positive $(1,-m)$-direction.
 \item
 $U_4(1)$ is preserved if only $X_0\,{\neq}\,0$.
 Since $q_1(X_0)=-n$, $U_1(1)\to\ZZ_n$,
 $q_2(X_0)=m{-}2$ implies $U_2(1)\to\ZZ_{m{-}2}$ and
 $q_3(X_0)=-(n{-}1)m{-}2$ implies $U_3(1)\to\ZZ_{(n{-}1)m{+}2}$.
 Now~\eqref{e:VEVD2} and~\eqref{e:VEVD1} imply: $(m{-}2)t_1{+}n\,t_2=0$
 along the positive $\big({-}n,(m{-}2)\big)$-direction.
\end{enumerate}

Summarizing, we have found demarcation rays in the Fayet--Iliopoulos $(t_1,t_2)$-space generated by:
\begin{equation}
\begin{array}{@{}r|cccc@{}}
  &(\textit{i\/}) &(\textit{ii\/}) &(\textit{iii\/}) &(\textit{iv\/}) \\ \toprule
 \pM{t_1\\[-2pt]t_2} &\pM{0\\1} &\pM{1\\0} &\pM{~~1\\-m} &\pM{-n\\m-2} \\ \bottomrule
\end{array}
\end{equation}
which are {\em\/precisely\/} the distinct (vertically read) 2-vectors appearing in the $(q_1,q_2)$-rows in the tabulation~\eqref{e:q1-4}! The preceding $U(1)^2$-stratification in Figure~\ref{f:EvsP} thus {\em\/precisely\/} reproduces the vev-analysis\cite{rPhases, rAGM01, rAGM06, rAGM04, rMP0, rBH-gB} and the phase-diagram in Figure~\ref{f:PhDiag}, also known as the GKZ decomposition\cite{Oda:1991aa} {for $\cKs{\FF{m}}$, since $X_0$ is included}.
\begin{figure}[htb]
$$
\hspace{-28pt}
\vC{\TikZ{[very thick, scale=.75]
      \path[use as bounding box](-3,-3.3)--(3.8,2.3);
      \corner{(0,0)}{0}{90}{2}{green};
      \corner{(0,0)}{90}{{90+atan(3/1)}}{2}{red};
      \corner{(0,0)}{{90+atan(3/1)}}{{270+atan(1/3)}}{2}{blue};
      \corner{(0,0)}{{270+atan(1/3)}}{360}{2}{yellow};
      \draw[-stealth](0,0)--(1,0)node[above]{\scriptsize$(1,0)$};
       \path(2,0)node{(\textit{ii})};
      \draw[-stealth](0,0)--(0,1)node[right]{\scriptsize$(0,1)$};
       \path(0,1.8)node{(\textit{i})};
      \draw[-stealth](0,0)--(-3,1)node[above]{\scriptsize$(-n,m-2)$};
       \draw[thin, -stealth](-2,.667)to[out=-220, in=0]++(-.5,-.2)
        node[left=-2pt]{(\textit{iv})};
      \draw[-stealth](0,0)--(1,-3)node[above right]{\scriptsize$(1,-m)$};
       \draw[thin, -stealth](.667,-2)to[out=210, in=0]++(-.5,-.2)
        node[left=-2pt]{(\textit{iii})};
      \path(45:2)node{\Large\bf I};
      \path(-40:2)node{\Large\bf II};
      \path(225:2)node{\Large\bf III};
      \path(130:2)node{\Large\bf IV};
      \filldraw[fill=white](0,0)circle(1mm);
            }}
\parbox[c]{70mm}{\raggedright\noindent
$\bullet$~Only the ({\it iv}) demarcation depends on $n=\dim(\FF{m})$.\\[2mm]
$\bullet$~Only the ({\it iii}) and ({\it iv}) demarcations depend on the
 ``twist,'' $m$.\\[2mm]
$\bullet$~The ({\it ii}) and ({\it iii}) demarcations coincide for $m=0$,
 removing phase~II.
}
$$
 \caption{The $\FF{m}[c_1]$ GLSM ``phase diagram,'' plotted here for $n=3$ and $m=3$.}
 \label{f:PhDiag}
\end{figure}
With the specified demarcations, Equations~\eqref{e:VEVD1} and \eqref{e:VEVD2} determine the actual values of the nonzero vevs (omitting the usual 
$\vev{{\cdots}}$ notation and correcting\cite{rBH-gB} (Figure~1)); see Table~\ref{t:PhValues},
where two of the vevs, in phases~II and~VI, are:
\begin{equation}\textstyle
   |X_1|_{\rm II}=\sqrt{t_1-\sum_{i=2}^n|X_i|^2}~\overset!{\geqslant}0,\qquad
   |X_0|_{\rm IV}=\sqrt{\frac{|X_{n+1}|^2+|X_{n+2}|^2-t_1}{n}}.
 \label{e:|x1|}
\end{equation}
\begin{table}[htb]
\caption{The vacuum expectation (background) values of the various fields in the various phases of the $\FF{m}[c_1]$ GLSM\cite{rBH-gB}.}
\centering\vspace*{2mm}
\begin{tabular}{r@{\qquad}ccccccc}
     &\boldmath{$|X_0|$}&\boldmath{$|X_1|$}&\boldmath{$|X_2|$}&\boldmath{$\cdots$}&\boldmath{$|X_n|$}&\boldmath{$|X_{n+1}|$}&\boldmath{$|X_{n+2}|$} \\ 
\toprule
     \textbf{{{\itshape i}}\,}
                  & 0 & 0 & 0 &$\cdots$& 0 & * &  * \\
     \textbf{I\,}
                  & 0 & * & * &$\cdots$& * & * &  * \\
     \textbf{{\itshape ii}\,}
                  & 0 & 0 & * &$\cdots$& * & 0 &  0 \\
     \textbf{II\,}
                  & 0 & \text{see~\eqref{e:|x1|}} & * &$\cdots$& * & * &  * \\
     \textbf{{\itshape iii}\,}
                  & 0  & $\sqrt{t_1}$
                              & 0 &$\cdots$& 0 & 0 &  0 \\[1mm]
     \textbf{III\,}
                  & $\sqrt{\frac{-mt_1{-}t_2}{(n{-}1)m{+}2}}$
                        & $\sqrt{\frac{(2{-}m)t_1{-}nt_2}{(n{-}1)m{+}2}}$
                              & 0 &$\cdots$& 0 & 0 &  0 \\[1mm]
     \textbf{{\itshape iv}\,}
                  & $\sqrt{-t_1/n}$
                        & 0 & 0 &$\cdots$& 0 & 0 &  0 \\[1mm]
     \textbf{IV\,}
                  & \text{see~\eqref{e:|x1|}}
                        & 0 & 0 &$\cdots$& 0 & * &  * \\ 
                        \bottomrule
 \MC8l{\footnotesize{``\,*\,'' denotes nonzero values; all vevs vanish at  
 $(t_1,t_2)\,\!=\!\,(0,0)$.}}
                        \end{tabular}
\label{t:PhValues}
\end{table}%
The vevs change continuously throughout the $(t_1,t_2)$-plane{, and so does the physical system modeled by the GLSM}.
The particular $q_a$-choices in~\eqref{e:q1-4} are thus seen as particularly convenient for determining the $U(1;\IC)^2$ gauge-symmetry breaking pattern throughout the phase diagram in Figure~\ref{f:PhDiag}, as induced by the values provided in Table~\ref{t:PhValues}.
Since $q_3$ and $q_4$ are linear combinations of $q_1$ and $q_2$, the above analysis can be re-done using any of the $\binom42=6$ bases of any two of $q_1,\cdots q_4$, resulting in the same phase diagram, up to a $\GL(2;\ZZ)$-transformation. Quite literally then, all $\binom42=6$ $q_a$-bases are physically equivalent, whereas built-in positivity requirements in complex-algebraic toric geometry single out the Mori vectors $(q_1,q_2)${; see Remark~\ref{r:Mori+}}.

\begin{figure}[htb]
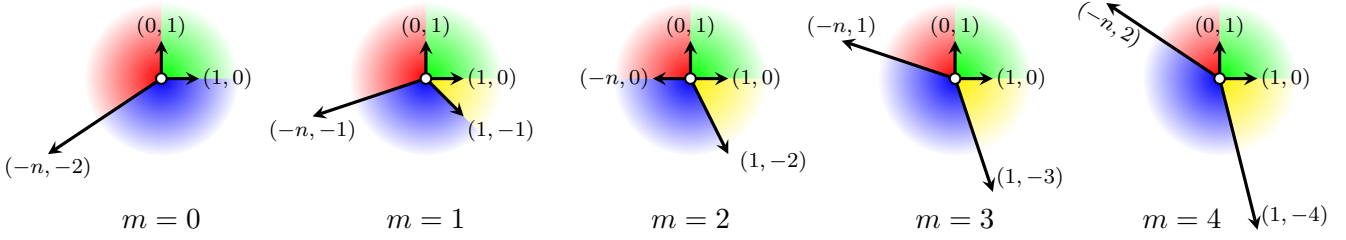

$$
\hspace{-25pt}
\vC{\TikZ{[thick, scale=.5]
      \path[use as bounding box](-3,-4)--(28,2.3);
      \corner{(0,0)}{0}{90}{2}{green};
      \corner{(0,0)}{90}{{180+atan(2/3)}}{2}{red};
      \corner{(0,0)}{{180+atan(2/3)}}{360}{2}{blue};
      \draw[very thick, -stealth](0,0)--(1,0)node[right=-3pt]
          {\scriptsize$(1,0)$};
      \draw[very thick, -stealth](0,0)--(0,1)node[above=-3pt]
          {\scriptsize$(0,1)$};
      \draw[very thick, -stealth](0,0)--(-3,-2)node[below=-3pt]
          {\scriptsize$(-n,-2)$};
      \path(0,-3.7)node{$m=0$};
      \filldraw[fill=white](0,0)circle(4pt);
      \begin{scope}[xshift=7cm]
      \corner{(0,0)}{0}{90}{2}{green};
      \corner{(0,0)}{90}{{180+atan(1/3)}}{2}{red};
      \corner{(0,0)}{{180+atan(1/3)}}{315}{2}{blue};
      \corner{(0,0)}{315}{360}{2}{yellow};
      \draw[very thick, -stealth](0,0)--(1,0)node[right=-3pt]
          {\scriptsize$(1,0)$};
      \draw[very thick, -stealth](0,0)--(0,1)node[above=-3pt]
          {\scriptsize$(0,1)$};
      \draw[very thick, -stealth](0,0)--(-3,-1)node[below=-3pt]
          {\scriptsize$(-n,-1)$};
      \draw[very thick, -stealth](0,0)--(1,-1)node[below right=-3pt]
          {\scriptsize$(1,-1)$};
      \path(0,-3.7)node{$m=1$};
      \filldraw[fill=white](0,0)circle(4pt);
      \end{scope}
      \begin{scope}[xshift=14cm]
      \corner{(0,0)}{0}{90}{2}{green};
      \corner{(0,0)}{90}{180}{2}{red};
      \corner{(0,0)}{180}{{270+atan(1/2)}}{2}{blue};
      \corner{(0,0)}{{270+atan(1/2)}}{360}{2}{yellow};
      \draw[very thick, -stealth](0,0)--(1,0)node[right=-3pt]
          {\scriptsize$(1,0)$};
      \draw[very thick, -stealth](0,0)--(0,1)node[above=-3pt]
          {\scriptsize$(0,1)$};
      \draw[very thick, -stealth](0,0)--(-1,0)node[left=-3pt]
          {\scriptsize$(-n,0)$};
      \draw[very thick, -stealth](0,0)--(1,-2)node[right, yshift=-3pt]
          {\scriptsize$(1,-2)$};
      \path(0,-3.7)node{$m=2$};
      \filldraw[fill=white](0,0)circle(4pt);
      \end{scope}
      \begin{scope}[xshift=21cm]
      \corner{(0,0)}{0}{90}{2}{green};
      \corner{(0,0)}{90}{{90+atan(3/1)}}{2}{red};
      \corner{(0,0)}{{90+atan(3/1)}}{{270+atan(1/3)}}{2}{blue};
      \corner{(0,0)}{{270+atan(1/3)}}{360}{2}{yellow};
      \draw[very thick, -stealth](0,0)--(1,0)node[right=-3pt]
          {\scriptsize$(1,0)$};
      \draw[very thick, -stealth](0,0)--(0,1)node[above=-3pt]
          {\scriptsize$(0,1)$};
      \draw[very thick, -stealth](0,0)--(-3,1)node[above=-3pt]
          {\scriptsize$(-n,1)$};
      \draw[very thick, -stealth](0,0)--(1,-3)node[above right=-3pt]
          {\scriptsize$(1,-3)$};
      \path(0,-3.7)node{$m=3$};
      \filldraw[fill=white](0,0)circle(4pt);
      \end{scope}
      \begin{scope}[xshift=28cm]
      \corner{(0,0)}{0}{90}{2}{green};
      \corner{(0,0)}{90}{{90+atan(3/2)}}{2}{red};
      \corner{(0,0)}{{90+atan(3/2)}}{{270+atan(1/4)}}{2}{blue};
      \corner{(0,0)}{{270+atan(1/4)}}{360}{2}{yellow};
      \draw[very thick, -stealth](0,0)--(1,0)node[right=-3pt]
          {\scriptsize$(1,0)$};
      \draw[very thick, -stealth](0,0)--(0,1)node[above=-3pt]
          {\scriptsize$(0,1)$};
      \draw[very thick, -stealth](0,0)--(-3,2)node
          [below=-2pt, rotate=-30, xshift=6pt]{\scriptsize$(-n,2)$};
      \draw[very thick, -stealth](0,0)--(1,-4)node[above right=-3pt]
          {\scriptsize$(1,-4)$};
      \path(-1,-3.7)node{$m=4$};
      \filldraw[fill=white](0,0)circle(4pt);
      \end{scope}
            }}
$$
 \caption{The phase diagrams of the first {five} $\FF[3]m$-GLSMs. 
  Including $q_a(X_0)$, they also depict the anticanonical bundles 
  $\mathcal{K}^*_{\smash{\FF{m}}}$, sections of which define 
  Ricci-flat hypersurfaces as their zero locus.}
 \label{f:3F0-4}
\end{figure}

The infinite sequence ($m=0,1,2,\dots$) of Hirzebruch surface (complex two-dimensional) scrolls, $\FF[2]m$, is well-known to provide only two diffeomorphism classes\cite{rH-Fm, rGrHa, rKC-Fm}: the even-twisted $\FF[2]{2k}$ are all diffeomorphic to each other (``same, as smooth real manifolds''), as are the odd-twisted $\FF[2]{2k+1}$. We may write 
$\FF[2]{m}\<{\approx_{\sss\IR}}\FF[2]{m\,\smash{(\mathrm{mod}\,2)}}$, 
and we find that
$\FF{m}\<{\approx_{\sss\IR}}\FF{m\,\smash{(\mathrm{mod}\,n)}}$ analogously\cite{rBH-gB}. In turn, the $\FF{m}$ for differing $m$ are all distinct {\em\/as complex manifolds,} and this is clearly seen from the phase diagrams in Figure~\ref{f:PhDiag}, reproduced in Figure~\ref{f:3F0-4} for $n=3$ and for $m=0,1,\cdots 4$. The sequence of Hirzebruch scrolls is infinite, with the 
$\IP^{n-1}\!\hookrightarrow\!\FF{m}\!\twoheadrightarrow\!\IP^1$
bundle twist, $m$, unbounded. In fact, as the next section shows, there is an additional diversity for $n\geqslant3$ that is growing combinatorially with $m$.

The $m$-dependent characteristic that unequivocally distinguishes the various Hirzebruch scrolls, $\FF{m}$, is their hallmark {\em\/exceptional divisor\/} (complex co-dimension-1 irreducible {submanifold}) 
$\sS\!\subset_\IC\!\FF{m}$, called the {\em\/directrix\/}\cite{rGrHa}, which is identified by its maximally negative self-intersection:
$[\sS]^n={-}(n{-}1)m$. 
The Ricci-flat (Calabi--Yau) hypersurfaces 
$\sX\subset\FF{m}$ themselves being of complex co-dimension-1 therefore intersect this $\sS$:
\begin{equation}
  \dim\big(\sX\subset\FF{m}\big)=n{-}1, \quad\Rightarrow\quad
  \dim\big((\sX\cap\sS)\subset\FF{m}\big)=n{-}2,
\end{equation}
and so ``inherit'' the distinguishing features stemming from $\sS\subset\FF{m}$; those loci naturally turn up in $H_{n-2}(\sX,\ZZ)$, and so provide a relevant distinction in stringy applications.

In contradistinction, the phase diagrams in Figure~\ref{f:3F0-4} quite clearly {\em\/do not\/} exhibit the diffeomorphism 3-cycle, $\FF[3]m\approx_{\sss\IR}\FF[3]{m\,\smash{{(}\mathrm{mod}\,3{)}}}$. Since these diagrams are defined only up to $\GL(2;\ZZ)$ basis transformations, one may wonder if perhaps their acyclic nature is a mirage. That this is not the case should be evident on comparing the $m=0$ and $m=3$ diagrams: the two cannot possibly be $\GL(2;\ZZ)$-equivalent, since the former lacks phase~II.
It may be a little less evident that this in fact persists for any pair, 
$\FF[2]m,\FF[2]{m'}$ with $m'\<\neq m$. Requiring the candidate $\GL(2;\ZZ)$-transformation $\boldsymbol{g}$ to map
$\{\pM{1\\0},\pM{0\\1},\cdots\}_m \to \{\pM{1\\0},\pM{0\\1},\cdots\}_{m'}$ 
fixes $\boldsymbol{g}\<=\Ione$, which then cannot possibly map the remaining two demarcations,
$\{\pM{~~1\\-m},\pM{-n\\2-m}\} \not\to 
 \{\pM{~~1\\[-1pt]-m'},\pM{-n\\[-1pt]2-m'}\}$.
This exhibits the inequivalence of $\FF{m}[c_1]$ GLSMs even at this semi-classical stage\cite{rPhases}!
These differences only increase {(and highly nontrivially)} upon including also the cumulative worldsheet instanton effects, following\cite{rMP0}.\footnote{I am indebted to Per Berglund for collaboration on this result, which will be reported elsewhere{, as well as extensive and helpful discussions on the analysis presented here.}}

However, the phase diagrams (also known as {\em\/secondary fans,} $\pFn[\,\prime\prime]{\sX}$) partition the Fayet--Iliopoulos $t_a$-space (parametrizing the $U(1;\IC)^r$ gauge symmetry {and the ``fully enlarged K\"ahler moduli space''\cite{rAGM01, rAGM06, rAGM04}}), and so they parametrize not the complex structure but the (complexified) {\em\/K{\"a}hler structure\/}\cite{rPhases, rAGM01, rAGM06, rAGM04, rMP0} of the ambient scrolls, $\FF{m}$, and thus (properly restricted) also of the Ricci-flat hypersurfaces therein. By including the $q_a(X_0)=\pM{-n\\m-2}$ demarcation, they also refer to (the fiber of) the anticanonical class, $\cKs{\FF{m}}$, and so they pertain to the deformation family {of} anticanonical hypersurfaces, $\FF{m}[c_1]$.
\begin{corl}\label{C:S''m}
The $\pFn[\,\prime\prime]{\FF{m}}\<{\not\approx}\pFn[\,\prime\prime]{\FF{m'}}$ differences exhibited in Figure~\ref{f:3F0-4} characterize (the GLSM dependence on) the K{\"a}hler structure of the Hirzebruch scrolls, their anticanonical bundles, and so also their Ricci-flat (anticanonical, Calabi--Yau) hypersurfaces throughout
$\FF{m}[c_1]$.
\end{corl}
{%
\begin{corl}\label{C:LG/CY}
The continuous connections between the distinct ``phases''\cite{rPhases, rAGM01, rAGM06, rAGM04, rMP0}, as reconstructed and analyzed in 
Sections~\ref{s:GO} and \ref{s:VEVs} and shown in Figures~\ref{f:PhDiag} and~\ref{f:3F0-4}, also relate
the transposition-mirror construction\cite{rBH, rBH-LGO+EG, Krawitz:2010FJR} (proven in Refs.\cite{rC+R-MirrBH, rLB-MirrBH}), originally formulated in the LGO framework, and
Batyrev's toric mirror duality\cite{Batyrev:1993oya, rLB-MirrBH, rBatyBor3}. The two formulations overlap\cite{Candelas:1994bu} but have diverse applications and provide different insights. Recall that in the stringy-complexified $t_a$-space (induced by including ``$B$-fields,'' often related to geometric torsion) the demarcation rays as in \mbox{Figures~\ref{f:PhDiag} and~\ref{f:3F0-4}} in fact have real co-dimension 2 and present no obstruction to this connectivity\cite{rPhases}.
\end{corl}%
}

\section{Deformation Family of Ambient Spaces}
\label{s:AmbiDefo}
The study of Calabi--Yau hypersurfaces in Hirzebruch scrolls benefits greatly from being able to provide a concrete realization of the hallmark exceptional divisor. This uses the novel technique of generalized hypersurfaces and their complete \mbox{intersections\cite{rgCICY1, rBH-Fm, rGG-gCI}}, and it then translates that into the complex-algebraic toric geometry framework. 
The (unbounded in $m$) sequence of Hirzebruch scrolls, $\FF{m}$, also harbors an additional diversity for $n\geqslant3$ that is growing combinatorially with $m$, which is exhibited in the biprojective embedding, and which is then organized and cataloged using the toric identifications.

\subsection{Biprojective vs.\ Toric Renditions}
\label{s:2P=T}
\paragraph{Biprojective Rendition:}
Following\cite{Berglund:2022dgb} (Construction~2.1) and\cite{Hubsch:2025teh}, in every particular hypersurface
\begin{alignat}9
  \FF{m;\e} &:=
  \{p_{\e\,}(x,y)=0\} \in \K[{r||c}{\IP^n&1\\\IP^1&m}], \label{e:pexy1}\\
  p_{\e\,}(x,y) &:= 
     x_0y_0^m+x_1y_1^m +
     \sum_{i=0}^n\sum_{\ell=1}^{m-1} \epsilon_{i\ell}\,
                                     x_i\,y_0^{m-\ell}y_1^\ell,\quad
  \epsilon_{i\ell}\in\IC \label{e:pexy2}
\end{alignat}
one finds its collection of mutually {\em\/algebraically independent\,}\footnote{In this context, that $\Fs_\a(x,y)$ (defining $\sS_\a$) is ``algebraically independent'' from $\Fs_\b(x,y),\Fs_\g(x,y)$ means that there exists no algebraic relationship of the form 
	$\Fs_\a(x,y) = y_0^{b_0}y_1^{b_1}\Fs_\b(x,y) +y_0^{c_0}y_1^{c_1}\Fs_\g(x,y)$.} irreducible hypersurfaces
$\big\{ \sS_\a:=\{\Fs_\a(x,y)\,\!=\!\,0\}\subset\FF{m} \big\}$, 
each of maximally negative self-inter\-sec\-tion and explicitly constructed as generalized hypersurfaces.
This collection then also allows writing explicit deformation families of Calabi--Yau hypersurfaces, and identifying their intersections with the $\sS_\a$, to compute the corresponding ``inherited'' characteristics and so be of direct relevance to the ultimate goal of describing the Calabi--Yau hypersurfaces 
$\sX\in\FF{m}[c_1]$.

For example, Hirzebruch's original $p_0(x,y)=x_0y_0^m+x_1y_1^m$ hypersurface\cite{rH-Fm} has its directrix identified as the zero locus of
the following {\em\/equivalence class\/}(!) of sections:
\begin{equation}
  \Fs_0(x,y):=
  \bigg[ \Big(\frac{x_0}{y_1^m}-\frac{x_1}{y_0^m}\Big) 
              +\l \frac{p_0(x,y)}{(y_0y_1)^m} \bigg]
  =\bigg\{ \begin{array}{@{}l@{\quad}r@{\,}l@{\quad\text{when}~}r@{\,}l@{}}
          +2x_0/y_1^m, &\l&={+}1, &y_1&\neq0 ,\\[1mm]
          -2x_1/y_0^m, &\l&={-}1,&y_0&\neq0.\\
          \end{array}
 \label{e:s(x,y)}
\end{equation}
The two $\IP^1_{\!\sss y}$-chartwise holomorphic local representatives specify a single section, since\,\footnote{Viewed this way, the mechanism~\eqref{e:s(x,y)}--\eqref{e:2charts} insuring that $\Fs_0(x,y)$ is in fact a holomorphic section is {\em\/precisely\/} the same that enables the Wu-Yang construction of a magnetic monopole\cite{rWY-MM}, and so it turns out to be completely standard in QFT!}
\begin{equation}
  \Fs_0(x,y)\big|_{\l={+}1}-\Fs_0(x,y)\big|_{\l={-}1}
  =2\frac{p_0(x,y)}{(y_0y_1)^m} \6{\sss\rm id}= 0,\quad
  \text{on}~~\{p_0(x,y)\<=0\}\eqco\FF{m}.
 \label{e:2charts}
\end{equation}
The directrix $\Fs_0(x,y)$ is holomorphic on $\FF{m}\subset\IP^n\!\times\!\IP^1$, although not on $\IP^n\!\times\!\IP^1$.
It is also transverse, since
$\rd\Fs_0=\big(\frac{\rd x_0}{y_1^m},\frac{\rd x_1}{y_0^m},\cdots\big)$ cannot vanish anywhere over $\IP^1_{\!\sss y}$, so its zero locus in $\FF{m}$ is irreducible and smooth.
Calabi--Yau hypersurfaces in $\FF{m}$ must have bi-degree $(n,2{-}m)$ in
$\IP^n_{\!x}\!\times\!\IP^1_{\!y}$, and one finds {(invariably, but with no formal proof)} that
\begin{equation}
  f_0(x,y) \in 
  \Big(\!\bigoplus\nolimits_{k=0}^2\big(\f_k(x)\,y_0^{2-k}y_1^k\big)\!\Big)\,\Fs_0(x,y),\quad
  \deg_{\IP^n}[\f_k(x)]=n{-}1,
 \label{e:CYf(x,y)}
\end{equation}
indeed provide the full complement of regular anticanonical sections for $m\ge3$ when $c_1(\FF{m})$ is negative over the base-$\IP^1_{\!\sss y}$ and provide the ``missing'' sections for the marginal case when $c_1(\FF{2})$ is null over the base-$\IP^1_{\!\sss y}$\cite{rBH-Fm}. 
For $m\<=0,1$, there is no meaningful distinction between~\eqref{e:CYf(x,y)} and standard deg-$(n,2{-}m)$ $(x,y)$-polynomials.

\paragraph{Toric translation:}
Consider now the simple change of variables:
\begin{alignat}9
 (x_0,x_1,\cdots x_n;y_0,y_1) &\in  \IP^n\!\times\!\IP^1 \label{e:2P-T1}\\*
   \to
 \big( p_\epsilon(x,y),\, \Fs_0(x,y)\!=\!X_1,\, 
       \{x_i\!=\!X_i\}_{i=2,\cdots n},\,
       \{y_j\!=\!X_{n+j+1}\}_{j=0,1} \big) &\xrightarrow{p_\epsilon(x,y)=0}\FF{m}. \label{e:2P-T2}
\end{alignat}
The Jacobian of the variable change being a constant, this (QFT-standard field redefinition) provides a 1--1 ``direct translation'' from the biprojective embedding~\eqref{e:pexy1}--\eqref{e:pexy2} to the toric definition~\eqref{e:nFmFan}, where the $U(1;\IC)^2$-charges $q_1,q_2$ in~\eqref{e:q1-4} are indeed directly inherited from the $\IP^n\!\times\!\IP^1$ homogeneity degrees.

This ``direct translation'' extends throughout the $\epsilon_{i\ell}$-deformation family~\eqref{e:pexy1}--\eqref{e:pexy2}\cite{Berglund:2022dgb}.
In hypersurfaces $\FF{m;\e}$~\eqref{e:pexy1}--\eqref{e:pexy2} with more than one directrix, the biprojective-to-toric variable change~\eqref{e:2P-T1}--\eqref{e:2P-T2} replaces the $\IP^n_{\!\ss x}$-coordinates with directrices, in order and by increasing $\IP^1_{\!\sss y}$ degree, while verifying that:
\begin{enumerate}[labelsep=13pt]
 \item all directrices found as in Ref.\cite{Berglund:2022dgb} (Construction~2.1) 
  are transverse, so the zero-locus of each is an irreducible divisor;
 \item the Jacobian of the variable change is constant;
 \item the \eqref{e:CYf(x,y)}-like linear combination of the directrices provides
  the full complement of regular anticanonical sections on $\FF{m;\e}$.
\end{enumerate}

To showcase the so-constructed family, and adapting from Ref.\cite{Hubsch:2025teh}, consider the $(n;m)=(3;5)$ case of~\eqref{e:pexy1}--\eqref{e:pexy2}, several relevant members of which are provided in Table~\ref{t:3F5cases}.\footnote{This also corrects {several typographical errors in}\cite{Berglund:2022dgb}.}

Differently deformed biprojective hypersurface scrolls~\eqref{e:pexy1}--\eqref{e:pexy2} may well result in the same toric specification.
Indeed, the explicit deformation family is far from {\em\/effective\/}. 
All hypersurfaces along a ray generated by 
$\IC^*$-scaling any particular choice of finite and nonzero $\e_{i\ell}$s in~\eqref{e:pexy1}--\eqref{e:pexy2} end up corresponding to the same toric specification, and they are thereby ``infinitesimally near'' the central scroll, 
$\{p_0(x,y)=0\}$. Also, some of those $\IC^*$-rays turn out to provide the same toric specification, such as
the symmetric deformation $p_{3}(x,y)$ and
the asymmetric deformation $p_{4}(x,y)$ in Table~\ref{t:3F5cases};
both of these produce $\FF[3]{\sss3,1,1}$, and we return to this equivalence below.
In turn, experimenting for $m<10$ and hand-picked deformations seems to suggest---but by no means prove---the following contention:
\begin{conj}\label{C:biP=all}
The family~\eqref{e:pexy1}--\eqref{e:pexy2} is {\bfseries\/complete,} in that it contains all the ``cousins'' to the central scroll, $\FF[3]m$, identified by:
\begin{enumerate}[labelsep=13pt]
 \item partitioning the $m$-twist into an $n$-tuple, 
 $\ora{m}=(m_1,m_2,\cdots m_n)$, with $m_k\<\geqslant0$ and any particular but
 fixed ``taxicab''-magnitude $|\ora{m}|=\sum_{k=1}^n m_k=m$;
 \item which is tantamount to having ``distributed'' the deg-$\pM{~~1\\-m}$
 directrix into several deg-$\pM{~~1\\-m_k}$ directrices with $\sum_k m_k=m$ 
 for the $m_k\<>0$.
\end{enumerate}
\end{conj}
From this admittedly small sample, but knowing that all compact complex-algebraic toric varieties can be embedded in a (sufficiently large) projective space, one may---{\em\/perhaps\/}---{speculate}:
\begin{conj}\label{C:T=CICY}
All compact, complex-algebraic toric varieties can be embedded as complete intersections of hypersurfaces in some product of projective spaces.
\end{conj}
\begin{table}[h!]
\footnotesize
\caption{Several particular Hirzebruch scrolls within the $(n;m)=(3;5)$ deformation family~\eqref{e:pexy1}--\eqref{e:pexy2}.}
\label{t:3F5cases}
\centering\vspace*{3mm}
\resizebox{\textwidth}{!}{%
$\def\arraystretch{.5}
\begin{array}{c@{:~~}r@{\,=\,}ll}
\MC1c{\textbf{Deg.}} &\MC2{c}{\textbf{Biprojective Defining Section}}
 &\textbf{Name, GLSM Charges} \\
\toprule
 \MC1c{}
 &p_0(x,y) &x_0\,y_0^5 +x_1\,y_1^5 
           &\FF[3]5 \\ \midrule
 \pM{~~1\\-5}
 &\Fs_{0}(x,y)
 &\Big[\Big(\frac{x_0}{y_1^5} -\frac{x_1}{y_0^5}\Big)\Big/p_0(x,y)\Big]
 &\scriptsize
  \begin{array}{@{}ccccc}
    X_1 &X_2 &X_3 &X_4 &X_5 \\[1pt] \toprule\noalign{\vglue-1pt}
    ~~~1&1   &1   &0   &0 \\
    -5  &0   &0   &1   &1 \\
  \end{array} \\
\midrule[1pt]
 \MC1c{}
 &p_1(x,y) &p_0(x,y) +x_2\,y_0^4 y_1 
           &\FF[3]{\sss4,1} \\ \midrule
 \pM{~~1\\-2}
 &\Fs_{1a}(x,y)
 &\Big[\Big( \frac{x_0\,y_0}{y_1^5} 
             -\frac{x_1}{y_0^4}
             +\frac{x_2}{y_1^4} \Big)\Big/p_1(x,y) \Big]
 &\MR2*{\scriptsize$\begin{array}{@{}ccccc}
               X_1 &X_2 &X_3 &X_4 &X_5 \\[1pt] \toprule\noalign{\vglue-1pt}
               ~~~1&~~~1&1   &0   &0 \\
               -4  &-1  &0   &1   &1 \\
         \end{array}$} \\[2mm]
 \pM{~~1\\-2}
 &\Fs_{1b}(x,y)
 &\Big[\Big( \frac{x_0}{y_1}
             -\frac{x_1\,y_1^4}{y_0^5}
             -\frac{x_2}{y_0} \Big)\Big/p_1(x,y) \Big] \\
  \midrule[1pt]
 \MC1c{}
 &p_2(x,y) &p_0(x,y) +x_2\,y_0^3\,y_1^2 
           &\FF[3]{\sss3,2} \\ \midrule
 \pM{~~1\\-3}
 &\Fs_{2a}(x,y)
 &\Big[\Big( \frac{x_0\,y_0^2}{y_1^5}  
            -\frac{x_1}{y_0^3}
            +\frac{x_2y_0}{y_1^4} 
            +\frac{x_3}{y_1^3} \Big)\Big/p_3(x,y)  \Big]
 &\MR2*{\scriptsize$\begin{array}{@{}ccccc}
               X_1 &X_2 &X_3 &X_4 &X_5 \\[1pt] \toprule\noalign{\vglue-1pt}
               ~~~1&~~~1&1   &0   &0 \\
               -3  &-2  &0   &1   &1 \\
         \end{array}$} \\[2mm]
 \pM{~~1\\-1}
 &\Fs_{2b}(x,y)
 &\Big[\Big( \frac{x_0}{y_1^2} 
            -\frac{x_1 y_1^3}{y_0^5} 
            -\frac{x_2}{y_0^2} \Big)\Big/p_2(x,y) \Big] \\
  \midrule[1pt]
 \MC1c{}
 &p_3(x,y) &p_0(x,y) +x_2 y_0^4 y_1^1 +x_3 y_0^1 y_1^4
           &\FF[3]{\sss3,1,1a}\approx_{\sss\IR}\FF[3]2 \\ \midrule
 \pM{~~1\\-3}
 &\Fs_{3a}(x,y)
 &\Big[\Big( \frac{x_0 y_0}{y_1^4}
             -\frac{x_1 y_1}{y_0^4}
             +\frac{x_2}{y_1^3}
             -\frac{x_3}{y_0^3} \Big)\Big/p_3(x,y) \Big]
 &\multirow{8}{*}{\scriptsize$\begin{array}{@{}ccccc}
               X_1 &X_2 &X_3 &X_4 &X_5 \\[1pt] \toprule\noalign{\vglue-1pt}
               ~~~1&~~~1&~~~1&0   &0 \\
               -3  &-1  &-1  &1   &1 \\
         \end{array}$} \\[2mm]
 \pM{~~1\\-1}
 &\Fs_{3b}(x,y)
 &\Big[\Big( \frac{x_0}{y_1}
             -\frac{x_1 y_1^4}{y_0^5}
             -\frac{x_2}{y_0}
             -\frac{x_3 y_1^3}{y_0^4} \Big)\Big/p_3(x,y) \Big] \\[2mm]
 \pM{~~1\\-1}
 &\Fs_{3c}(x,y)
 &\Big[\Big( \frac{x_0 y_0^4}{y_1^5}
             -\frac{x_1}{y_0}
             +\frac{x_2 y_0^3}{y_1^4}
             +\frac{x_3}{y_1} \Big)\Big/p_3(x,y) \Big] \\[1pt]
  \midrule[1pt]
 \MC1c{}
 &p_4(x,y) &p_0(x,y) +x_2\,y_0^4 y_1 +x_3\,y_0^3\,y_1^2 
           &\FF[3]{\sss3,1,1b}\approx_{\sss\IR}\FF[3]2 \\ \midrule
 \pM{~~1\\-3}
 &\Fs_{4a}(x,y)
 &\Big[\Big( \frac{x_0\,y_0^2}{y_1^5}  
             -\frac{x_1}{y_0^3}
             +\frac{x_2y_0}{y_1^4} 
             +\frac{x_3}{y_1^3} \Big)\Big/p_4(x,y) \Big]
 &\MR3*{\scriptsize\shortstack[l]{
              $\begin{array}{@{}ccccc}
               X_1 &X_2 &X_3 &X_4 &X_5 \\[1pt] \toprule\noalign{\vglue-1pt}
               ~~~1&~~~1&~~~1&0   &0 \\
               -3  &-1  &-1  &1   &1 \\
         \end{array}$\\
         equivalent to $p_3(x,y)$
         }} \\[2mm]
 \pM{~~1\\-1}
 &\Fs_{4b}(x,y)
 &\Big[\Big( \frac{x_0}{y_1} 
             -\frac{x_1 y_1^4}{y_0^5} 
             -\frac{x_2}{y_0}
             -\frac{x_3 y_1}{y_0^2} \Big)\Big/p_4(x,y) \Big] \\[2mm]
 \pM{~~1\\-1}
 &\Fs_{4c}(x,y)
 &\Big[\Big( \frac{x_0 y_0}{y_1^2} 
             -\frac{x_1 y_1^3}{y_0^4} 
             +\frac{x_2}{y_1} 
             -\frac{x_3}{y_0}\Big)\Big/p_4(x,y) \Big] \\
  \midrule[1pt]
 \MC1c{}
 &p_5(x,y) &p_0(x,y) +x_2y_0^3y_1^2 +x_3y_0^2y_1^3
           &\FF[3]{\sss2,2,1}\approx_{\sss\IR}\FF[3]{\sss1,1} \\ \midrule
 \pM{~~1\\-2}
 &\Fs_{5a}(x,y)
 &\Big[\Big( \frac{x_0}{y_1^2}
             -\frac{x_1 y_1^3}{y_0^5}
             -\frac{x_2}{y_0^2}
             -\frac{x_3 y_1}{y_0^3} \Big)\Big/p_5(x,y) \Big]
 &\MR8*{\scriptsize$\begin{array}{@{}ccccc}
               X_1 &X_2 &X_3 &X_4 &X_5 \\[1pt] \toprule\noalign{\vglue-1pt}
               ~~~1&~~~1&~~~1&0   &0 \\
               -2  &-2  &-1  &1   &1 \\
         \end{array}$} \\[2mm]
 \pM{~~1\\-2}
 &\Fs_{5b}(x,y)
 &\Big[\Big( \frac{x_0 y_0^3}{y_1^5}
              -\frac{x_1}{y_0^2}
              +\frac{x_2 y_0}{y_1^3}
              +\frac{x_3}{y_1^2} \Big)\Big/p_5(x,y) \Big] \\[2mm]
 \pM{~~1\\-1}
 &\Fs_{5c}(x,y)
 &\Big[\Big( \frac{x_0 y_0^2}{y_1^3}
            -\frac{x_1 y_1^2}{y_0^3}
            +\frac{x_2}{y_1}
            -\frac{x_3}{y_0} \Big)\Big/p_5(x,y) \Big] \\
\bottomrule
\end{array}
$}
\end{table}
If proven, the claim of Conjecture~\ref{C:T=CICY} would imply that all hypersurfaces in complex-algebraic toric varieties\cite{Kreuzer:2000xy, wKS-CY} in fact are particular cases of complete intersections in products of projective spaces (CICYs)\cite{rH-CY0, rGHCYCI, rCYCI1}! The latter collection has been recorded to contain Calabi--Yau 3-folds with Euler characteristic and Hodge numbers\cite{rGHL-Hog, rGHCYCI} that are but a rather small subset of that data found among the hypersurfaces in toric varieties\cite{Kreuzer:2000xy, wKS-CY}---but the CICY data has only ever been calculated for the {\em\/generic\/} members in the deformation classes of CICYs, which would have to be extended {\em\/in two ways\/}:
\begin{enumerate}[labelsep=13pt]
 \item by identifying all particular cases within each deformation class that give rise to {\em\/discretely distinct\/} complex manifolds\cite{rBH-Fm}---akin to the differing deformations in Table~\ref{t:3F5cases},
 \end{enumerate}
 
    and
 \begin{enumerate}[resume, labelsep=13pt]
 \item by including ``generalized CICYs,''
where some hypersurfaces have negative degrees of homogeneity with respect to some of the ambient projective spaces (as does the directrix~\eqref{e:s(x,y)})\cite{rgCICY1, rBH-Fm, rGG-gCI}. Here, the hypersurfaces with all non-negative degrees define the complete intersections in Conjecture~\ref{C:T=CICY}, and these include non-weak-Fano algebraic varieties---exemplified by $\FF{\sss\ora{m}}$ with $|\ora{m}|\<\geqslant3$.
\end{enumerate}

\paragraph{Veronese twist:}
Table~\ref{t:3F5cases} also exhibits three instances of the $\FF{m}\approx_{\sss\IR}\FF{m\,\smash{(\mathrm{mod}\,n)}}$ diffeomorphism\cite{rBH-Fm}
$\FF[3]{\sss3,1,1a}\approx_{\sss\IR}\FF[3]2\approx_{\sss\IR}\FF[3]{\sss3,1,1b}$ and 
$\FF[3]{\sss2,2,1}\approx_{\sss\IR}\FF[3]{\sss1,1}$, which we now explore. 

In all three instances, the biprojective construction implies the $U(1;\IC)^2$-charges for the GLSM as indicated in the rightmost column of Table~\ref{t:3F5cases}. For the underlying worldsheet QFT, this indicated basis is fully equivalent to the basis using $\Tw{q}_2=q_2{+}q_1$ instead of $q_2$:
\begin{alignat}9
{\small
 \begin{array}{r|ccccc}
 \makebox[0pt][r]{$\FF[3]{\sss3,1,1a}{\sim}\FF[3]{\sss3,1,1b}$}
          &X_1 &X_2 &X_3 &X_4 &X_5 \\ \toprule \noalign{\vglue-2pt}
      q_1 &~~~1&~~~1&~~~1&0   &0 \\
      q_2 &-3  &-1  &-1  &1   &1 \\
 \end{array}}
&\too{~\Tw{q}_2\,=\,q_2{+}q_1~}
{\small
 \begin{array}{r|ccccc}
 \FF[3]2  &X_1 &X_2 &X_3 &X_4 &X_5 \\ \toprule \noalign{\vglue-2pt}
      q_1 &~~~1&1  &1  &0   &0 \\
 \Tw{q}_2 &-2  &0  &0  &1   &1 \\
\end{array}}
 \label{e:311-2}\\[1mm]
{\small
 \begin{array}{r|ccccc}
 \FF[3]{\sss2,2,1}
          &X_1 &X_2 &X_3 &X_4 &X_5 \\ \toprule \noalign{\vglue-2pt}
      q_1 &~~~1&~~~1&~~~1&0   &0 \\
      q_2 &-2  &-2  &-1  &1   &1 \\
 \end{array}}
&\too{~\Tw{q}_2\,=\,q_2{+}q_1~}
{\small
 \begin{array}{r|ccccc}
 \FF[3]{1,1}
          &X_1 &X_2 &X_3 &X_4 &X_5 \\ \toprule \noalign{\vglue-2pt}
      q_1 &~~~1&~~~1&1  &0   &0 \\
 \Tw{q}_2 &-1  &-1  &0  &1   &1 \\
\end{array}}
 \label{e:221-11}
\end{alignat}
The latter of these $U(1;\IC))^2$-charge assignments are straightforwardly identified as the less-twisted Hirzebruch scrolls on the right-hand side in~\eqref{e:311-2}--\eqref{e:221-11}.

The $\FF{m}$ and $\FF{m-n}$ scrolls are distinct complex manifolds although they are diffeomorphic to each other; that is, they are ``the same real manifold'' endowed, however, with differing complex structures\cite{rBH-Fm}. This 
{generalizes Gross' and Ruan's {\em\/non-algebraic deformation equivalence\/} example ($\FF[4]{\!\sss2,1,1,0}\<{\approx_{\sss\IR}}\FF[4]0$ in the present notation)\cite{Gross:1994The, Ruan:1996top} and} 
exhibits the higher-dimensional generalization of the well-known {\em\/discrete\/} ({\em\/jumping\/}) deformation of the complex structure in classic two-dimensional Hirzebruch surfaces; for relevance to characterizing Calabi--Yau hypersurfaces, see Refs.\cite{rGHSAR, rBeast2,rBH-Fm}.

Following Ref.\cite{Hubsch:2025teh}, we identify the Veronese-twisted mapping:
\begin{alignat}9
 (x_0 y_0^3, x_1 y_1^3, x_2 y_0^2 y_1, x_3 y_0 y_1^2; y_0, y_1)
   &\6{\tau}\dto (\x_0, \x_1, \x_2, \x_3; \h_0, \h_1);
 \label{e:Veronese1}\\*[1mm]
  x_0 y_0^5 +x_1 y_1^5 +x_2 y_1 y_0^4 +x_3 y_1^4 y_0
 &\6{\tau}\dto (\x_0+\x_2)\h_0^2 +(\x_1+\x_3)\h_1^2;
 \label{e:Veronese2}\\*
\mkern-12mu
 \ssK[{r||c}{\IP^3&1\\[2pt] \IP^1&5}]\smash{\left\{\rule{0pt}{10mm}\right.}
 x_0 y_0^5 +x_1 y_1^5 +x_2 y_0^4 y_1 +x_3 y_0^3 y_1^2
 &\6{\tau}\dto (\x_0{+}\x_2{+}\x_3)\h_0^2 +\x_1\h_1^2;
 &&\smash{\left.\rule{0pt}{10mm}\right\}}\!\ssK[{r||c}{\IP^3&1\\[2pt] \IP^1&2}]
 \label{e:Veronese3}\\*
 x_0 y_0^5 +x_1 y_1^5 +x_2 y_0^3 y_1^2 +x_3 y_0^2 y_1^3
 &\6{\tau}\dto \x_0\h_0^2 +\x_1\h_1^2 +(\x_2+\x_3)\h_0\h_1;
 \label{e:Veronese4}
\end{alignat}
as an evident candidate for the $\FF[3]5\approx_{\sss\IR}\FF[3]2$ diffeomorphism.
The non-constant Jacobian of this variable change,
$\det\big[\frac{\vd(\x,\h)}{\vd(x,y)}\big]=(y_0y_1)^6$,
indicates that it is ill-defined at the poles of $\IP^1_{\!\sss y}$ as a holomorphic mapping---which is as expected, since $\FF[3]5$ and $\FF[3]2$ are distinct complex manifolds. Aiming only for a {\em\/diffeomorphism,} 
Ref.\cite{Hubsch:2025teh} reminds us that the mapping~\mbox{\eqref{e:Veronese1}--\eqref{e:Veronese4}} can in fact be {\em\/smoothed\/}---non-holomorphically(!)---using partitions of unity (``bump functions'') localized to within a small neighborhood over each of the poles of 
$\IP^1_{\!\sss y}$.

In a perhaps surprising twist (pun intended), the Veronese-twisted mapping~\eqref{e:Veronese1}--\eqref{e:Veronese4}, however, {\em\/can\/} be modified to a constant-Jacobian variant (which is thereby better-suited for the purposes of QFT in general and GLSMs in {particular}):
\begin{alignat}9
  \Big( x_0 y_0^3{+}x_2 y_0^2 y_1,~ x_1 y_1^3{+}x_3 y_0 y_1^2,~ 
        \frac{x_2}{y_0^3},~ \frac{x_3}{y_1^3};~ y_0,~ y_1)
   &\6{\tau'}\dto (\x_0, \x_1, \x_2, \x_3; \h_0, \h_1),\label{e:rat1}\\*
 x_0 y_0^5 +x_1 y_1^5 +x_2 y_0^4 y_1 +x_3 y_0 y_1^4
 &\6{\tau'}\dto \x_0\h_0^2 +\x_1\h_1^2;\\
  \Big( x_0 y_0^3{+}x_2 y_0^2 y_1,~ x_1 y_1^3{+}x_3 y_0^3,~ 
        \frac{x_2}{y_0^3},~ \frac{x_3}{y_1^3};~ y_0,~ y_1)
   &\6{\tau''}\dto (\x_0, \x_1, \x_2, \x_3; \h_0, \h_1),\label{e:rat2}\\*
 x_0 y_0^5 +x_1 y_1^5 +x_2 y_0^4 y_1 +x_3 y_0^3 y_1^2
 &\6{\tau''}\dto \x_0\h_0^2 +\x_1\h_1^2;
\intertext{and}
  \Big( x_0 y_0^3{+}x_2 y_0 y_1^2,~ x_1 y_1^3{+}x_3 y_0^2y_1,~ 
        \frac{x_2}{y_0^3},~ \frac{x_3}{y_1^3};~ y_0,~ y_1)
   &\6{\tau'''}\dto (\x_0, \x_1, \x_2, \x_3; \h_0, \h_1),\label{e:rat3}\\*
 x_0 y_0^5 +x_1 y_1^5 +x_2 y_0^3 y_1^2 +x_3 y_0^2 y_1^3
 &\6{\tau'''}\dto \x_0\h_0^2 +\x_1\h_1^2;
\end{alignat}
at the obvious price of now involving rational monomials in~\eqref{e:rat1}, \eqref{e:rat2} and~~\eqref{e:rat3}---which should not surprise us, since the inverse maps of {\eqref{e:Veronese1}--\eqref{e:rat3}} already involve rational monomials. It is as if the zeros and poles of the Jacobian of~\eqref{e:Veronese1}--\eqref{e:Veronese4} have been moved into a balancing act of the zeros and poles in the individual components of the rational map~\eqref{e:rat1}--\eqref{e:rat3}.
\begin{remk}\label{r:311-200}
The $U(1;\IC)^2$-transformation~\eqref{e:311-2}--\eqref{e:221-11} is neither a flip nor a flop\cite{rCLS-TV} but corresponds to the diffeomorphism 
$\FF{m}\<{\approx_{\sss\IR}}\FF{m-n}$,\footnote{The mapping~\eqref{e:Veronese1}--\eqref{e:Veronese4} incorporates the Veronese mapping,
$\IP^1_{\!\sss y}\ni(y_0,y_1)\to(y_0^n,y_0^{n-1}y_1,\cdots y_1^n)\in\IP^n_{\!\sss y}$, to then twist $\FF{m}$ as a $\IP^n_{\!\sss x}$-fibration over $\IP^1_{\!\sss y}$ into a $\diag[\IP^n_{\!\sss y}{\times}\IP^n_{\!\sss x}]$-fibration over the same base. Thus, the degree of the twist-equivalence, $m\pmod{\2n}$, is canonically fixed to be multiples of the dimension of the fiber-$\IP^n$.} with two coordinate-level realizations provided in~\eqref{e:Veronese1}--\eqref{e:rat3}, exhibiting non-algebraic deformation \mbox{equivalence\cite{Gross:1994The, Ruan:1996top}}. These maps are straightforward to specify in the biprojective realization of $\FF{m}$ but their precise {and complete}
({\bf a})~local geometry (blow-up/down combination?) and 
({\bf b})~toric geometry 
interpretations remain to be determined.
An analogous Veronese-like twisting of the Cox variables $\{X_1,X_2,X_3\}$ by the $\{X_4,X_5\}$ would have degree-2, which would not reproduce the $q_2\to\Tw{q}_2$ change~\eqref{e:311-2}--\eqref{e:221-11}. In turn, no linear twisting of 
$\{X_1,X_2,X_3\}$ by the $\{X_4,X_5\}$, as suggested by~\eqref{e:311-2}--\eqref{e:221-11}, can be uniform over $\{X_1,X_2,X_3\}$. {In the ``phase~I'' description~\eqref{e:nFmQuot}, at least, the changes in the
$U(1;\IC)^2$-transformations~\eqref{e:311-2}--\eqref{e:221-11} are simply the result of the 
$\IP^2_\text{fiber}$-projectivity,
$(X_1,X_2,X_3)\simeq(\l_1X_1,\l_1,X_2,\l_1X_3)$ with $\l_1=\l_2$.}
\end{remk}
Similar constant-Jacobian (albeit rational) variable changes can be found throughout the $\e_{i\ell}$-defor\-ma\-tion family, {and not only} when the $q_2(X_i)$ become all nonzero and of the form shown at the left-hand side of the equations~\eqref{e:311-2}--\eqref{e:221-11}. {For example,
\begin{alignat}9
  \Big( x_0 y_0^3{+}x_2y_0^2y_1,~ x_1 y_1^3,~ 
        \frac{x_2}{y_0 y_1^2},~ \frac{x_3}{y_0^2y_1};~ y_0, y_1)
   &\6{\Tw\tau}\dto (\x_0, \x_1, \x_2, \x_3; \h_0, \h_1),\\*
 \FF[3]{\sss4,1,0}\,{:}\quad
 x_0 y_0^5 +x_1 y_1^5 +x_2 y_0^4 y_1
 &\6{\Tw\tau}\dto \x_0\h_0^2 +\x_1\h_1^2 \quad{:}\,\FF[2]2,\\
  \Big( x_0 y_0^3{+}x_2y_0y_1^2,~ x_1 y_1^3,~ 
        \frac{x_2}{y_0 y_1^2},~ \frac{x_3}{y_0^2y_1};~ y_0, y_1)
   &\6{\Tw\tau}\dto (\x_0, \x_1, \x_2, \x_3; \h_0, \h_1),\\*
 \FF[3]{\sss3,2,0}\,{:}\quad
 x_0 y_0^5 +x_1 y_1^5 +x_2 y_0^3 y_1^2
 &\6{\Tw\tau}\dto \x_0\h_0^2 +\x_1\h_1^2 \quad{:}\,\FF[2]2,
\end{alignat}
{\em cannot\/} be the result of the $\IP^2_\text{fiber}$-projectivity nor are these simple basis-changes in the gauge $U(1;\IC)^2$-transformation of the (toric) $\FF[3]{\sss4,1}$ GLSM. They are, however, simple, constant-Jacobian coordinate field redefinitions in the biprojective embedding
$\{p_1(x,y)\<=0\}\<\subset\IP^3{\times}\IP^1$ and their GLSM rendition,
and they are close analogues of Gross' and Ruan's {\em\/non-algebraic deformation equivalence\/} example, 
$\FF[4]{\!\sss2,1,1,0}\<{\approx_{\sss\IR}}\FF[4]0$\cite{Gross:1994The, Ruan:1996top}.
}

\subsection{Toric Classification}
\label{s:TC}
Given the correlation between the {gauged} $U(1;\IC)^2$-charges and the generators of the toric fan {(the non-gauged $U(1)^3$-symmetry with charges in the first three rows of~\eqref{e:q1-4})}, as detailed in Equations~\eqref{e:q-nu} and \eqref{e:q1-4} and the passage containing them, we can reconstruct the toric fans for the examples in Table~\ref{t:3F5cases}. The generators of this fan being determined up to overall $\GL(3;\ZZ)$ lattice redefinitions, it is gratifying to find that varying $q_1,q_2$, as in Table~\ref{t:3F5cases}, $\n_0,\cdots\n_4$ in~\eqref{e:q1-4} may be held fixed, with only $\n_5$ having to vary; see Table~\ref{t:n5}.
\begin{table}[h!]
\small
\caption{The various ``cousins'' of $\FF[3]5$, as defined by the biprojective embedding in $A=\IP^3{\times}\IP^1$, which defines the $q_a$-charges, which then determine $\n_5$ for the toric realization, which, in turn, reduces $q_2\to\Tw{q}_2$.}
\label{t:n5}
\centering\vspace*{2mm}
\resizebox{\textwidth}{!}{%
$\setlength{\arraycolsep}{3pt}
\begin{array}{rcccccc}
\boldsymbol{\FF[3]{\sss\ora{m}}}
 &\boldsymbol{\FF[3]5} &\boldsymbol{\FF[3]{\sss4,1}} &\boldsymbol{\FF[3]{\sss3,2}} &\boldsymbol{\FF[3]{\sss3,1,1a}} &\boldsymbol{\FF[3]{\sss3,1,1b}} &\boldsymbol{\FF[3]{\sss2,2,1}}  \\
\toprule
 p_\e(x,y):
 &p_0(x,y) &p_1(x,y) &p_2(x,y) &p_3(x,y) &p_4(x,y) &p_5(x,y) \\ 
\midrule
 \boldsymbol{q_2}|_A
 &(\75,0,0,1,1) &(\74,\71,0,1,1) &(\73,\72,0,1,1) &(\73,\71,\71,1,1) &(\73,\71,\71,1,1) &(\72,\72,\71,1,1) \\ 
\midrule
 \boldsymbol{\to\nu_5}
 &\pM{-5\\-5\\~~1} &\pM{-3\\-4\\~~1} &\pM{-1\\-3\\~~1} &\pM{-2\\-2\\~~1} &\pM{-2\\-2\\~~1} 
 &\pM{~~0\\-1\\~~1} \\ 
\midrule
 {\boldsymbol{\Tw{q}_2}}|_{\n_5}
 &(\75,0,0,1,1) &(\74,\71,0,1,1) &(\73,\72,0,1,1) &(\72,0,0,1,1) &(\72,0,0,1,1) &(\71,\71,0,1,1) \\
\bottomrule
\MC7l{\text{\footnotesize For compactness in the $q_2$ specification, ``$\7n$'' is used to denote ``$-n$.''}} 
\end{array}
$}
\end{table}
With this choice of $\n_5$ completing the definition of the fan $\pFn{\FF[3]{5,\e}}$, the Mori vectors for this toric specification re-compute the $q_a$-charges of the $U(1;\IC)^2$ gauge symmetry and perfectly reproduce the overall (``taxicab''-magnitudes) $m\to (m{-}n)$ twist reduction, as shown in the last-row entries in Table~\ref{t:n5}. This, of course, is consistent with this reduction being implemented by the $\GL(2;\ZZ)$ transformation,
$(q_1,\Tw{q}_2)=(q_1,q_2){\cdot}\pM{1&1\\0&1}$.

The foregoing then implies that various choices of $\n_5$ characterize the distinct versions of the 5-twisted Hirzebruch 3-folds, resulting in the following generalization of~\eqref{e:q1-4}:
\begin{equation}
{\small
  \begin{array}{r|@{~}r|rrr@{~}|@{~}rr@{~}|}
 \boldsymbol{n=3} & \n_0 &\nu_1 & \nu_2 & \nu_3 & \nu_4 & \nu_5\\ \toprule
  \multirow3*
  {\raisebox{13pt}{$\pFn{\FF[3]{\ora{m}}}\left\{\rule{0mm}{6mm}\right.$}}
      &0 &-1 & 1 & 0 & 0 & -m_1 \\[-2pt]
      &0 &-1 & 0 & 1 & 0 & -m_2 \\[-2pt]
      &0 & 0 & 0 & 0 & 1 & -1 \\[-2pt] \hline
      &1 & 0 & 0 & 0 & 0 &  0 \\[-1pt] \midrule
  q_1 & -3 & 1 & 1 & 1 & 0 & 0 \\[-2pt]
  q_2 &m_1{-}2(m_2{+}1) & m_2 & m_2{-}m_1 & 0 & 1 & 1\\[-2pt]
  q_3 &m_2{-}2(m_1{+}1) & m_1 & 0 & m_1{-}m_2 & 1 & 1 \\[-2pt]
  q_4 & m_2{+}m_1{-}2 & 0 &-m_1 &-m_2 & 1 & 1 \\[-2pt]
  q_5 &0 & m_2{+}m_1{-}2 & m_2{-}2(m_1{+}1) & m_1{-}2(m_2{+}1) & 3 & 3 \\[-1pt] \bottomrule
 \makebox[0pt][r]{\textbf{Cox variables:}}
     & X_0 & X_1 & X_2 & X_3 & X_4 & X_5\\
  \end{array}
}
 \label{e:q1-5}
\end{equation}
The increased list of possible choices of a basis of $U(1;\IC)^2$-charges exhibit a symmetry generated by the simultaneous swap 
$(m_1\<\iff m_2,\;\n_2\<\iff\n_3,\;q_2\<\iff q_3)$. Furthermore, whereas 
$\FF[3]m\<\coeq\FF[3]{-m,-m}$ holds by definition,
$\FF[3]{0,m}$ and $\FF[3]{m,0}$ are isomorphic to them by virtue of the 
$\GL(3,\ZZ)$-equivalence of their respective fans:
\begin{enumerate}[labelsep=13pt]
 \item $(-m,-m,-1)$ co-planar with $(0,0,1)$, $(-1,-1,0)$ in $\FF[3]{-m,-m}\<\eqco\FF[3]m$;
 \item $(m,0,-1)$ co-planar with $(0,0,1)$, $(1,0,0)$ in $\FF[3]{m,0}\<\simeq\FF[3]m$;
 \item $(0,m,-1)$ co-planar with $(0,0,1)$, $(0,1,0)$ in $\FF[3]{0,m}\<\simeq\FF[3]m$.
\end{enumerate}

The relating $\GL(3,\ZZ)$-transformation is the same that in the fiber plane 
($\n_1,\n_2,\n_3$) permutes the three vertices of the sub-fan of the fiber-$\IP^2$, providing a triality of descriptions.

\begin{figure}[htb]
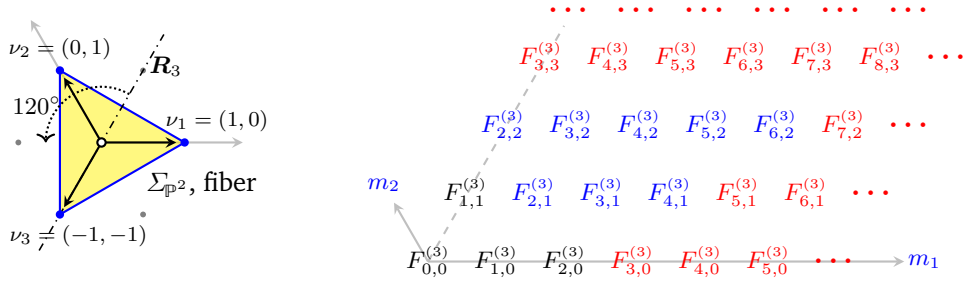

$$
 \vC{\TikZ{[scale=1.1, thick]
            \path[use as bounding box](-1.2,-1.5)--(2,1.5);
            \draw[gray!50, -stealth](0,0)--(0:1.7);
            \draw[gray!50, -stealth](0,0)--(120:1.7);
            \foreach\x in{1,3,5} \fill[gray]({60*\x}:1)circle(.35mm);
            \fill[yellow,opacity=.5](1,0)--(120:1)--(240:1);
            \draw[blue,line join=round](1,0)--(120:1)--(240:1)--cycle;
            \draw[semithick, dash dot](-120:1.5)--(60:1.5);
             \path(50:1.2)node{\fnSz$\bS{R}_3$};
            \draw[densely dotted,thick,->](60:.67)arc(60:180:.67);
             \path(150:.9)node{\fnSz$120^\circ$};
            \draw[stealth-stealth](120:.9)--(0,0)--(.9,0);
            \draw[-stealth](0,0)--(240:.9);
             \path(0:1.2)node[above, xshift=6pt]{\scSz$\n_1=(1,0)$};
             \path(120:1)node[above]{\scSz$\n_2=(0,1)$};
             \path(240:1)node[below, xshift=7pt]{\scSz$\n_3=(-1,-1)$};
            \fill[blue](1,0)circle(.5mm);
            \fill[blue](120:1)circle(.5mm);
            \fill[blue](240:1)circle(.5mm);
            \filldraw[fill=white](0,0)circle(.5mm);
             \path(1.2,-.5)node{\small $\pFn{\IP^2}$, fiber};
            }}
\qquad
 \vC{\TikZ{[scale=.9, thick]
            \path[use as bounding box](-1.5,-.5)--(8,4);
            \draw[gray!50, -stealth](0,0)--(120:1);
             \path[blue](120:1.3)node{\fnSz$m_2$};
            \draw[gray!50, dashed](0,0)--(60:4);
            \draw[gray!50, -stealth](0,0)--(0:7);
             \path[blue](0:7.3)node{\fnSz$m_1$};
            \path(0.55,1)node{\fnSz$\FF[3]{1,1}$};
            \path(0.,0)node{\fnSz$\FF[3]{0,0}$};
            \path(1.,0)node{\fnSz$\FF[3]{1,0}$};
            \path(2.,0)node{\fnSz$\FF[3]{2,0}$};
            \path[blue](1.1,2)node{\fnSz$\FF[3]{2,2}$};
            \path[blue](2.1,2)node{\fnSz$\FF[3]{3,2}$};
            \path[blue](3.1,2)node{\fnSz$\FF[3]{4,2}$};
            \path[blue](4.1,2)node{\fnSz$\FF[3]{5,2}$};
            \path[blue](5.1,2)node{\fnSz$\FF[3]{6,2}$};
            \path[blue](1.55,1)node{\fnSz$\FF[3]{2,1}$};
            \path[blue](2.55,1)node{\fnSz$\FF[3]{3,1}$};
            \path[blue](3.55,1)node{\fnSz$\FF[3]{4,1}$};
            \foreach\x in{2,...,7}\path[red]({\x+.1},3.7)node{$\bS\cdots$};
            \path[red](1.65,3)node{\fnSz$\FF[3]{3,3}$};
            \path[red](2.65,3)node{\fnSz$\FF[3]{4,3}$};
            \path[red](3.65,3)node{\fnSz$\FF[3]{5,3}$};
            \path[red](4.65,3)node{\fnSz$\FF[3]{6,3}$};
            \path[red](5.65,3)node{\fnSz$\FF[3]{7,3}$};
            \path[red](6.65,3)node{\fnSz$\FF[3]{8,3}$};
            \path[red](7.65,3)node{$\bS\cdots$};
            \path[red](6.1,2)node{\fnSz$\FF[3]{7,2}$};
            \path[red](7.1,2)node{$\bS\cdots$};
            \path[red](4.55,1)node{\fnSz$\FF[3]{5,1}$};
            \path[red](5.55,1)node{\fnSz$\FF[3]{6,1}$};
            \path[red](6.55,1)node{$\bS\cdots$};
            \path[red](3.,0)node{\fnSz$\FF[3]{3,0}$};
            \path[red](4.,0)node{\fnSz$\FF[3]{4,0}$};
            \path[red](5.,0)node{\fnSz$\FF[3]{5,0}$};
            \path[red](6.,0)node{$\bS\cdots$};
            }}
$$
\caption{Left: the triangle spanning the fan of $\IP^2$, presented in a $120^\circ$-oblique coordinate system; right: the induced dihedral fundamental domain of $(m_1,m_2)$-twisted Hirzebruch 3-fold scrolls}
 \label{f:3FkmChart}
\end{figure}

The effects of such equivalences are easiest visualized and traced by plotting the $(\hat{e}_1,\hat{e}_2)$-plane in the $N$-lattice in a $120^\circ$-oblique coordinate system: cyclic permutations of the fiber-$\IP^2$ {vectors $\n_1,\n_2,\n_3$} show as $120^\circ$-rotations. Accompanied by the reflections across {$\hat{e}_1$ and $\hat{e}_2$}, these generate the dihedral symmetry group $D_3$ acting along the $(\hat{e}_1,\hat{e}_2)$-plane in the $N$-lattice. This induces a corresponding $D_3$-action in the $\ora{m}\coeq(m_1,m_2)$-space, so that
$\FF[3]{\ora{m}}\approx\FF[3]{\ora{m}'}$ with $\ora{m}'=\ID{\cdot}\ora{m}$, where $\ID$ is an element of the dihedral group $D_3$ acting as {a} {\em\/mapping class group.} This permits limiting the $(m_1,m_2)$-range, e.g., to
\begin{equation}
  \big\{\,\FF[3]{m_1,m_2}:~ 0\<\leqslant m_1\<\leqslant m_1\,\big\},
   \qquad\text{where}\qquad
  \FF[3]{m,0}\define\FF[3]{m},~~\text{and}~~\FF[3]{m,m}=\FF[3]{-m}.
\end{equation}
This specifies a fundamental domain of $\ZZ^2/D_3$ in the $(m_1,m_2)$-{plane}, as shown in Figure~\ref{f:3FkmChart}.
This shows the starting corner of the dihedral{ly reduced} range
$0\<\leqslant m_2\<\leqslant m_1$ for convenience. The remainder of the 
$(m_1,m_2)$-plane is populated by the dihedral group action:
$\bS{R}_3$ reflects this upper-right $60^\circ$-wedge
displayed in Figure~\ref{f:3FkmChart}
to the upward vertical $60^\circ$-wedge,
and the two are then jointly $120^\circ$-rotated \textsc{ccw}, and once more to fill the plane.

\subsection{The Anticanonical System}
\label{s:AC}
With the $U(1;\IC)^2$-charges for any given GLSM, such as presented in Table~\ref{t:3F5cases}, one can address the choice of appropriate superpotentials, following the template in item~\ref{i:S(X)}, discussing the key requirement~\eqref{e:CY} and reorganizing the deformation-theoretic development in Ref.\cite[\SS\:2]{Hubsch:2025teh}.

We start with the anticanonical monomials of the $p_0(x,y)$-defined ``standard'' Hirzebruch scroll $\FF[3]m$, and we then specialize to $m=5$ to {discuss} its variants in Table~\ref{t:3F5cases}.

The multiply motivated (see Remark~\ref{r:key}) key condition~\eqref{e:CY} always has an obvious solution, the so-called ``fundamental monomial''\cite{rHY-SL2} 
$\Pi X\,{:=}\,\prod_iX_i$, so that the ``defining function'' $f(X)$ in the superpotential $X_0\,f(X)$ may be regarded as a deformation of $\Pi X$. This, the simplest of all $\FF[3]m[c_1]$ superpotentials $W(X)=X_0X_1X_2X_3X_4X_5$, produces the $F$-term in~\eqref{e:U} that is combinatorially and geometrically very simple, being the {(indexed)} union of 
$X_i$-hyperplanes, ${\bigsqcup_i}\{X_i=0\}\subset\FF{m}$ 
(in full generality).
However, this space is also singular, and already in co-dimension-1, where two distinct hyperplanes intersect, which location is itself singular where three distinct hyperplanes intersect.
In our three-dimensional showcasing examples, this is where the hierarchy stops; in the physics-wise more interesting four-dimensional case there is one more stratum of singularity in the so-obtained {\em\/polyhoron.} Since multiple intersections are restricted by the Stanley--Reisner ideal\cite{rCLS-TV}, which for $\FF{m}$ (in full generality) differs from that of $\IP^n$, this is not the literal but the conceptual analogue of the familiar regular pentahoron at infinity of the so-called Dwork pencil of $\IP^4[5]$ hypersurfaces\cite{Dwork:1969p-a}; for relevance to string theory {and to ``enumerative mirror symmetry''}, see\cite{Candelas:1990qd, Candelas:1990rm, Berglund:1993ax}.

Whereas standard non-renormalization properties of worldsheet $(2,2)$-supersymmetric GLSMs protect the superpotential from being changed by renormalization, we seek to {\em\/smooth\/} the zero-locus of its $F$-term by
following the standard practice in QFT: we {\em\/deform\/} $\Pi X$ in $W(X)=X_0\,\Pi X$ by replacing {one-by-one} each $X_i$ in $\Pi X$ with a monomial, $\Fm_i(X)$, independent of $X_i$ but such that $q_a(\Fm_i(X))=q_a(X_i)$.
Formally, these replacements are implemented by the action of the operators
$\d_i\define\Fm_i(X)\vd_i$ (no sum on $i$).
From the QFT side, the independence of $\Fm_i(X)$ from $X_i$ is motivated by analogy with the requirement in quantum mechanics perturbation theory, where changes of any state $\ket*$ must be orthogonal to 
$\ket*$ itself, so as to preserve the norm and unitarity.
{Also, $X_i$-independent deformations of $\Pi X$ are those that remain nonzero and so preserve the effective role of $|f(X)|^2$ in~\eqref{e:U}, when 
$\vev{X_i}\to0$ in various directions in the  $(t_1,t_2)$-plane ``phase diagram''; see Table~\ref{t:PhValues}.}
From the algebraic geometry side, such replacements are generators of nontrivial $U(1;\IC)^2$-equivariant coordinate reparametrizations, also cited as (Demazure) ``automorphisms stemming from roots'' (here, the $X_i$)\cite{rCK} (p.\,48).\footnote{Each such automorphism is generated by a relation of the form $X_i\!-\!\l\Fm_i(X)\sim0$, effectively substituting $X_i$ by a multiple of $\Fm_i(X)$---which is {\em\/precisely\/} how the $\Fm_i(X)\vd_i$ act.}

Given the simple form of the starting point, $\Pi X$, the following observations follow:
\begin{enumerate}[labelsep=13pt]\raggedright
 \item 
   The 1st-order deformations are all of the form
   $\d_i\Pi X=\Fm_i(X)\,(\vd_i\Pi X)$, with no summation on $i$.
 \item 
   Since both $(\vd_i\Pi X)$ and $\Fm_i(X)$ are $X_i$-independent,
   so is the collection $\d_i\Pi X$.
 \item 
   Two distinct collections, $\d_i\Pi X$ and  $\d_j\Pi X$, have in common
   the subset of monomials that are independent of {\em\/both\/} $X_i$ and $X_j$.
 \item 
   Three distinct collections have in common their subset of monomials 
   that are independent of three of $X_i,X_j,X_k$'s, etc.
 \item \label{i:[diPX]}
   Each collection is delimited (bounded) by the monomials it has in common
   with another collection; write $[\d_i\Pi X]$ for the so-delimited 
   collection.
 \item \label{i:no2nd}
   Since $\vd_i^2\Pi X\equiv0$, there are no {(quadratic)} 2nd-order deformations.
\end{enumerate}

The union of these collections with the fundamental monomial,
\begin{equation}
  (\Pi X) ~{\sqcup~ \textstyle\bigsqcup_i} [\delta_i\Pi X]
 \label{e:allK*}
\end{equation}
has, astoundingly, an overall combinatorial structure that is well-nigh identical to the zero locus of $\Pi X$ itself, the polyhoron mentioned above! 

In this three-dimensional lattice of monomials with the requisite charge~\eqref{e:CY}, 
each $X_i$-independent subset forms a ``pane,''
each $X_i,X_j$-independent subset forms a ``line'' common to the $X_i$- and $X_j$-``pane,''
and 
each $X_i,X_j,X_k$-independent subset is a corner that is common to those three ``panes.''
In turn, each ``planar'' subset of monomials is delimited by the monomials that it has in common with another ``planar'' subset, and so on.\footnote{This in particular implies that any non-lattice (``fractional'') intersections of such lattice ``panes'' do not in fact delimit any of the combinatorial strata of the poset.} 
This structure is generated by the $X_i$-replacement operators, $\d_i$, which are as follows:
\begin{equation}
 \begin{array}{r@{:\quad}c@{~~\to~~}c}
   \boldsymbol{i} & \boldsymbol{\Fm_i(X)} & \boldsymbol{\d_i\Pi X} \\
 \toprule
 1& X_2^{\ell-1} X_3^{2-\ell} X_4^{r-1} X_5^{1-m-r}
  & X_2^\ell X_3^{3-\ell} X_4^r X_5^{2-m-r} \\ 
 2\text{~or~}3
  & X_1^{k-1} X_{3|2}^{2-k}  X_4^{r-1} X_5^{1+(k-1)m-r}
  & X_1^k X_{3|2}^{3-k} X_4^r X_5^{2+(k-1)m-r} \\ 
 4\text{~or~}5
  & X_1^{k-1} X_2^{\ell-1} X_3^{2-k-\ell} {X}_{5|4}^{1+(k-1)m}
  & X_1^k X_2^\ell X_3^{3-k-\ell} X_{5|4}^{2+(k-1)m} \\ 
 \bottomrule
 \end{array}
 \label{e:planes}
\end{equation}
where linear combinations over varying $k,\ell,r$ are understood{, and 
where ``$X_{i|j}$'' stands for ``$X_i$ or $X_j$'' corresponding to the choice in the header column}.

Two ``planar'' subsets have in common the monomials found by seeking the condition that makes the $\d_i\Pi X$-monomials also $X_j$-independent. For example,
\begin{alignat}9
  [\d_1\Pi X]\cap[\d_2\Pi X] &= 
  \big\lfloor \Ht{X_2^\ell} X_3^{3-\ell} X_4^r X_5^{2-m-r} \big\rfloor_{\ell=0}
  =X_3^3 X_4^r X_5^{2-m-r};\\
  [\d_1\Pi X]\cap[\d_4\Pi X] &= 
  \big\lfloor X_2^\ell X_3^{3-\ell} \Ht{X_4^r} X_5^{2-m-r} \big\rfloor_{r=0}
  =X_2^\ell X_3^{3-\ell} X_5^{2-m}.
\intertext{Then, looking for an $X_4$-and-$X_5$-independent ``line'' yields:}
  [\d_4\Pi X]\cap[\d_5\Pi X] &= 
  \big\lfloor X_1^k X_2^\ell X_3^{3-k-\ell} \Ht{X_5^{2+(k-1)m}} \big\rfloor
  ~~\To~~ k=1-\frc2m.
\intertext{This can produce an integral exponent only for $m=1,2$.
For $m=2$, omitting $X_5$ forces $k=0$ yields:}
  [\d_4\Pi X]\cap[\d_5\Pi X] &= 
  X_2^\ell X_3^{3-\ell} =(X_2\<\oplus X_3)^3, \label{e:m=2}
\intertext{which is correct for the exceptional ``weak Fano'' case
$\FF{2}=\Bl_{\sss\rm MPCP}[\IP^n_{\sss\!(2:\cdots:2:1:1)}]$.
In turn, for $m=1$, omitting $X_5$ forces $k=-1$ yields:}
  [\d_4\Pi X]\cap[\d_5\Pi X] &= 
  X_1^{-1} X_2^\ell X_3^{4-\ell} = X_1^{-1}(X_2\<\oplus X_3)^4. \label{e:m=1}
\end{alignat}
Recalling that $X_1$ was in~\eqref{e:2P-T1}--\eqref{e:2P-T2} identified with the directrix~\eqref{e:s(x,y)} $\Fs_0(x,y)$ in the biprojective realization, its negative powers cannot satisfy the ``mod~$p_0(x,y)$'' equivalence and so cannot be used in anticanonical sections akin to~\eqref{e:CYf(x,y)}.
\begin{corl}\label{C:no45}
The foregoing rules out the existence of a monomial common to the $[\d_4\Pi X]$ and $[\d_5\Pi X]$ anticanonical monomials---except~\eqref{e:m=2}, for the exceptional ``weak Fano'' $\FF2$.
\end{corl}
\begin{remk}\label{R:no123}
With the $q_a$-charges~\eqref{e:q1-4}, it is immediate that there cannot exist any $X_1,X_2,X_3$-independent deg-$\pM{3\\2-m}$ monomial. Together with the preceding Corollary~\ref{C:no45}, this reproduces the effect of the so-called Stanley--Reisner ideal\cite{rCLS-TV} on anticanonical sections.
\end{remk}

In fact, the so-organized structure of anticanonical monomials~\eqref{e:allK*} is described in detail as a {\em\/poset,} as depicted in Figure~\ref{f:defPiX}:
\begin{figure}[htb]
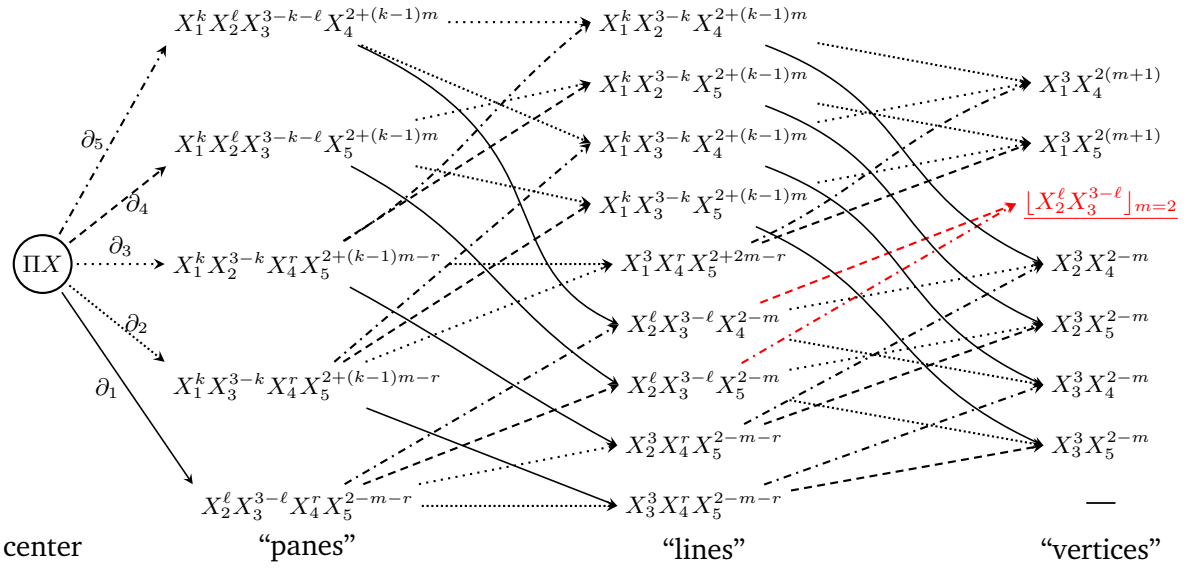

\centering
\TikZ{[thick, xscale=1.75, yscale=.8,
       every node/.style={inner sep=0,outer sep=3pt}]
 \path[use as bounding box](-.3,-6.5)--(8.5,3.3);
 \node[inner sep=2pt, outer sep=2pt, circle, draw](0)
            at(0,-1) {\fnSz$\Pi X$};
 \path(0,-5.7)node{center};
 \node(5)   at(2, 3) {\fnSz$X_1^k X_2^\ell X_3^{3-k-\ell} X_4^{2+(k-1)m}$};
 \node(4)   at(2, 1) {\fnSz$X_1^k X_2^\ell X_3^{3-k-\ell} X_5^{2+(k-1)m}$};
 \node(3)   at(2,-1) {\fnSz$X_1^k X_2^{3-k} X_4^r X_5^{2+(k-1)m-r}$};
 \node(2)   at(2,-3) {\fnSz$X_1^k X_3^{3-k} X_4^r X_5^{2+(k-1)m-r}$};
 \node(1)   at(2,-5) {\fnSz$X_2^\ell X_3^{3-\ell} X_4^r X_5^{2-m-r}$};
 \path(2,-5.7)node{``panes''};
 \node(35)  at(5, 3) {\fnSz$X_1^k X_2^{3-k} X_4^{2+(k-1)m}$};
 \node(34)  at(5, 2) {\fnSz$X_1^k X_2^{3-k} X_5^{2+(k-1)m}$};
 \node(25)  at(5, 1) {\fnSz$X_1^k X_3^{3-k} X_4^{2+(k-1)m}$};
 \node(24)  at(5, 0) {\fnSz$X_1^k X_3^{3-k} X_5^{2+(k-1)m}$};
 \node(23)  at(5,-1) {\fnSz$X_1^3 X_4^r X_5^{2+2m-r}$};
 \node(15)  at(5,-2) {\fnSz$X_2^\ell X_3^{3-\ell} X_4^{2-m}$};
 \node(14)  at(5,-3) {\fnSz$X_2^\ell X_3^{3-\ell} X_5^{2-m}$};
 \node(13)  at(5,-4) {\fnSz$X_2^3 X_4^r X_5^{2-m-r}$};
 \node(12)  at(5,-5) {\fnSz$X_3^3 X_4^r X_5^{2-m-r}$};
 \path(5,-5.7)node{``lines''};
 \node(235) at(8, 2) {\fnSz$X_1^3 X_4^{2 (m+1)}$};
 \node(234) at(8, 1) {\fnSz$X_1^3 X_5^{2 (m+1)}$};
 \node(145) at(8, 0)[red] {\fnSz$\2{\lfloor X_2^\ell X_3^{3-\ell}\rfloor_{m=2}}$};
 \node(135) at(8,-1) {\fnSz$X_2^3 X_4^{2-m}$};
 \node(134) at(8,-2) {\fnSz$X_2^3 X_5^{2-m}$};
 \node(125) at(8,-3) {\fnSz$X_3^3 X_4^{2-m}$};
 \node(124) at(8,-4) {\fnSz$X_3^3 X_5^{2-m}$};
 \node(123) at(8,-5) {---};
 \path(8,-5.7)node{``vertices''};
 \draw[dash dot, -stealth](0)--node[left]{\fnSz$\vd_5$}(5.south west);
 \draw[densely dashed, -stealth](0)--node[right]{\fnSz$\vd_4$}(4.south west);
 \draw[dotted, -stealth](0)--node[above]{\fnSz$\vd_3$}(3);
 \draw[densely dotted, -stealth](0)--node[right]{\fnSz$\vd_2$}(2.north west);
 \draw[semithick, -stealth](0)--node[left]{\fnSz$\vd_1$}(1.north west);
 \draw[dash dot, -stealth](1)--(15.west);
 \draw[dash dot, -stealth](2)--(25.west);
 \draw[dash dot, -stealth](3)to[out=50,in=240](35.west);
 \draw[densely dashed, -stealth](1)--(14.west);
 \draw[densely dashed, -stealth](2)--(24.west);
 \draw[densely dashed, -stealth](3)--(34.west);
 \draw[dotted, -stealth](1)--(13.west);
 \draw[dotted, -stealth](2)--(23.west);
 \draw[dotted, -stealth](4)--(34.west);
 \draw[dotted, -stealth](5)--(35.west);
 \draw[densely dotted, -stealth](1)--(12.west);
 \draw[densely dotted, -stealth](3)--(23.west);
 \draw[densely dotted, -stealth](4)--(24.west);
 \draw[densely dotted, -stealth](5)--(25.west);
 \draw[semithick, -stealth](2)--(12.west);
 \draw[semithick, -stealth](3)to[out=-45,in=130](13.west);
 \draw[semithick, -stealth](4)to[out=-45,in=125](14.west);
 \draw[semithick, -stealth](5)..controls++(2,-2)and++(-1,1)..(15.west);
 \draw[dash dot, -stealth](12)--(125.west);
 \draw[dash dot, -stealth](13)--(135.west);
 \draw[dash dot, -stealth, red](14)--(145.west);
 \draw[dash dot, -stealth](23)to[out=40,in=220](235.west);
 \draw[densely dashed, -stealth](12)--(124.west);
 \draw[densely dashed, -stealth](13)--(134.west);
 \draw[densely dashed, -stealth, red](15)--(145.west);
 \draw[densely dashed, -stealth](23)--(234.west);
 \draw[dotted, -stealth](14)--(134.west);
 \draw[dotted, -stealth](15)--(135.west);
 \draw[dotted, -stealth](24)--(234.west);
 \draw[dotted, -stealth](25)--(235.west);
 \draw[densely dotted, -stealth](14)--(124.west);
 \draw[densely dotted, -stealth](15)--(125.west);
 \draw[densely dotted, -stealth](34)--(234.west);
 \draw[densely dotted, -stealth](35)--(235.west);
 \draw[semithick, -stealth](24)to[out=-40](124.west);
 \draw[semithick, -stealth](25)to[out=-35](125.west);
 \draw[semithick, -stealth](34)to[out=-35](134.west);
 \draw[semithick, -stealth](35)to[out=-35](135.west);
}
\caption{The complete poset of $\Pi X$-deformations, organized by decreasing $X_i$-dependence; the exponents simplify for $m\!=\!2$ \big(``weak Fano,''
$c_1(\FF{2})\!\ge\!0$\big). The underlined terms are included only in the exceptional $m=2$ ``weak Fano'' case, and they form a ``line''; see text.
}
 \label{f:defPiX}
\end{figure}
This poset structure identifies the collection of monomials~\eqref{e:allK*}, detailed in the discussion subsequent to the Corollary~\ref{C:no45} and Remark~\ref{R:no123}, with the Newton {\em\/multitope\,}\footnote{\label{fn:mT} The Latin/Greek	heterotic portmanteau of (possibly) {multi}-layered+polytope denotes a real $n$-dimensional {bounded} polyhedral complex of finitely many convex $n$-dimensional polytopes, 
	$\pD\coeq\sqcup_\a\pD_\a$, which contains each face of each $\pD_\a$; 
	$\pD_\a\cap\pD_\b$ is a face of both, and 
	$\pD$ is a poset, such as in Figure~\ref{f:defPiX}. Moreover, each $\pD_\a$ has lattice-$\L$-primitive vertices, and the complex is homeomorphic to an $n$-ball\cite{Hibi:1995aa}, although $\pD$ need only immerse in $\IR^n\supset\L$. Thus, polytopes are multitopes that do embed in $\IR^n$; their image in $\IR^n$ does not self-cross.} $\pDN{\FF[3]5}$ and the {\em\/multifan\/}\cite{rM-MFans, Masuda:2000aa, rHM-MFs, Masuda:2006aa, rHM-EG+MF, rH-EG+MFs2, Nishimura:2006vs, Davis:1991uz, Ishida:2013aa, Ishida:2013ab, buchstaber2014toric, Jang:2023aa} it is defined to span\cite{rBH-gB, Berglund:2022dgb, Berglund:2024zuz}.

In turn, and as presented in Ref.\cite{Hubsch:2025teh}, each generator
$\d_i\propto\vd_i$ by definition acts transversally to the $X_i$-independent ``pane'' of anticanonical monomials, so that the collection $\{\d_i\}$ spans the face-wise polar, i.e., {\em\/transpolar\,}\footnote{\label{fn:tP}The {\em\/transpolar operation\/} is the face-wise {\em\/polar operation\/}\cite{rO-TV, rF-TV, rGE-CCAG, rCLS-TV} applied iteratively throughout the poset of a multitope and subject to the universal inclusion-reversing nature of the polar operation\cite{rBH-gB, Berglund:2022dgb, Berglund:2024zuz}; on convex polytopes, it is identical to the standard polar operation.} multitope, 
$\pDs{\FF[3]5}\<=(\pDN{\FF[3]5})^\wtd$ and the multifan it spans, 
$\pFn{\FF[3]5}\smt\pDs{\FF[3]5}$---again all in perfect agreement with Refs.\cite{rBH-gB, Berglund:2022dgb, Berglund:2024zuz}. In particular, the poset automatically includes the monomials that are rational for $m\geqslant3$:
\begin{equation}
  \oplus_{\ell,r} X_2^\ell X_3^{3-\ell} X_4^r X_5^{2-m-r}
  =\Bigg\{
  \begin{array}{@{}l@{\quad\text{for}~~}r@{\:}l}
  (X_2\<\oplus X_3)^3 \big( X_4\<\oplus X_5\big)^{2-m}, &m&\leqslant2;\\[1mm]
  (X_2\<\oplus X_3)^3 \big( \frc1{X_4}\<\oplus\frc1{X_5}\big)^{m-2}, &m&\geqslant3.
  \end{array}
 \label{e:2m2}
\end{equation}
They are omitted in the complex-algebraic standard toric geometry but are germane to smoothing the Tyurin-degenerate Calabi--Yau hypersurfaces in $\FF{m}$, which are for $m\geqslant3$ standardly deemed ``unsmoothable''\cite{Berglund:2022dgb}. {Precisely these rational monomials are also crucial to the Euler and Hodge number computations\cite{rBH-gB} as well as to characteristic class computations and the Chern--Todd--Hirzebruch identities\cite{Berglund:2024zuz}. The Newton polytope ``extension'' to which they correspond is transpolar\,\footref{fn:tP} to the non-convexity in the fan-spanning polytope of the non-weak-Fano $\FF{m}$ and so necessary for perfect \mbox{duality\cite{rBH-gB, Berglund:2022dgb, Berglund:2024zuz}}.}
As the smoothing deformation-motivated construction~\eqref{e:allK*}--\eqref{e:2m2} and especially the {precluding of ``run-away'' directions in
$|f(X)|^2$ in~\eqref{e:U}} imply, the inclusion of precisely these rational monomials fits well within the GLSM framework; see also\cite[\SS\:4]{Berglund:2022dgb}.

\begin{remk}\label{r:IPp}
The 1-cone generators of a fan of a (complex-algebraic) toric variety, $V$, define the vertices of the fan-spanning lattice polytope, 
$(\pDs{V}\<\lat\pFn{V})\subset(L\<\coeq\ZZ^n{\otimes_{\sss\IR}}\IR^n)$, and they define the complex Cox variables, $x_i\<\mapsto \n_i\<\in\pFn[\sss(1)]{V}$,
the zero locus of which are the characteristic toric divisors $D_i\subset V$.
Anticanonical sections are linear combinations of monomials,
$\G(\cKs{V})=\big\{ \prod_{\n_i\in\pFn[\sss(1)]{V}} x_i^{\vev{\n_i,u}+1},~
                     u\<\in L^\vee\<\cap\pDN{V} \big\}$,
corresponding to the dual lattice points in the Newton polytope, $\pDN{V}=(\pDs{V})^\circ$\cite{rD-TV, rO-TV, rF-TV, rGE-CCAG, rCLS-TV, rCK}.
The correlation between 
the generic polynomial being transverse ($\pD$-regular\cite{Batyrev:1993oya})
and 
the convex hull of the lattice points enclosing the origin%
---as will be shown below for $\pDN{\FF[3]5}$, $\pDN{\FF[3]{\sss4,1}}$, 
$\pDN{\FF[3]{\sss3,2}}$, $\pDN{\FF[3]{\sss3,1,1}}$ and 
$\pDN{\FF[3]{\sss2,2,1}}$---%
was first noted by Kreuzer and Skarke\cite{rKreSka98}, who called it the ``IP property.''
It arguably provides a combinatorial rendition of Arnold's original classification\cite{rAGZV-Sing1}; see also\cite{Kreuzer:1992bi}.
In the deformation-generated collection~\eqref{e:allK*} of anticanonical monomials, this ``0-enclosing'' means that $\Pi X$ is fully surrounded (in every $\d_i$-direction) by its 1st-order deformations.
\end{remk}
\begin{remk}\label{r:rfX-VEX}
Defined for convex polytopes, Batyrev's mirror duality\cite{Batyrev:1993oya} requires the polar operation to be involutive, 
so polytopes in mutually polar pairs are {reflexive.}
The GLSM-deforming collection of anticanonical monomials~\eqref{e:allK*}
consistently reproduces\,\footnote{\label{fn:exp}This is an ``experimental fact'' obtained for many dozens of $n=2,3$ examples (including some infinite sequences, such as starting with~{\eqref{e:q1-4} and}~\eqref{e:pexy1}--\eqref{e:pexy2} for arbitrary $m$) {and with no known counter-example}---but I am not aware of a formal, rigorous proof {either}.} $\G(\cKs{V})$ for toric varieties but straightforwardly extends to pairs of mutually transpolar\,\footref{fn:tP} VEX multitopes\,\footref{fn:mT}\cite{rBH-gB, Berglund:2022dgb, Berglund:2024zuz}, reproducing the exact complete poset structure of $\G(\cKs{})$---for both; see Section~\ref{s:Mirrors}, below.
\end{remk}%

\begin{figure}[htb]
\centering\setbox9=\hbox{$\FF[3]{\sss\ora{m}}$}
\TikZ{[thick]
 \path[use as bounding box](-.2,-1.5)--(17,10);
 \path(1,5)node{\includegraphics[height=100mm]{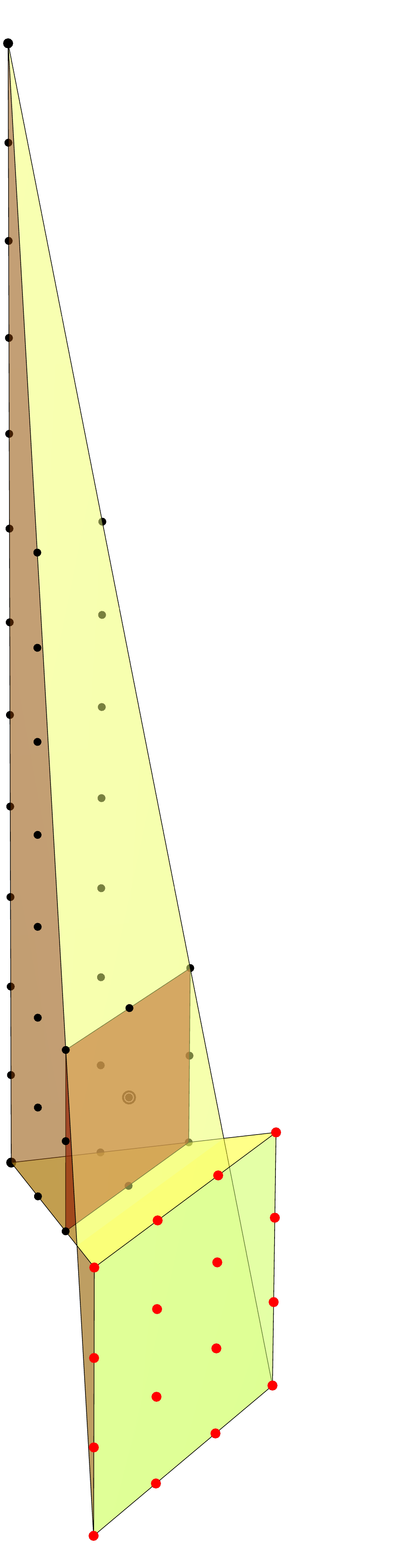}};
 \path(2,-.6)node{\includegraphics*[width=40mm, viewport=0mm 33mm 200mm 80mm]
  {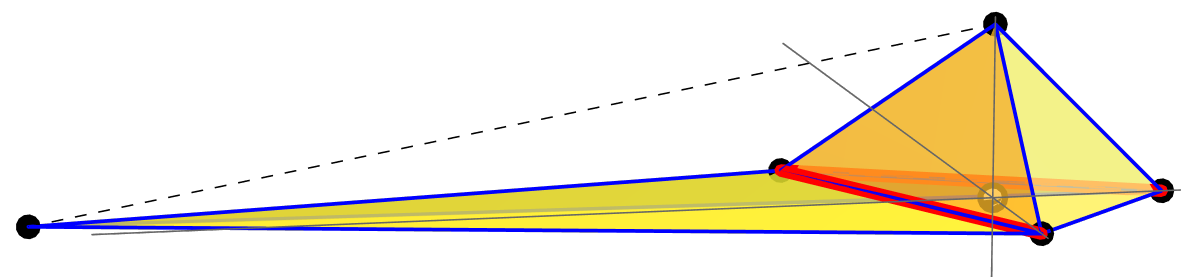}};
  \draw[densely dashed, stealth-stealth](0,2)..controls++(0,-2.3)..
   node[above right=-1mm]{$\SSS\wtd$}++(1.7,-2.6);
 \path(.5,7.5)node{$\pDN{\FF[3]5}$};
 \path(2.5,-.2)node{$\pDs{\FF[3]5}$};
 \begin{scope}[xshift=40mm]
 \path(0,5.05)node{\includegraphics[height=100mm]{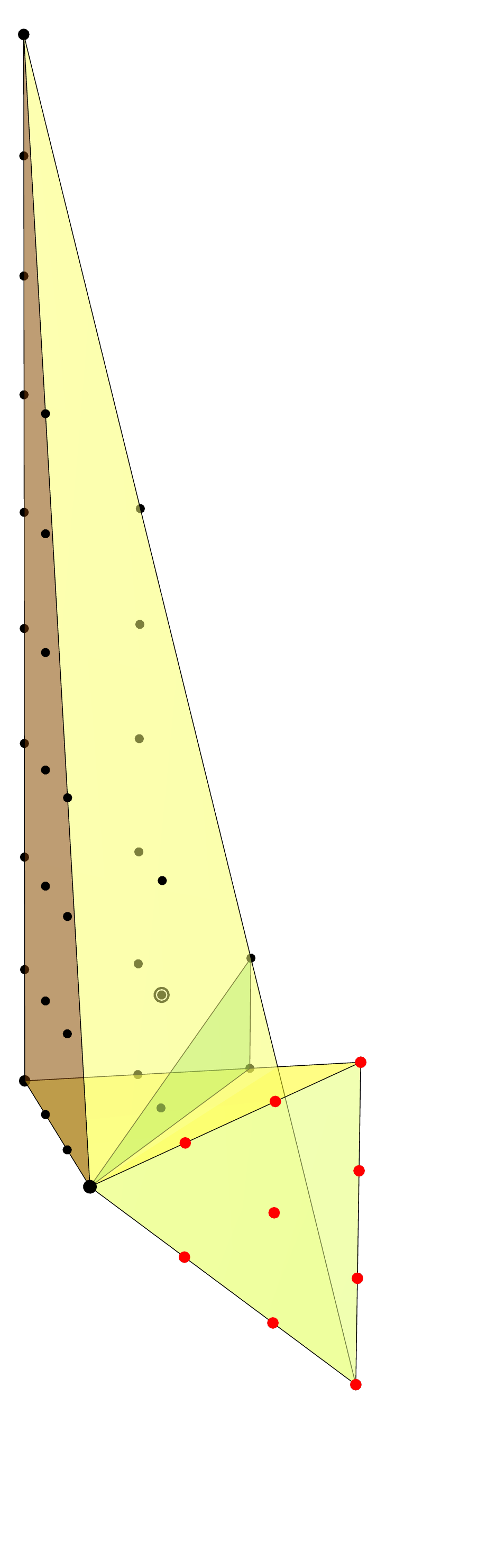}};
 \path(0.3,.7)node{\includegraphics[width=35mm]{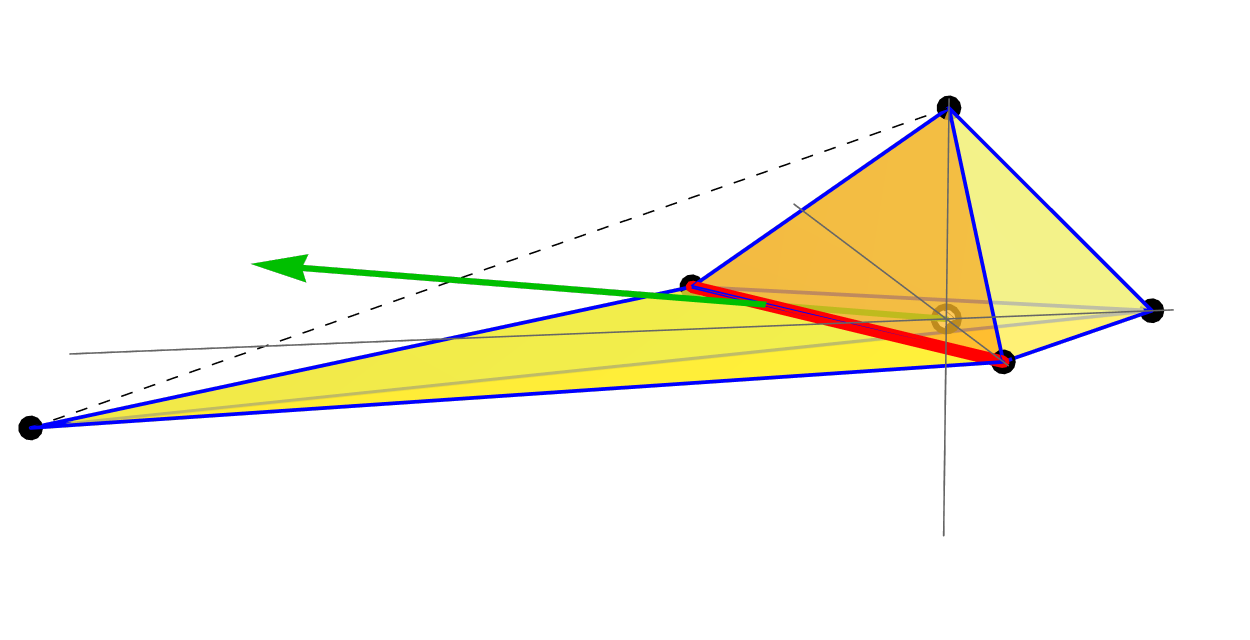}};
 \draw[densely dashed, stealth-stealth](-1,2.2)..controls++(0,-.7)..
   node[above right=-1mm]{$\SSS\wtd$}++(1.4,-1.2);
 \draw[green!60!gray, densely dotted, stealth-stealth](-.4,3.1)--++(0,-2.25);
 \path(-.5,7.5)node{$\pDN{\FF[3]{\sss4,1}}$};
 \path(1.5,.3)node{$\pDs{\FF[3]{\sss4,1}}$};
 \end{scope}
 \begin{scope}[xshift=75mm]
 \path(0,5.53)node{\includegraphics[height=90mm]{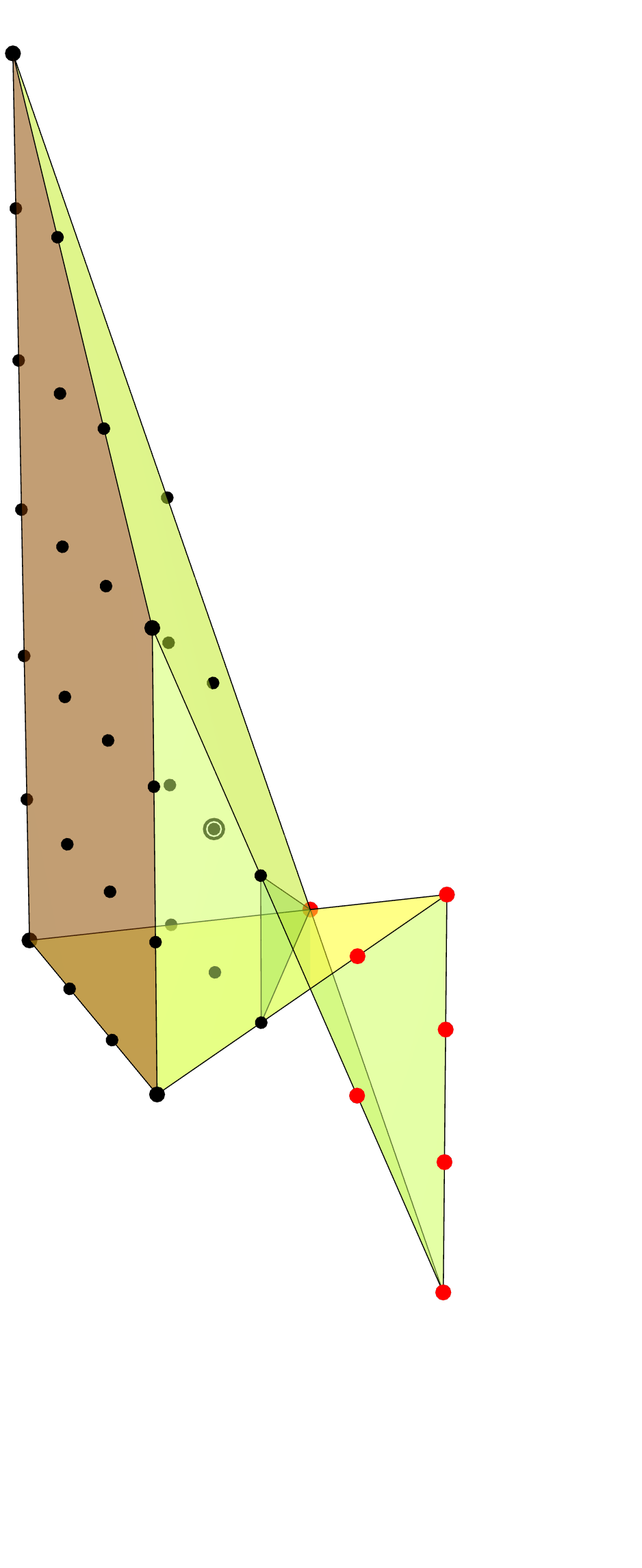}};
 \path(0.5,1.8)node{\includegraphics[width=35mm]{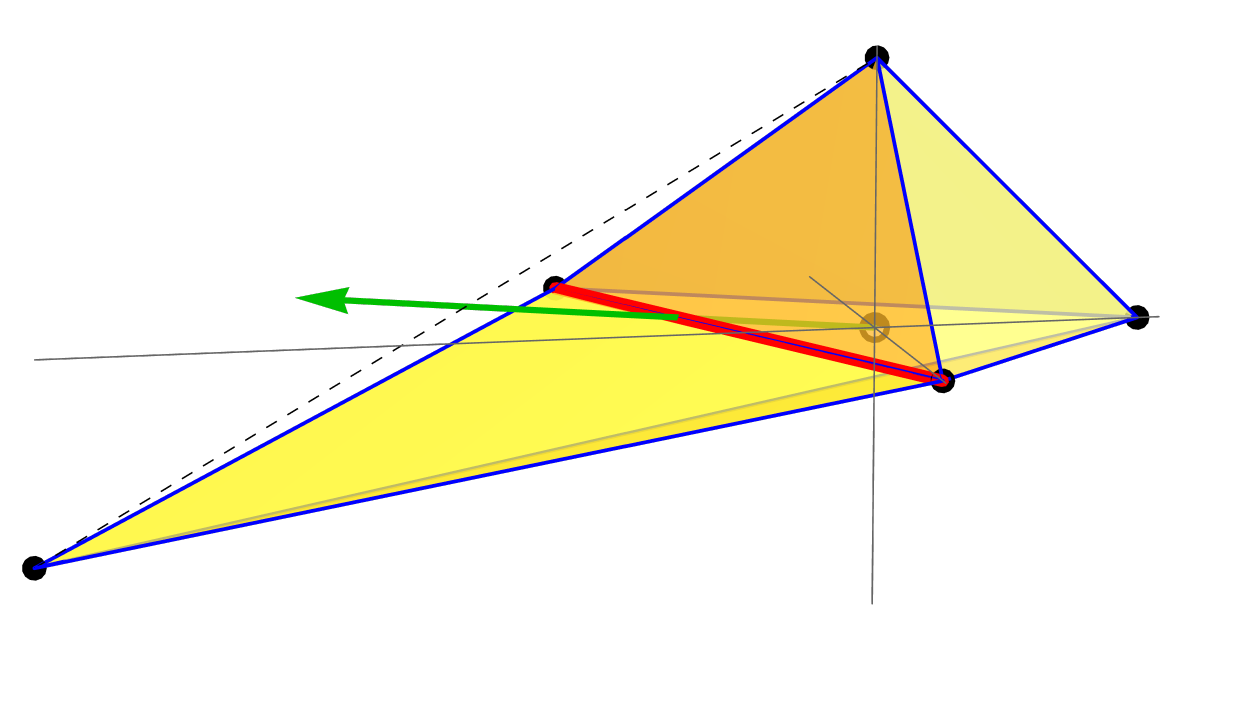}};
 \draw[densely dashed, stealth-stealth](-1.5,4.2)..controls++(0,-1.5)..
   node[above right=-1mm]{$\SSS\wtd$}++(2,-2);
 \draw[green!60!gray, densely dotted, stealth-stealth](-.25,4.6)--++(0,-2.6);
 \path(-.6,7.5)node{$\pDN{\FF[3]{\sss3,2}}$};
 \path(1.2,1.3)node{$\pDs{\FF[3]{\sss3,2}}$};
 \end{scope}
 \begin{scope}[xshift=115mm]
 \path(0,6.96)node{\includegraphics[height=60mm]{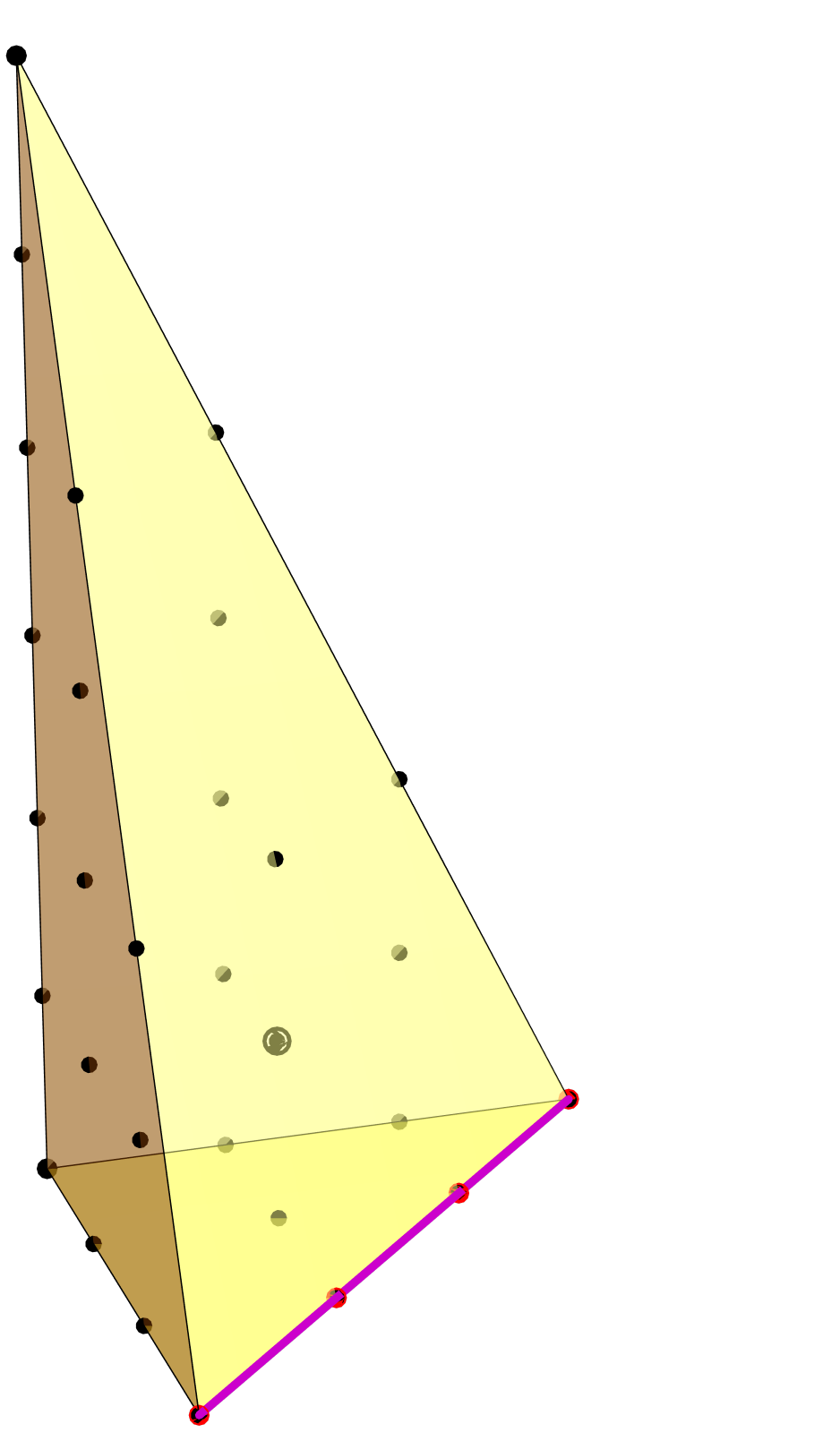}};
 \path(0.4,3.3)node{\includegraphics[width=35mm]{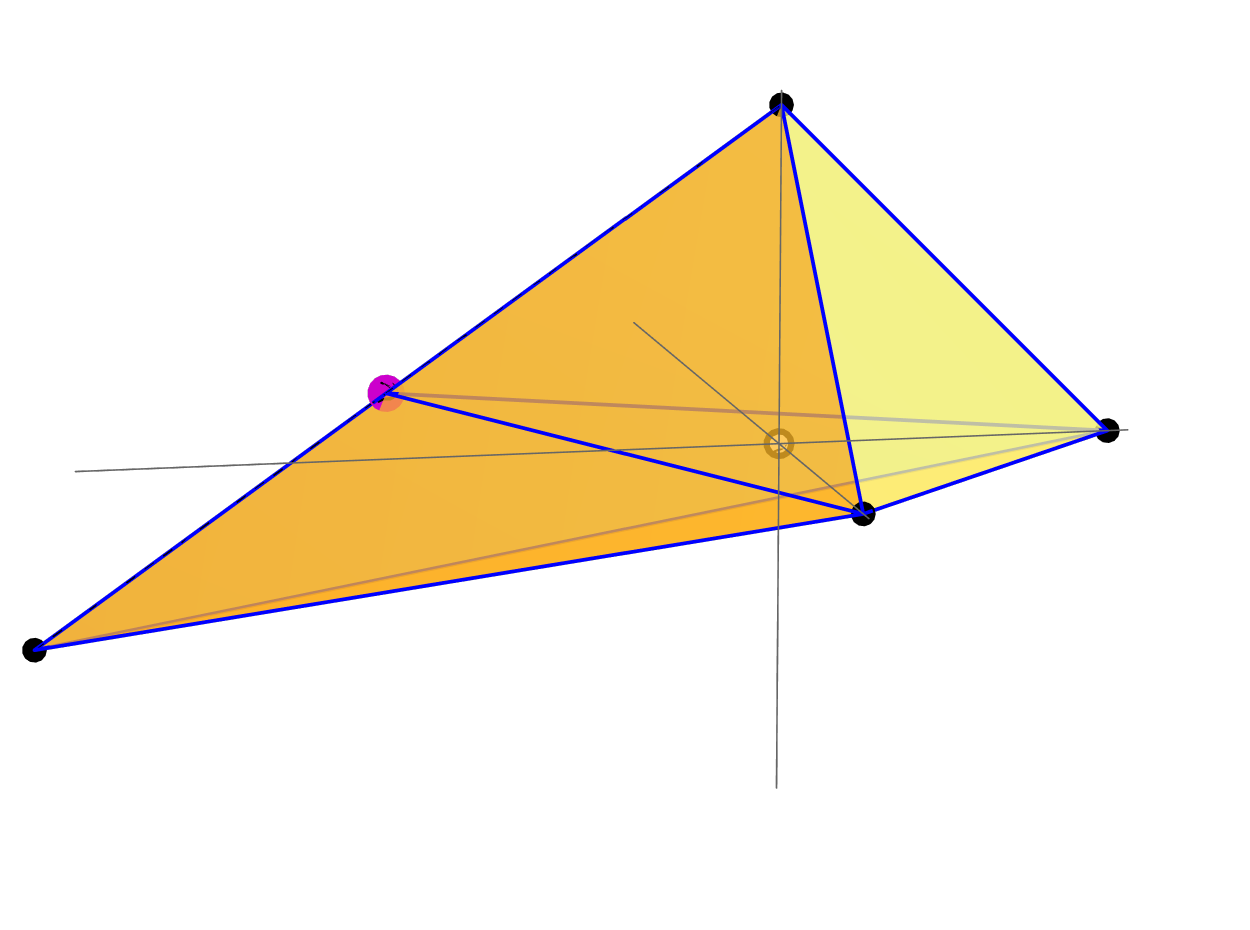}};
 \draw[densely dashed, stealth-stealth](-1.4,4.6)
   ..controls++(-.2,-.9)and++(0,-.1)..
   node[above right=-1mm]{$\SSS\wtd$}++(.9,-1.2);
 \draw[magenta, densely dotted, stealth-stealth](-.48,4.4)--++(.2,-.85);
 \path(-.1,7.5)node{$\pDN{\FF[3]{\sss3,1,1}}$};
 \path(1,2.7)node{$\pDs{\FF[3]{\sss3,1,1}}$};
 \end{scope}
 \begin{scope}[xshift=155mm]
 \path(0,7.8)node{\includegraphics[height=40mm]{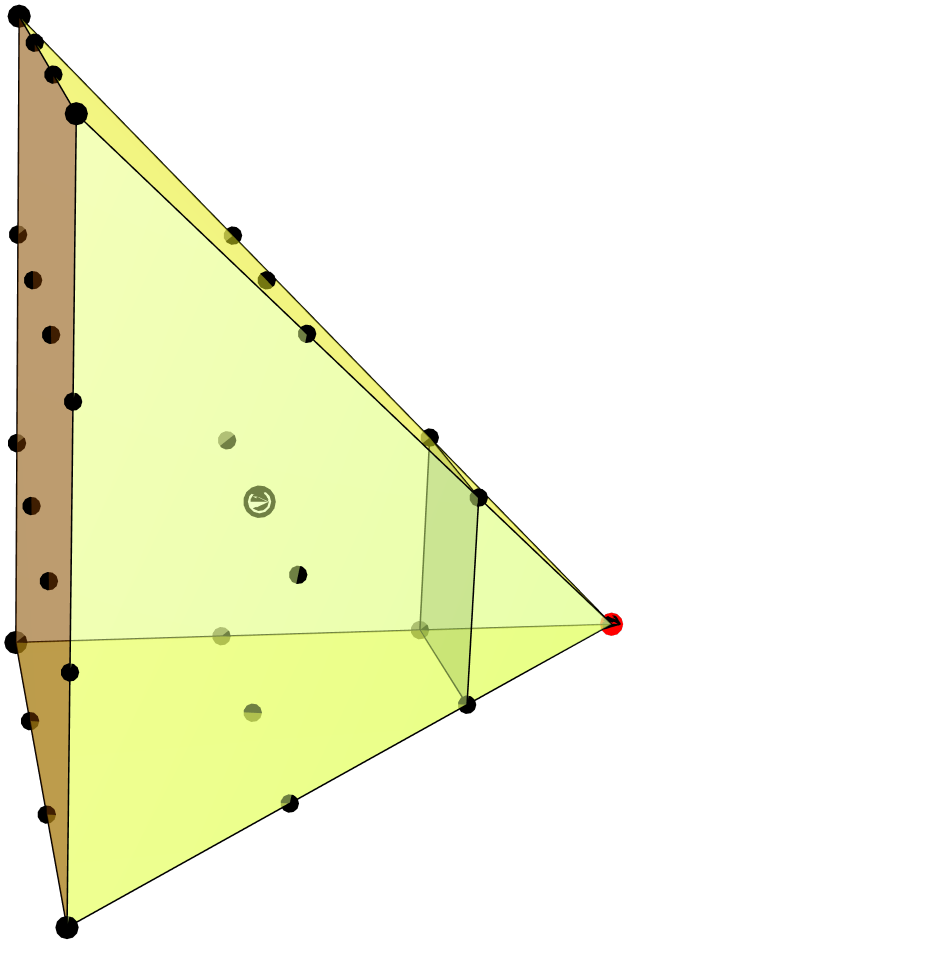}};
 \path(0.1,5)node{\includegraphics[width=30mm]{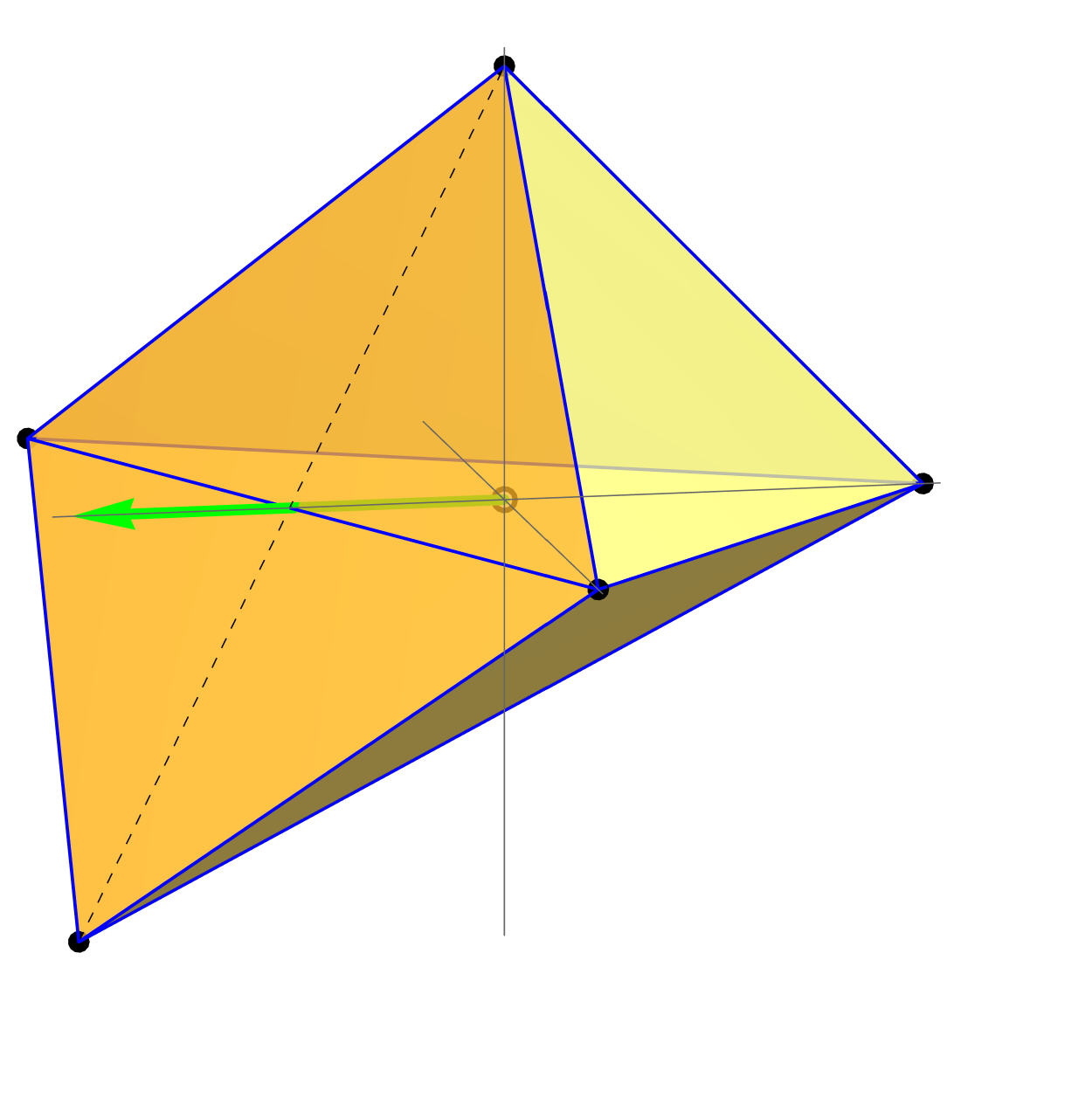}};
 \draw[densely dashed, stealth-stealth](-1.8,6.2)
   ..controls++(-.2,-.7)and++(0,-.1)..
   node[above right=-1mm]{$\SSS\wtd$}++(.4,-1);
 \draw[green!60!gray, densely dotted, stealth-stealth](-.1,7.4)--++(-.7,-2.25);
 \path(-.1,8.5)node{$\pDN{\FF[3]{\sss2,2,1}}$};
 \path(.3,4.3)node{$\pDs{\FF[3]{\sss2,2,1}}$};
 \end{scope}
 \path(10.5,1.4)node[right]{\parbox[t]{55mm}{\small\raggedright$\bullet$~The
 toric $\FF[3]{\sss3,1,1a}$ and $\FF[3]{\sss3,1,1b}$ are not distinguishable and
 encode the same ``weak Fano'' toric variety.}};
 \path(7,-.6)node[right]{\parbox[t]{82mm}{\small\raggedright$\bullet$~Newton 
 multitope points (with the extension, red points) represent anticanonical 
 monomials.}};
}
\caption{The progression of Newton and spanning multitopes for the deformations in Table~\ref{t:3F5cases}. The two multitopes are related by the involutive {\em\/transpolar\/} ($^\wtd$) operation\cite{rBH-gB, Berglund:2022dgb, Berglund:2024zuz}. Truncating the Newton multitope to a convex polytope at the drawn-in (green-shaded) facet encodes a blowup of that \copy9, indicated by the  new vertex-arrow in the spanning polytope; see text. The (magenta-highlighted) lower-right edge in $\pDN{\FF[3]{3,1,1}}$ may be seen as the collapsed lower-right rectangle in $\pDN{\FF[3]5}$. The 
 $\vd_i$-generated fan and polytope\cite{Hubsch:2025teh} equals precisely the 
 transpolar
 $\pDs{\sX}=(\pDN{\sX})^\wtd$\cite{rBH-gB, Berglund:2022dgb, Berglund:2024zuz}.}
 \label{f:3F5cases}
\end{figure}
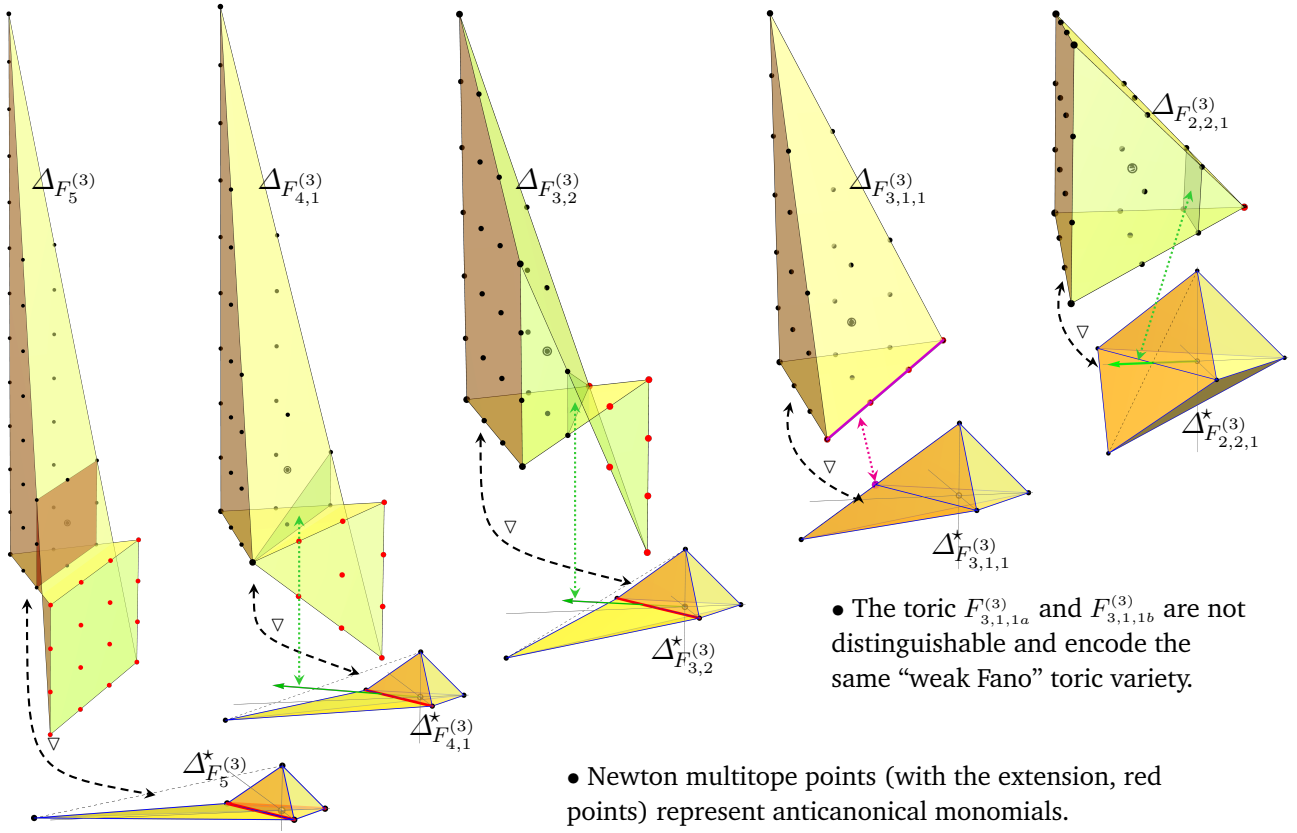

\paragraph{Deformations of $\FF[3]5$:}
We now turn to the deformations of $\FF[3]5$, as listed in Table~\ref{t:3F5cases}, foregoing, however, the detailed analysis and argumentation as done above for
$\FF[3]5$, and presenting instead the corresponding Newton and spanning multitopes, $\pDN{\FF[3]{\sss\ora{m}}}$ and $\pDs{\FF[3]{\sss\ora{m}}}$; see Figure~\ref{f:3F5cases}.
This makes manifest the overall combinatorial structure of the collection {of} anticanonical monomials~\eqref{e:allK*} for each Hirzebruch scroll,
$\FF[3]{\sss\ora{m}}$. Generally, in the $\pDN{\FF[3]{\sss\ora{m}}}$-plots in Figure~\ref{f:3F5cases} the power of $X_1$ drops from 3 in the back to 0 in the front; the powers of $X_2,X_3$ ``trade {one for another}'' left--right, and the powers of $X_4,X_5$ ``trade'' up--down.
In the (smaller) spanning polytopes the concave edges are (red) highlighted and are the transpolar images of the ``extensions'' in the  Newton multitopes, comprising the frontmost ``hanging'' segments (red points). The spanning polytope
$\pDs{\FF[3]{3,1,1}}\approx_{\sss\IR}\pDs{\FF[3]2}$ exceptionally has a lattice point along its (upper back, slanted) edge, which is the transpolar image of the frontmost edge in $\pDN{\FF[3]{3,1,1}}\approx_{\sss\IR}\pDN{\FF[3]2}$, which, in turn, may be seen (progressing from left to right in Figure~\ref{f:3F5cases}) as the result of the collapsed ``extension.''

\begin{figure}[htb]
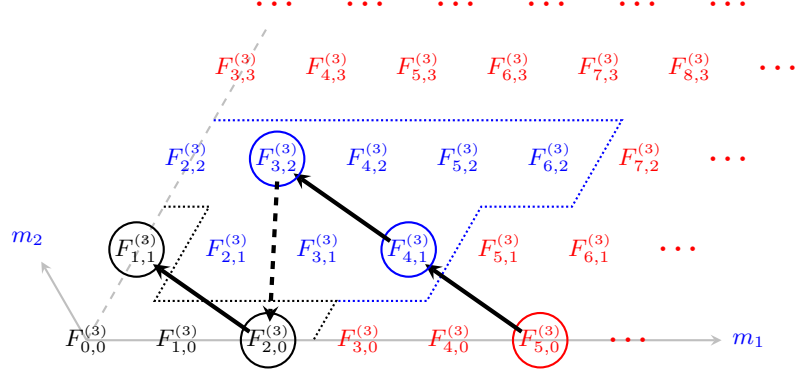

$$
 \vC{\TikZ{[scale=1.2, thick]
            \path[use as bounding box](-1.4,-.5)--(8,3.8);
            \draw[gray!50, -stealth](0,0)--(120:1);
             \path[blue](120:1.3)node{\fnSz$m_2$};
            \draw[gray!50, dashed](0,0)--(60:4);
            \draw[gray!50, -stealth](0,0)--(0:7);
             \path[blue](0:7.3)node{\fnSz$m_1$};
            \path(0.55,1)node{\fnSz$\FF[3]{1,1}$};
            \path(0.,0)node{\fnSz$\FF[3]{0,0}$};
            \path(1.,0)node{\fnSz$\FF[3]{1,0}$};
            \path(2.,0)node{\fnSz$\FF[3]{2,0}$};
            \path[blue](1.1,2)node{\fnSz$\FF[3]{2,2}$};
            \path[blue](2.1,2)node{\fnSz$\FF[3]{3,2}$};
            \path[blue](3.1,2)node{\fnSz$\FF[3]{4,2}$};
            \path[blue](4.1,2)node{\fnSz$\FF[3]{5,2}$};
            \path[blue](5.1,2)node{\fnSz$\FF[3]{6,2}$};
            \path[blue](1.55,1)node{\fnSz$\FF[3]{2,1}$};
            \path[blue](2.55,1)node{\fnSz$\FF[3]{3,1}$};
            \path[blue](3.55,1)node{\fnSz$\FF[3]{4,1}$};
            \foreach\x in{2,...,7}\path[red]({\x+.1},3.7)node{$\bS\cdots$};
            \path[red](1.65,3)node{\fnSz$\FF[3]{3,3}$};
            \path[red](2.65,3)node{\fnSz$\FF[3]{4,3}$};
            \path[red](3.65,3)node{\fnSz$\FF[3]{5,3}$};
            \path[red](4.65,3)node{\fnSz$\FF[3]{6,3}$};
            \path[red](5.65,3)node{\fnSz$\FF[3]{7,3}$};
            \path[red](6.65,3)node{\fnSz$\FF[3]{8,3}$};
            \path[red](7.65,3)node{$\bS\cdots$};
            \path[red](6.1,2)node{\fnSz$\FF[3]{7,2}$};
            \path[red](7.1,2)node{$\bS\cdots$};
            \path[red](4.55,1)node{\fnSz$\FF[3]{5,1}$};
            \path[red](5.55,1)node{\fnSz$\FF[3]{6,1}$};
            \path[red](6.55,1)node{$\bS\cdots$};
            \path[red](3.,0)node{\fnSz$\FF[3]{3,0}$};
            \path[red](4.,0)node{\fnSz$\FF[3]{4,0}$};
            \path[red](5.,0)node{\fnSz$\FF[3]{5,0}$};
            \path[red](6.,0)node{$\bS\cdots$};
            \draw[densely dotted](60:1.7)--++(.5,0)--++(-120:1.2)--++(2,0)
              --++(-120:.5);
            \draw[densely dotted, blue](60:2.8)--++(4.5,0)--++(-120:1.1)
              --++(-1,0)--++(-120:1.2)--++(-1,0);
            \draw[red](5,0)circle(3mm);
             \draw[ultra thick, -stealth](5,-.05)++(145:.25)--++(145:1.3);
            \draw[blue](3.55,1)circle(3mm);
             \draw[ultra thick, -stealth](3.55,.95)++(145:.25)--++(145:1.3);
            \draw[blue](2.1,2)circle(3mm);
             \draw[ultra thick, dashed, -stealth](2.1,1.75)--++(-93:1.55);
            \draw(2,0)circle(3mm);
             \draw[ultra thick, -stealth](2,-.05)++(145:.25)--++(145:1.3);
            \draw(0.55,1)circle(3mm);
            }}
$$
\caption{The iterative deformations connecting five models from Table~\ref{t:3F5cases} and Figure~\ref{f:3F5cases} in the dihedral fundamental domain of $(m_1,m_2)$-twisted Hirzebruch 3-fold scrolls, and the partitioning of this domain described in the text}
 \label{f:3F5trip}
\end{figure}

The deformation sequence in Table~\ref{t:3F5cases} and Figure~\ref{f:3F5cases} may thus also be identified in the plot of {scrolls} in Figure~\ref{f:3FkmChart}, as shown in Figure~\ref{f:3F5trip},
which jointly help to explain the key differences between the models along this {continuous} deformation path, as well as the partitioning of the starting corner of this (wedge-shaped) $D_3$-primitive domain in the space of all $\FF[3]{\sss\ora{m}}$ Hirzebruch scrolls and also their Calabi--Yau hypersurfaces:

\paragraph{$\FF[3]5$:}
This initial scroll has a (complete, extended) Newton multitope with its ``extension'' presented as the frontmost $3{\times}3$ pane of {sixteen} (red) points. 
While all other points represent {\em\/regular\/} (non-negative $X_i$-powers) anticanonical monomials, the (red) ``extension'' points represent rational monomials 
$(X_2{\oplus}X_3)^3\big(\frac1{X_4}{\oplus}\frac1{X_5}\big){}^3$ listed in the bottommost ``pane'' in the poset in Figure~\ref{f:defPiX}; see~\eqref{e:2m2}.
As indicated by the darker-shaded vertical $2{\times}2$ square centered at the origin, there is no way to trim away the extension of $\pDN{\FF[3]5}$ {(i.e., restrict to the non-rational anticanonical monomials)} without exposing the origin into a facet{: t}he entire ``regular'' part of the Newton {polytope} occupies only a half-space (the back, as depicted in Figure~\ref{f:3F5cases}, leftmost). Thereby, no {\em\/regular\/} anticanonical $X_i$-polynomial can be transverse, all Calabi--Yau hypersurfaces are Tyurin degenerate and standardly deemed ``unsmoothable''---but they are readily smoothed by including the ``extension'' rational monomials{, and they are in agreement with various computational requirements}\cite{rBH-gB, Berglund:2022dgb, Berglund:2024zuz}.

\paragraph{$\FF[3]{\sss4,1}$:}
The situation improves already with the first deformation: 
the $\pDN{\FF[3]{\sss4,1}}$ Newton multitope, second from left in Figure~\ref{f:3F5cases}, {\em\/can\/} be trimmed at the (green) shaded 3-point triangle, and so trim away the nine ``extension'' (red point) monomials. The remaining convex polytope still encloses the origin, so that generic anticanonical (regular) polynomials are transverse, and they define smooth Calabi--Yau hypersurfaces with {\em\/no need to\/} include the nine rational monomials.

In turn, the so-reduced Newton polytope also corresponds to another, related underlying ambient space, encoded by the introduction of the new facet (the 3-point triangular ``pane''), which has its own normal,
\begin{equation}
   \Big[\pM{~~2\\[-1pt]-1\\[-1pt]-1},\,\pM{-1\\[-1pt]~~1\\[-1pt]-1},\,
         \pM{-1\\[-1pt]~~1\\~~0}\Big]
   \fif{~\wtd~} \pM{-2\\[-1pt]-3\\[-1pt]~~0}\<=\n_6 \iff \vd_6,
\end{equation}
and which introduces a new vertex {in the fan-spanning polytope 
$\pDs{\FF[3]{4,1}}$}.
The original $\n_1$ turns out to be in the relative interior of the new
$[\n_3,\n_5,\n_6,\n_4]$ facet, and the polytope spanned by 
$\{\n_2,\n_3,\n_4,\n_5,\n_6\}$ is convex (reflexive,\footnote{\label{fn:rflx}That is, the {\em\/standard polar\/} \mbox{operation\cite{rF-TV, rGE-CCAG, rCLS-TV}} produces a lattice polytope and squares to the identity.} moreover), 
and it encodes 
$\Bl[\FF[3]{\sss4,1}]$, a blowup of $\FF[3]{\sss4,1}$. This introduces a new $(1,1)$-form, which is then inherited by the Calabi--Yau hypersurface{s}, which {are thereby} substantially changed.

\paragraph{$\FF[3]{\sss3,2}$:}
This deformation of $\FF[3]5$ is qualitatively similar to $\FF[3]{\sss4,1}$, in that the Newton multitope also can be trimmed at the (green) shaded 3-point triangle, and so trim away the six ``extension'' (red point) monomials. This leaves a convex polytope that still encloses the origin, so that generic anticanonical (regular) polynomials are transverse, and they define smooth Calabi--Yau hypersurfaces with {\em\/no need to\/} include the six rational monomials. 
The so-reduced Newton polytope again also corresponds to a blowup of
$\FF[3]{\sss3,2}$, which again substantially changes the Calabi--Yau hypersurfaces.

\paragraph{$\FF[3]{\sss3,1,1}\approx_{\sss\IR}\FF[3]2$:}
This deformation is diffeomorphic to the marginal ``weak Fano'' $\FF[3]2$, with a strongly convex Newton polytope. This equivalence appears to be realized by a simple change of the $U(1;\IC)^2$-charge basis~\eqref{e:311-2}, and so it {is} a straightforward equivalence relation between the corresponding GLSMs{ and the result of the $\IP^2_\text{fiber}$-rescaling in ``phase~I''}. This is further supported by the explicit coordinate-level mapping~\eqref{e:Veronese1}--\eqref{e:Veronese4} and its constant-Jacobian variant~\eqref{e:rat1}--\eqref{e:rat3}. Both $\pDN{\FF[3]2}$ and 
$\pDs{\FF[3]2}$ are convex and reflexive, and they are each other's {standard polar images}\cite{Batyrev:1993oya}.

\paragraph{$\FF[3]{\sss2,2,1}\approx_{\sss\IR}\FF[3]{\sss1,1}$:}
This final deformation is diffeomorphic, and it is related by~\eqref{e:221-11}--\eqref{e:rat3} to the scroll of the same overall $m=2$ twist but has two distinct deg-$\pM{~~1\\-1}$ directrices instead of a single deg-$\pM{~~1\\-2}$ {one}. Now both the Newton and the spanning polytopes are strongly convex and reflexive, and they are each other's polars\cite{Batyrev:1993oya}.\bigskip

These cases exemplify the characteristics of all Hirzebruch scrolls in the map in Figure~\ref{f:3F5trip},
which may then be summarized as follows:
\begin{corl}\label{C:cases}
{Taken m}odulo the {fiber-$\IP^2$-induced} $D_3$-action, the $\ora{m}$-plane of Hirzebruch 3-folds 
$\FF[3]{\sss\ora{m}}$ contains three {equivalence} classes:
\begin{enumerate}[labelsep=13pt]

 \item
  The four 3-folds, $\FF[3]{\sss0,0}\<\define\FF[3]0$,~ 
  $\FF[3]{\sss1,0}\<\define\FF[3]1$,~ $\FF[3]{\sss2,0}\<\define\FF[3]2$ and 
  $\FF[3]{\sss1,1}$ have convex, in fact reflexive, spanning and Newton polytopes that are each other's polars.
\end{enumerate}

The remaining infinite collection of 3-folds all have non-convex but VEX spanning {polytopes} and Newton {multitopes}\cite{rBH-gB, Berglund:2022dgb, Berglund:2024zuz}. These, however, belong to two classes:

\begin{enumerate}[labelsep=13pt, resume]

 \item\label{i:whyXt}
  The convex integral hull of the standard part of the Newton {multitope} of the eight special {3-folds}
, ${\FF[3]{\sss2,1}}$, ${\FF[3]{\sss3,1}}$, 
  ${\FF[3]{\sss4,1}}$, ${\FF[3]{\sss2,2}}$, ${\FF[3]{\sss3,2}}$, 
  ${\FF[3]{\sss4,2}}$, ${\FF[3]{\sss5,2}}$ and ${\FF[3]{\sss6,2}}$ 
  {\bfseries\/does\/} enclose the origin.

 \item The convex integral hull of the standard part of the Newton {multitope} of the remaining infinitely many 3-folds {\bfseries\/does not\/} enclose the origin; {some monomials from} the extension must be included {in transverse sections}. The first few in this infinite class are indicated in Figure~\ref{f:3F5trip}
by red ink. 
\end{enumerate}

The same explicit deformation family~\eqref{e:pexy1}--\eqref{e:pexy2} contains all Hirzebruch scrolls, $\FF{\sss\ora{m}}$, with ``taxicab''-magnitudes
$|\ora{m}|=\sum_{\,i}m_i=m{-}kn$, where $k=0,\dots \lfloor\frac{m}n\rfloor$.
\end{corl}
\begin{remk}\label{r:mapFm}
This classification/partitioning,~\eqref{e:q1-5}
and Figures~\ref{f:3FkmChart} and~\ref{f:3F5trip},
readily generalizes to higher dimensions, and it is based on the symmetries of the spanning polytope of fiber-$\IP^{n-1}$: viewed as an equilateral simplex, its symmetries are generated by reflections and rotations. In turn, for the original Hirzebruch (complex two-dimensional) surfaces this classification reduces to the well-known fact that $\FF[2]{-m}=\FF[2]m$ and that $m=0,1,2,\dots$.
It is worth noting that, overall, case~1 in Corollary~\ref{C:cases} is the best-studied and understood, case~{2} has room to explore, and case~{3} is the least studied, least understood---yet, the most abundant.
\end{remk}
{%
\begin{remk}\label{r:allFm}
The distinct toric varieties displayed in Figure~\ref{f:3F5cases} are equivalently given by the biprojective embeddings in Table~\ref{t:3F5cases}. These occur within the same $(n;m)=(3;5)$ deformation family~\eqref{e:pexy1}--\eqref{e:pexy2} of biprojective embeddings of Hirzebruch scrolls, making {\em\/non-algebraic deformation equivalence\/} (Refs.\cite{Gross:1994The, Ruan:1996top} and Remark~\ref{r:311-200}) explicitly continuous. In this sense, the explicit $\epsilon_{ak}$-deformation family~\eqref{e:pexy2} is somewhat akin to the likewise continuous ``phase diagram'' in Figure~\ref{f:PhDiag}.
\end{remk}
\begin{remk}\label{r:defoLim}
Only the central $\epsilon_{i\ell}=0$ model $\FF[3]5$ in the deformation family~\eqref{e:pexy2} requires rational (Laurent) deformations for smoothing anticanonical hypersurfaces. This suggests\cite{Hubsch:2025sph} (Conjecture\:2.2) that the rational smoothing deformations in $\G(\cKs{\FF[3]5})$ are limits of regular (non-rational) smoothing deformations in the system of anticanonical sections of other members of the deformation family~\eqref{f:3F5cases}. This conjecture evidently applies equally to all $n,m\geq2$ cases and is consistent with the GLSM analysis followed here.
\end{remk}%
}

\begin{figure}[htb]
$$
\vC{\TikZ{[xscale=1.5]
  \path[use as bounding box](-.2,-.2)--(4.5,1.8);
  \path(1.9,.7)node{\includegraphics*[width=65mm, viewport=0mm 33mm 200mm 80mm]
  {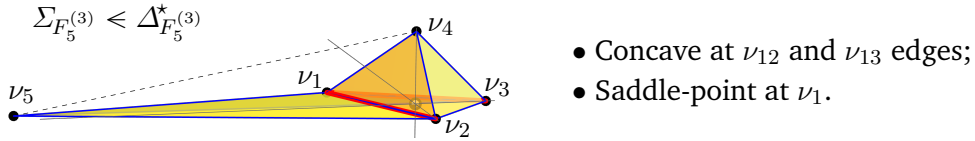}};
  \path(-.1,1.5)node[right]{$\pFn{\FF[3]5}\smt\pDs{\FF[3]5}$};
  \path(2.45, .70)node{$\n_1$};
  \path(3.75,0.05)node{$\n_2$};
  \path(4.10, .60)node{$\n_3$};
  \path(3.60,1.40)node{$\n_4$};
  \path(-.10, .50)node{$\n_5$};
}}\quad
\parbox[c]{60mm}{\raggedright
  $\bullet$~Concave at $\n_{12}$ and $\n_{13}$ edges;\\
  $\bullet$~Saddle-point at $\n_1$.}
$$
\caption{The fan-spanning non-convex polytope of $\FF[3]5$ and its key characteristics}
 \label{f:3F5S}
\end{figure}

In turn, a closer look at the {\em\/fan-spanning polytopes} 
$\pDs{\FF[3]{\sss3,2}}$, $\pDs{\FF[3]{\sss4,1}}$ and $\pDs{\FF[3]5}$, re-displayed here from Figure~\ref{f:3F5cases} in greater detail, provides additional {information}, as shown in Figure~\ref{f:3F5S}.
The dotted $\n_{4,5}$ line indicates the edge from the {\em\/convex hull\/} of the vertices; indeed, $\pDs{\FF[3]5}$ may be regarded as obtained from this convex hull by ``VEXing'': by trimming away the ``non-star (not adjacent to the origin) simplices'' $\n_{1524}$ and $\n_{1435}$---dubbed {\em\/divots\/} in Ref.\cite{MacFadden:2025ssx}. Modifying a polytope by trimming away non-star simplices leaves a ``star-triangulated'' possibly non-convex (as here) VEX polytope\cite{rBH-gB, Berglund:2022dgb, Berglund:2024zuz}. In turn, the convex hull encodes the singular weighted projective {space}, $\IP^3_{\sss(5:5:1:1)}$.
\begin{figure}[htb]
$$
\vC{\TikZ{[xscale=1.5]
  \path[use as bounding box](-.2,-.2)--(4.4,2.2);
  \path(1.9,1)node{\includegraphics*[width=65mm, viewport=0mm 33mm 200mm 95mm]
  {3F41S=.pdf}};
  \path(-.1,1.9)node[right]
      {$\pFn{\FF[3]{{\sss4,1}}}\smt\pDs{\FF[3]{\sss4,1}}$};
  \path(2.10,1.05)node{$\n_1$};
  \path(3.55, .25)node{$\n_2$};
  \path(4.05, .90)node{$\n_3$};
  \path(3.40,1.90)node{$\n_4$};
  \path(-.10, .35)node{$\n_5$};
  \path[green!75!black](0.7,1.10)node[left]
  {{\fnSz$\pM{-2\\[-1pt]-3\\[-1pt]~~0}$}$\,=\n_6$};
}}\quad
\parbox[c]{70mm}{\raggedright
  $\bullet$~Concave at the $\n_{12}$ edge;\\
  $\bullet$~The $\n_{13}$ edge is within the $\n_{435}$-plane;\\
  $\bullet$~The {\em\/blowup vertex\/}: $\n_6 \<\subset\n_{435}$-plane;\\
  $\bullet$~Saddle-point at $\n_1$.}
$$
\caption{The fan-spanning non-convex polytope of $\FF[3]{\sss4,1}$ and its key characteristics}
 \label{f:3F41S}
\end{figure}
The displayed configuration of the polytope in Figure~\ref{f:3F41S} was dubbed the ``hammock''\cite{Berglund:2024zuz}, as opposed to the ``tent'' configuration that includes the $\n_{1524}$-simplex and corresponds to the singular weighted projective space, $\IP^3_{\!\sss(1:1:3:4)}$. The former is non-convex but VEX, while the latter is convex but not reflexive\cite{Batyrev:1993oya}: its standard polar,
$(\Conv[\pDs{\IP^3_{\!(1:1:3:4)}}])^\circ$, has a fractional (non-lattice) vertex that would correspond to a radical monomial. 
Its desingularizing blowup coincides with the convex hull also enclosing a blowup vertex, $\n_6$, which also corresponds to the blowup 
$\Bl^\uA_{\n_6}[\FF[3]{\sss4,1}]$, which is both convex and {reflexive}.
\begin{figure}[htb]
$$
\vC{\TikZ{[xscale=1.5]
  \path[use as bounding box](-.2,-.2)--(3.7,2.5);
  \path(1.7,1)node{\includegraphics*[width=60mm]{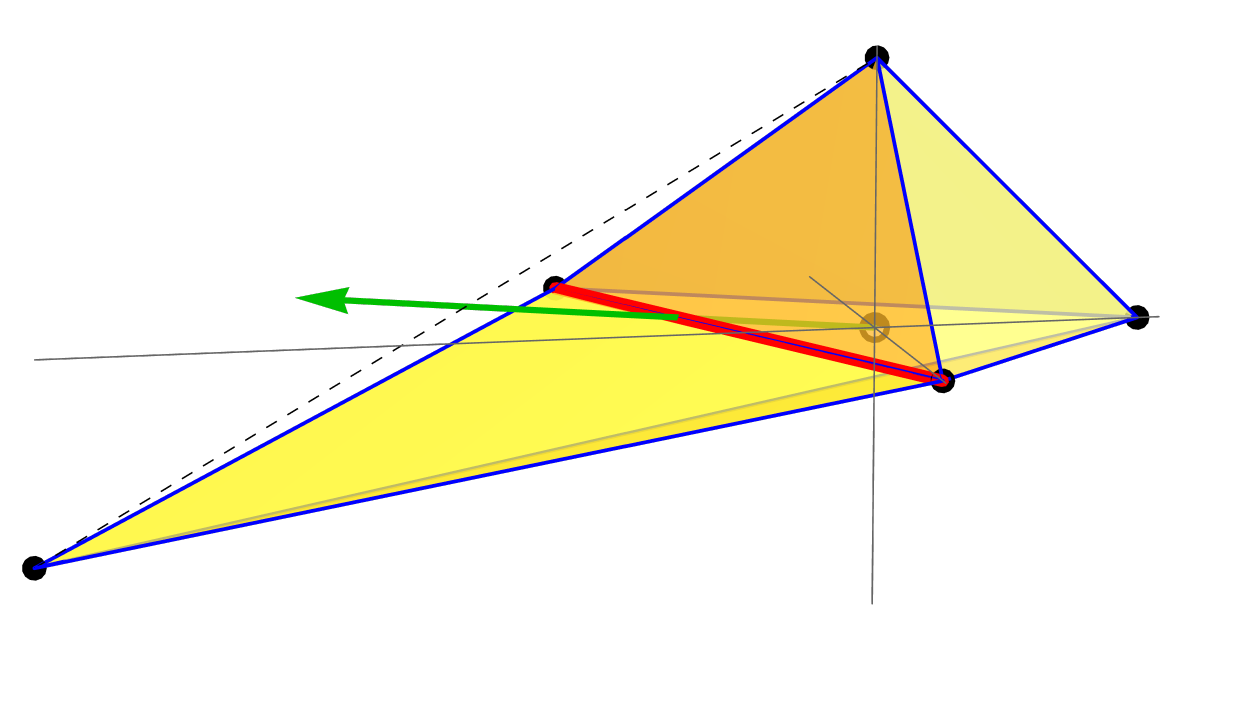}};
  \path(-.1,2.2)node[right]
      {$\pFn{\FF[3]{{\sss3,2}}}\smt\pDs{\FF[3]{\sss3,2}}$};
  \path[green!75!black](0.7,1.40)node[left]
  {{\fnSz$\pM{-1\\[-1pt]-2\\[-1pt]~~0}$}$\,=\n_6$};
  \path(1.33,1.60)node{$\n_1$};
  \path(2.85, .70)node{$\n_2$};
  \path(3.40, .95)node{$\n_3$};
  \path(2.70,2.50)node{$\n_4$};
  \path(-.10, .30)node{$\n_5$};
}}\quad
\parbox[c]{70mm}{\raggedright
  $\bullet$~Concave at the $\n_{12}$ edge;\\
  $\bullet$~The $\n_{13}$ edge is convex;\\
  $\bullet$~The {\em\/blowup vertex,} $\n_6\subset\n_{145}$-cone;\\
  $\bullet$~Saddle-point at $\n_1$.}
$$
\caption{The fan-spanning non-convex polytope of $\FF[3]{\sss3,2}$ and its key characteristics}
 \label{f:3F32S}
\end{figure}
The displayed (``hammock'') polytope in Figure~\ref{f:3F32S}change
is non-convex but VEX, while the ``tent'' configuration (including the $\n_{1524}$-simplex) is convex and reflexive and corresponds to the singular $\IP^3_{\!\sss(1:1:1:3)}$. The convex hull enclosing $\n_6$ is also convex and reflexive, and it corresponds to the blowup, 
$\Bl^\uA_{\n_6}[\FF[3]{\sss3,2}]$.

This analysis firmly connects the entire deformation family of each biprojective embedding of Hirzebruch scrolls to the well-studied constructions in weighted projective spaces and (complex-algebraic) toric varieties---while grounded in their application within the underlying worldsheet QFT and GLSM structure, relating this to toric geometry as needed---and extends them by including the non-Fano cases, 
$\FF{m}$ with $m\geqslant3$. This indicates an apparently unbounded collection of ambient spaces, $A=\FF{m}$, in which to embed the Calabi--Yau hypersurfaces of ultimate interest, and so it may be seen as a {sweeping} generalization of Gross' and Ruan's {\em\/non-algebraic deformation equivalence\/} example, $\FF[4]{\!\sss(2,1,1,0)}\<{\approx_{\sss\IR}}\FF[4]0$\cite{Gross:1994The, Ruan:1996top}.

Let us close with the following observations:
\begin{enumerate}[labelsep=13pt]
\item Trimming a two-dimensional convex polygon by removing non-star simplices of a particular triangulation to obtain a star-triangulated non-convex (but VEX) polygon is by now called {\em\/VEXing.} This {\em\/reduction\/} of a polygon necessarily implies an {\em\/expansion\/} in its transpolar/dual polygon, by introducing at least one new vertex.

\item VEXing a three-dimensional convex polytope need not introduce any new vertex in its transpolar/dual, as seen when trimming from the ``tent'' to the (displayed) ``hammock'' configurations in Figures~\ref{f:3F41S} and~\ref{f:3F32S}. The convex hull however is invariably found to be (see footnote~\ref{fn:exp} and Remark~\ref{r:rfX-VEX})
singular and requires some local {\em\/surgery\/} (typically, blowup) to desingularize the so-defined underlying toric space.

\item In four dimensions, one may analogously trim even {\em\/reflexive\/} polytopes, which was dubbed (lower-case styled) {\em\/vexing\/}\cite{huang2019fibration, Jefferson:2022ssj, MacFadden:2025ssx}, which also introduces co-dimension-2 (``pane'') non-convexity, but now neither the original reflexive polytope nor its vexed non-convex {star-}retriangulation need be singular\,\footnote{I thank Elijah Sheridan for verifying and informing me of this feature. In turn, a non-singular/smooth triangulation consists of unit-degree simplices (generated by a $\ZZ^n$-basis); in a star-triangulation, all simplices have the origin as one of their vertices. A non-singular/smooth star-triangulation is induced from a smooth underlying manifold with a full-dimensional $(\IC^*)^n$-action---corresponding to the non-gauged (global) $U(1)$-symmetries such as are specified by the upper four rows in~\eqref{e:q1-4} and~\eqref{e:q1-5}.} 
The so-obtained non-convex polytopes encode non-weak-Fano toric varieties and their Calabi--Yau hypersurfaces, which are combinatorially more abundant than the routinely studied (fine, regular and star-triangulated, ``FRST'') convex and reflexive polytopes, and increasingly rapidly so: already by $h^{1,1}\leqslant7$, 
the number of vex-triangulated (and so non-convex-trimmed) polytopes grows beyond twice as many as FRSTs\cite{MacFadden:2025ssx}. 
Since the Kreuzer--Skarke database\cite{Kreuzer:2000xy, wKS-CY} spans $1\leqslant h^{1,1}\leqslant491$, the likely very substantial preponderance of vex triangulations should be evident. 

VEX (but still ``ordinary,'' ``plain'') {\em\/polytopes\/} are more abundant yet, their convex hulls not being required to be reflexive. Furthermore, by including also self-crossing and otherwise multi-layered cases the full complement of VEX {\em\/multitopes\/} (four-dimensional analogues of $\pDN{\FF[3]5}$, $\pDN{\FF[3]{\sss4,1}}$ and $\pDN{\FF[3]{\sss3,2}}$ in Figure~\ref{f:3F5cases}; see also\cite{Berglund:2024zuz}) is vastly more plentiful.
\end{enumerate}

\section{Reflexions in the Mirror}
\label{s:Mirrors}
As argued above, the entire explicit deformation family~\eqref{e:pexy1}--\eqref{e:pexy2} of Calabi--Yau hypersurfaces in Hirzebruch scrolls may be realized within the GLSM framework\cite{rPhases, rAGM01, rAGM06, rAGM04, rMP0}---perhaps even more obviously so within the predecessing framework of constrained (nonlinear gauged) sigma models\cite{Eichenherr:1978SUN, rChaSM, rMargD, rUDSS08, rUDSS09}.
As shown in Section~\ref{s:AmbiDefo}, this deformation family~\eqref{e:pexy1}--\eqref{e:pexy2}, as showcased in Table~\ref{t:3F5cases}, contains several distinct Hirzebruch 3-folds exhibiting a hierarchy of five distinct types of combinatorial structures on display in Figure~\ref{f:3F5cases}:
\begin{enumerate}[labelsep=11pt, label={\Alph*.}]

 \item \label{i.A}
 $\FF[3]{\sss2,2,1}$: Both the Newton polytope $\pDs{\FF[3]{\sss2,2,1}}$ and the spanning polytope $\pDs{\FF[3]{\sss2,2,1}}$ are {\em\/reflexive} and, moreover, {\em\/strongly convex\/}; this is routine complex-algebraic toric geometry.

 \item \label{i.B}
 $\FF[3]{\sss3,1,1}$: Both polytopes are reflexive;
 the Newton polytope $\pDs{\FF[3]{\sss2,2,1}}$ is {\em\/strongly\/} convex and {\em\/reflexive,\/} but the spanning polytope $\pDs{\FF[3]{\sss2,2,1}}$ is convex {(but not strongly)}: in this deformation, the vertex $\n_1$ {lies in} the $\n_{45}$-line, now an edge of the fan-spanning polytope, which effectively encodes the ($\ZZ_2$-singular) weighted projective space 
$\IP^3_{\!\sss(1:1:2:2)}${, the lattice-point $\n_1\in\n_{45}$ encoding its MPCP-desingularization}.

 \item \label{i.C}
 $\FF[3]{\sss3,2}$: The spanning polytope $\pDs{\FF[3]{\sss3,2}}$ is {\em\/concave\/} at the $\n_{12}$ edge and its adjacent facets, and so it cannot be reflexive. 
{Dually,} the Newton multitope $\pDs{\FF[3]{\sss3,2}}$ is self-crossing flip-folded, but its regular part (without the extension rational monomials) suffices to provide transverse anticanonical sections.
The ``tent''-formed $\Conv[\pDs{\FF[3]{\sss3,2}}]$ is {not reflexive, as} its polar {image has a fractional point. Removing this single fractional point leaves a convex hull, the polar of which encodes the blowup of $\FF[3]{\sss3,2}$ at 
$\n_6$ shown in Figure~\ref{f:3F32S}.}

 \item \label{i.D}
 $\FF[3]{\sss4,1}$: The spanning polytope $\pDs{\FF[3]{\sss4,1}}$ is {\em\/concave\/} at the $\n_{12}$ edge and its adjacent facets. 
The Newton multitope $\pDs{\FF[3]{\sss4,1}}$ is self-crossing flip-folded, but its regular part (without the extension rational monomials) suffices to provide transverse anticanonical sections.
The ``tent''-formed $\Conv[\pDs{\FF[3]{\sss4,1}}]$ is {not reflexive, as its polar image has a fractional point. Removing this single fractional point leaves a convex hull, the polar of which encodes the blowup of $\FF[3]{\sss4,1}$ at 
$\n_6$ shown in Figure~\ref{f:3F41S}.}

 \item \label{i.E}
 $\FF[3]5$: The spanning polytope $\pDs{\FF[3]5}$ is 
(saddle-point) non-convex at the $\n_1$ vertex, 
concave at the adjacent edges, $\n_{12}$ and $\n_{13}$, 
{but convex at the adjacent edges $\n_{14}$ and $\n_{15}$}. 
The Newton multitope $\pDs{\FF[3]5}$ is self-crossing flip-folded, and now its extension {\em\/rational monomials are necessary\/} to provide transverse anticanonical sections.
Also, $\Conv[\pDs{\FF[3]5}]$ is {\em\/not reflexive\/}: its polar Newton polytope has {\em\/{two} fractional vertices.}
\end{enumerate}

The by-now standard, routine construction of mirror models\cite{Batyrev:1993oya} is defined only for models of type~\ref{i.A} and~\ref{i.B}, the {\em\/sufficiently generic\/} deformations within the family~\eqref{e:pexy1}--\eqref{e:pexy2}. 

The explicitly continuous and even complex-algebraically describable deformations throughout this family then {suggest} that mirror duality ought to extend throughout, also to the less generic deformations.
To this end, Refs.\cite{rBH-gB, Berglund:2022dgb, Berglund:2024zuz, Hubsch:2025sph, Hubsch:2025teh} continue verifying a direct generalization of the ``transposition{-}mirror construction''\cite{rBH} to the remaining (type~\ref{i.C}, \ref{i.D} and~\ref{i.E}) specializing deformations within explicit deformation families such as~\eqref{e:pexy1}--\eqref{e:pexy2}---as a showcasing template, indicating the need for a systematic and foundational study.

\subsection{Non-Convexity and Self-Crossing}
\label{s:stock}
\paragraph{Non-algebraic aspects:}
The fan-spanning polytopes, $\pDs{\FF{\ora{m}}}$, such as are shown in Figure~\ref{f:3F5cases}, are all (ordinary) polytopes, although several of them are non-convex.\footnote{Polytopes are multitopes that {\em\/embed\/} in $\IR^n$, while more general multitopes only immerse, such as the three self-crossing flip-folded Newton multitopes (see footnote~\ref{fn:mT}), $\pD(\FF[3]5)$, $\pD(\FF[3]{\!\sss4,1})$ and $\pD(\FF[3]{\!\sss3,2})$, in Figure~\ref{f:3F5cases}. Note that (complex-algebraic) toric geometry focuses on convex polytopes\cite{rO-TV, rF-TV, rGE-CCAG, rCLS-TV, rCK}, so practitioners tend to drop the ``convex'' qualifier for expediency; needless to say, {\em\/reflex\/} angles/cones and {\em\/non-convex\/} polygons and polytopes have not ceased to exist, and in fact they do turn up even in complex-algebraic toric geometry: see Figures~\ref{f:3F5S}--\ref{f:3F32S}.} 
Their transpolar Newton multitopes,
$\pDN{\FF{\ora{m}}}=(\pDs{\FF{\ora{m}}})^\wtd$, such as are shown in Figure~\ref{f:3F5cases}, motivated directly from the smoothing deformations of the GLSM superpotential, include self-crossing {bounded} polyhedral bodies, {such as}
$\pDN{\FF[3]5}$, $\pDN{\FF[3]{\!\sss4,1}}$ and 
$\pDN{\FF[3]{\!\sss3,2}}$.
These clearly do not belong to the (standard) complex-algebraic toric geometry framework, but they do turn up in studies of {\em\/pre-}{\em symplectic\/} $U(1)^r{=(S^1)^r}$-equivariant geometry{, named
{\em\/twisted polytopes\/}\cite{rK+T-pSympTM}; such bodies also appear (named
{\em\/virtual polytopes\/}) in Refs.\cite{rK+P-vrtPlytp0, rK+P-vrtPlytp},} 
and they fit the generalized notion {of} (bounded) {\em\/polyhedral complex{es}\/}\cite{Hibi:1995aa}.
Each reflexive polytope admits star-triangulations, each of which spans a central fan consisting of cones subtended by the facets of the polytope. Analogously, VEX {\em\/multitopes\/} {also} admit star-triangulations, each of which spans a central {\em\/multifan\/} of cones over its facets\,\footnote{``VEX multitopes'' were defined\cite{rBH-gB, Berglund:2022dgb, Berglund:2024zuz} as {bounded} polyhedral complexes that span complete multifans\cite{rM-MFans, Masuda:2000aa, rHM-MFs, Masuda:2006aa} and on which the transpolar operation is involutive.}---in the case of Newton multitopes,
$\pDN{\FF[3]5}$, $\pDN{\FF[3]{\!\sss4,1}}$ and $\pDN{\FF[3]{\!\sss{3,2}}}$, over the ``panes'' of the monomials shown in Figure~\ref{f:defPiX}.

\paragraph{Mirror mapping:}
The key operation in the complex-algebraic toric geometry framing of mirror duality\cite{Batyrev:1993oya}---which applies perfectly to the ``sufficiently generic'' of the deformations (type~\ref{i.A} and~\ref{i.B} in the above listing)---is the swapping of the r\^oles of the spanning and the Newton polytope. As detailed in Refs.\cite{Berglund:2024zuz, Hubsch:2025sph, Hubsch:2025teh}:
\begin{enumerate}[labelsep=13pt]
\item 
 Given a ``well-known'' ambient toric variety, $A$, encoded by the fan 
 that star-subdivides its {($\L\<\approx\ZZ^n$ lattice)} spanning polytope,
 $\pFn{A}\smt\pDs{A}{\subset(\L\<{\otimes_{\sss\IR}}\IR^n)}$,
\item \label{i:K*}
 the lattice points, $u_j\in{\L\!^\vee}\<\cap\pDN{A}$, 
 of its Newton polytope 
 $(\pDN{A}\<=(\pDs{A})^\circ)$ encode the sections of the anticanonical 
 line bundle, 
 $\G(\cKs{A})=\bigoplus_{u_j} a_j \prod_i x_i^{\vev{\n_i|u_j}+1}$,
 in terms of the Cox variables, 
 $x_i\mapsto\n_i\in\pFn[\sss(1)]{A}\fif{\sss\text{1--1}}\pDs[\sss(1)]{A}$, 
 one for each vertex of $\pDs{A}$ {and $a_j\in\IC$}.
\item
 The zero locus of a (transverse) section, $f(x)\<\in\cKs{A}$, 
 is a (smooth) Calabi--Yau hypersurface, $\sX\coeq\{f(x)\<=0\}\subset A$.
\item
 The mirror Calabi--Yau space $\chX$ is then the hypersurface
 $\chX\coeq\{g(y)\<=0\}\subset B$, where the toric variety $B$ is defined by
 swapping the r\^oles of $\pDs{A}$ and $\pDN{A}${\cite{Batyrev:1993oya}}:
  \begin{enumerate}[labelsep=13pt]
  \item\label{i:B=*A}
   The toric variety $B$ is encoded by the fan spanned by the $A$-Newton polytope,
   $\pFn{B}\<\smt\pDs{B}\<\coeq\pDN{A}$, so the $B$-Newton polytope is
   $\pDN{B}=\pDs{A}=(\pDN{A})^\circ=(\pDs{B})^\circ$.
  \item\label{i:*K*}
   Anticanonical sections are encoded again, encoded
   $\G(\cKs{B})=\bigoplus_{v_i} b_i \prod_j y_j^{\vev{\m_j|v_i}+1}$,
   where $v_i\in\ZZ^n\<\cap\pDN{B}$ are lattice points and the Cox variables are
   $y_j\mapsto\m_j\in\pFn[\sss(1)]{B}\fif{\sss\text{1--1}}\pDs[\sss(1)]{B}$,
   one for each vertex of $\pDs{B}$ {and $b_i\in\IC$}.
  \end{enumerate}
\end{enumerate}

These definitions expose the {\em\/direct\/} relation to the ``transposition mirror construction''\cite{rBH, rBH-gB}:
\begin{equation}
  f(x)\in
  \bigoplus_{\m_j\in\pDN[(1)]{A}} a_j
   \Big(\underbrace{\!\prod_{\n_i\in\pDs[(1)]{A}} x_i^{\vev{\n_i|\m_j}+1}}
        _{\text{step~\ref{i:K*}}}\Big)
  ~~\fif{\circ}~~
  \bigoplus_{\n_i\in\pDs[(1)]{A}} b_i 
   \Big(\underbrace{\!\prod_{\m_j\in\pDN[(1)]{A}} y_j^{\vev{\m_j|\n_i}+1}}
        _{\text{step~\ref{i:*K*}}}\Big)
  \ni g(y),
 \label{e:tM}
\end{equation}
using $\pDs{B}=\pDN{A}$ and $\pDN{B}=\pDs{A}$, and having restricted to the vertices in both cases. In fact, in each of the two polytopes any subset of lattice points the convex hull of which encloses the origin may be used\cite{Berglund:2024zuz} (Appendix~A), producing a web of multiple mirrors.

It is thus natural to ask, starting from a non-weak-Fano toric variety $A$ specified by a fan spanned by a non-convex polytope $\pFn{A}\<\smt\pDs{A}$,
what sort of space $B$ is prescribed by a ``minimal'' adaptation of the above construction, so that the relations:
\begin{equation}
 \begin{array}{r@{}l@{}r@{\quad}c@{\quad}l}
  B&{:} &\pFn{B}\<\smt\pDs{B}\<=\pDN{A} &\text{and} 
        &\pDN{B}\<=\pDs{A}\<\lat\pFn{A},\\*[2pt]
  \text{so}~~
  B&    &\supset\{g(Y)\<=0\}\<=\chX &\fif{~\text{mirror}~}
        &\sX\<=\{f(X)\<=0\}\subset A,
 \end{array}
 \label{e:B=*A1}
\end{equation}
continue to hold even when $\pDs{A}$ is non-convex, so that its transpolar $\pDN{A}$ is a self-crossing multitope rather than an ordinary polytope.

\subsection{The Transposed GLSM}
\label{s:tGLSM}
\paragraph{GLSM:}
At face value, prescription~\eqref{e:tM} provides the mirror-GLSM, requiring ``merely'' to read out the geometry. Given a left-hand side polynomial, we identify a transverse ``minimal'' part, 
$f_0(X)=\sum_{j=1}^{n+1}\big(\!\prod_{i=1}^{n+r} X_i^{e_{ij}}\big)$, 
with as many monomials as there are variables, and so the matrix of exponents,
$e_{ij}\<\coeq\vev{\m_j|\n_i}{+}1$, is of maximal rank{\cite{rBH-gB, Berglund:2022dgb, Berglund:2024zuz}; any complement,} $f(X)-f_0(X)$ is treated as a deformation. {Any such choice of $f_0(X)$ is ``invertible''\cite{Krawitz:2010FJR} and corresponds to a choice of anticanonical monomials in~\eqref{e:allK*} with a 0-enclosing simplex hull; see Remark~\ref{r:IPp}.}
The following procedure then constructs the {transposed} GLSM:
\begin{enumerate}[labelsep=13pt]
\item 
 Extending the ``transposition construction''\cite{rBH}, the mirror-defining polynomial is simply the ``transpose,''
$g_0(Y)=\sum_{i=1}^{n+r}\big(\!\prod_{j=1}^{n+\check{r}} Y_j^{e_{ji}}\big)$,
using $[\![e_{ji}]\!]=[\![e_{ij}]\!]^T$.
\item\label{e:tPq}
 This $g_0(Y)$ determines the $U(1;\IC)^{\check{r}}$-charges of the $Y_j$ by the double requirement
\begin{equation}
  \sum_{j=1}^{n+\check{r}} \check{q}_{aj}\,e_{ji} = -\check{q}_{a0} 
  \6*= \sum_{j=1}^{n+\check{r}} \check{q}_{aj},\qquad
  \check{q}_{aj}\<\coeq\check{q}_a(Y_j)~~\text{and}~~
  \check{q}_{a0}\<\coeq\check{q}_a(Y_0),
 \label{e:CY*}
\end{equation}
aiming for a mirror-superpotential of the general form $\widecheck{W}=Y_0\,g(Y)$, where {the} second, $*$-labeled equality stipulates the gauge-invariance, anomaly cancellation and Ricci-flatness condition{s} mirroring~\eqref{e:CY} on the transposed side{; see Remark~\ref{r:key}}.
\item\label{i:contq}
 Just like~\eqref{e:CY}, the gauge-invariance, anomaly-cancellation and Ricci-flatness conditions~\eqref{e:CY*} in the mirror model have a continuum of solutions. However, it is straightforward to find bases convenient for determining the $U(1;\IC)^{\check{r}}$ gauge-symmetry breaking patterns, mirroring~\eqref{e:q1-4} and~\eqref{e:q1-5}, and thus also the phase diagram {(gauge-orbits as in Section~\ref{s:GO}, i.e., GKZ decomposition)} of the mirror model, akin to the one in Figure~\ref{f:PhDiag}.
\item\label{i:posY}
 Given the charges, where $g(Y)\in(\Pi Y){\sqcup\bigsqcup_j}[\d_j\Pi Y]$ is chosen from a 
\eqref{e:allK*}-like collection of $Y$-monomials of charge $-\check{q}_{a0}$, 
 these 1st-order deformations of $\Pi Y$, $[\d_j\Pi{Y}]=[\Fm_j(Y)(\vd_j\Pi Y)]$ are again organized by their poset structure, mirroring {the one} in Figure~\ref{f:defPiX}.
\item\label{i:mu}
 From a chosen $\check{q}_{aj}$-basis, the mirror-analogue of~\eqref{e:q-nu},
\begin{equation}
  \sum_{i=1}^{n+2}\check{q}_a(Y_j)\,\m_j=-\check{q}_a(Y_0),
 \label{e:q*-mu}
\end{equation}
{ensures gauge-global mixing anomaly cancellations and} reverse-engineers the ``vertices'' {of the multitope with} the poset {structure} constructed in the previous step,~\ref{i:posY}.

\end{enumerate}

 It is reassuring to find (see footnote~\ref{fn:exp})
 that this procedure invariably reconstructs the generalized transposition-mirror relation~\eqref{e:B=*A1} and the results of the combinatorially defined and computed {\em\/transpolar\/} relationship with the $X$-model, such as the $\wtd$-indicated correspondences in Figure~\ref{f:3F5cases}---and for all non-convex and/or self-crossing VEX multitopes, just as it (of course) does for (convex) reflexive polytopes!

Consider, for proof-of-concept, the infinite sequence of $\FF[3]{m}[c_1]$ GLSMs, where retaining only the ``vertex''-monomials in Figure~\ref{f:defPiX} simplifies the superpotential to
\begin{equation}
  X_0\Big( X_1^3(X_4^{2+2m}+X_5^{2+2m})
          +X_2^3(X_4^{2-m}+X_5^{2-m})
          +X_3^3(X_4^{2-m}+X_5^{2-m}) \Big),
 \label{e:W3Fm}
\end{equation}
with the matrix of exponents (reading row-wise by monomials),
\begin{equation}
  [\![e_{ij}]\!]=
  \BM{ 3 & 0 & 0 & 2{+}2m & 0 \\[-2pt]
       3 & 0 & 0 & 0 & 2{+}2m \\[-2pt]
       0 & 3 & 0 & 2{-}m & 0 \\[-2pt]
       0 & 3 & 0 & 0 & 2{-}m \\[-2pt]
       0 & 0 & 3 & 2{-}m & 0 \\[-2pt]
       0 & 0 & 3 & 0 & 2{-}m \\ }.
 \label{e:mE}
\end{equation}
In step~\ref{e:tPq}, the twin conditions~\eqref{e:CY*} for the transpose matrix of exponents of~\eqref{e:W3Fm} then defines:
\begin{equation}
  \begin{array}{r|@{~}r|rrrrrr@{~}|}
     & Y_0 & Y_1 & Y_2 & Y_3 & Y_4 & Y_5 & Y_6\\ \toprule
{\check{q}_0}
            &{0}   &{2{-}m} &{m{-}2} &{0} &{0} 
                                      &{-2(m{+}1)} &{2(m{+}1)} \\[-2pt]
\check{q}_1 &9(m{-}2)  &0 & 3(m{-}2) &-2m{-}5 &5m{-}1 &2(m{-}2) &m{-}2 \\[-2pt]
\check{q}_2 &9(m{-}2)  &3(m{-}2) &0 &4m{+}1 &-m{-}7 &2(m{-}2) &m{-}2   \\[-2pt]
\check{q}_3 &12(m{+}1) &m{+}4 &3m &0 &4(m{+}1) &2(m{+}1) &2(m{+}1) \\[-2pt]
\check{q}_4 &6(m{+}2)  &2m &4 &2(m{+}2) &0 &2(m{+}1) &2          \\[-2pt]
\check{q}_5 &6 &3{-}m  &m{-}1 &-2m &2(m{+}1) &0 &2                 \\[-2pt]
\check{q}_6 &6(m{+}1)  &m{+}1 &m+1 &0 &2(m{+}1) &2(m{+}1) &0        \\[-1pt] \bottomrule
  \end{array}
 \label{e:*q1-6}
\end{equation}
As specified in step~\ref{i:contq},
these $\check{q}_a$-candidates are chosen so that $\check{q}_i(Y_i)\<=0$ {for at least one $Y_i$}---{akin} to the choices~\eqref{e:q1-4}---which then describes the phase-space of the model, following \mbox{Section~\ref{s:GO} and \ref{s:VEVs}}.
The rank of the ${7}{\times}7$ matrix of charges~\eqref{e:*q1-6} is 3, informing us that the transposed GLSM model has a $U(1;\IC)^3$ gauge symmetry, generated by any three of these $\check{q}_a$-charges, which provides for non-trivial differences as compared with the $(n;m)\<=(2;3)$ special case shown in Ref.\cite[\SS\:3.3]{Berglund:2022dgb}.
With these, the remaining two steps, \ref{i:posY}--\ref{i:mu}, are straightforward{, although the phase diagram is now three-dimensional, with seven distinct 1-cone generators indicated by the columns in~\eqref{e:*q1-6} and any choice of three rows of components}.

This generalizes considerably beyond the by-now well-studied transposition-mirror construction for ``invertible'' defining polynomials\cite{rBH, rBH-LGO+EG, rFJR-07b, Krawitz:2009aa, Krawitz:2010FJR, Ebeling:2012Mir, rLB-MirrBH, rACG-BHK, rA+P-MM, Favero:2016vhp, Parkhomenko:2022kju, Belavin:2023ldz, Parkhomenko:2024mxq, Cho:2024Ber}: the $5{\times}6$ matrix of exponents~\eqref{e:mE} is neither square nor of maximal rank. Incidentally,  $\rank[\![e_{ij}]\!]\<=4$ implies that restricting~\eqref{e:W3Fm} to four monomials (which in the Figure~\ref{f:3F5cases} plot span an origin-enclosing simplex) in four variables (by setting $X_1\to1$) does result in an ``invertible'' (albeit Laurent) special case. In fact, replacing~\eqref{e:W3Fm} with {\em\/any\/} other selection of four monomials from the poset in Figure~\ref{f:defPiX} that span an origin-enclosing 3-simplex in the $\pDN{\FF[3]5}$-point set also results in an ``invertible'' Laurent special case; for worked-out examples, see Ref.\cite{Berglund:2022dgb, Berglund:2024zuz}. This plethora of choices grows combinatorially with $(n;m)$ and indicates that a full 
$\FF{m}[c_1]$-GLSM may also be thought of as a {\em\/generating function\/} for such ``invertible'' Laurent models.

The foregoing discussion then justifies the expectation that the so-constructed {transposed}-GLSM can describe both the mirror Calabi--Yau model, $\chX$, and, owing to the observation in Remark~\ref{r:noW}, also (a choice of) the ambient space $B$ in which that mirror is the hypersurface,
\begin{equation}
  B\supset\chX\<\eqco\{g(Y)=0\} \fif{\sss~\text{transpose}~} 
  \{f(X)=0\}\<\coeq\sX\subset A,
 \label{e:B=*A2}
\end{equation}
justifying the interest in candidate ambient spaces $B$ corresponding to self-crossing multitopes, such as $\pDN{\FF[3]5}$, $\pDN{\FF[3]{\!\sss4,1}}$ and 
$\pDN{\FF[3]{\!\sss3,2}}$ in Figure~\ref{f:3F5cases}, and the multifans they span.

\subsection{Ambient Unitary Torus Manifolds}
Since the explicitly continuous deformation families of generalized complete intersections of Calabi--Yau manifolds in products of projective spaces\cite{rgCICY1, rBH-Fm, rGG-gCI} manifestly include the Hirzebruch $n$-fold scrolls \eqref{e:pexy1}--\eqref{e:pexy2} and $\FF{\!\sss\ora{m}}$ with arbitrarily high (``taxicab'') magnitude $|\ora{m}|$ as factors in the ambient space, ``the cat is out of the bag'': Discrete deformations of the $\FF{m}\<{\approx_{\sss\IR}}\FF{m-n}$ kind may well be widespread,\footnote{Whereas among the complex {one- and} two-dimensional candidates for a factor in a ``well-known'' ambient space only the Hirzebruch surfaces exhibit the discretely variable complex structure\cite{rGHSAR, rBeast2}; {one would expect} higher-dimensional varieties {to} be more amenable to such features.}
requiring the inclusion of ambient spaces (and their suitable hypersurfaces) corresponding to flip-folded or otherwise self-crossing multitopes, such as
$\FF[3]5$, $\FF[3]{\!\sss4,1}$ and 
$\FF[3]{\!\sss3,2}$ in Figure~\ref{f:3F5cases}.

Such self-crossing multitopes have been studied for considerable time---in {\em\/pre-}{\em symp\-lec\-tic geometry\/}\cite{rK+T-pSympTM}---which is {both fascinating and} encouraging, given that mirror duality maps complex structure to (special) symplectic structure {and vice versa}\cite{Kontsevich:1995wkA, Kontsevich:1994Mir, gross2016intrinsic}%
{, but exploring links to symplectic geometry and homological mirror symmetry will have to be left for a separate effort.}
{Self-crossing multitopes} span flip-folded {\em\/multifans\/} that also
have a long history\cite{rM-MFans, Masuda:2000aa, rHM-MFs, Masuda:2006aa, rHM-EG+MF, rH-EG+MFs2, Nishimura:2006vs, Davis:1991uz, Ishida:2013aa, Ishida:2013ab, buchstaber2014toric, Jang:2023aa}. Such fans are known to encode ``half-dimensional'' $(S^1)^n$-action (transformation) on the underlying real $2n$-dimensional {\em\/torus manifold,} $\sM$, which then contains multifan-encoded ``characteristic submanifolds,'' $M_i$, fixed under some subgroup of the ``compact torus'' group, $\rT\<\coeq(S^1)^n$.
These $M_i$ are analogous to the Cox divisors (in real co-dimension-2) and other submanifolds indicated by the various cones in the fan encoding the variety.

For the purposes of building a GLSM (Sections~\ref{s:TST} and~\ref{s:tGLSM}) with such a torus manifold as the ambient space, a subclass is needed that (to start with) accommodates:
\begin{enumerate}[labelsep=13pt]
 \item Complex-valued local coordinates (the $X_i$ and $Y_j$ chiral superfields are $\IC$-valued). 
 \item An anticanonical $\IC$-line bundle with $\IC$-valued sections the zero-locus of which is to define Calabi--Yau hypersurfaces of our ultimate interest.
\end{enumerate}

Both of these can be satisfied by so-called {\em\/unitary torus manifolds} (UTMs)\cite{rM-MFans, Masuda:2000aa, rHM-MFs}. 

By definition, the {\em\/unitary\/} (i.e., {\em\/weakly almost complex\/}) structure on a {\em\/real\/} $2n$-dimensional manifold, $\sM$, is a complex structure on its {\em\/stable tangent bundle,} 
$\sT_{\!\sM}\<\oplus\2{\IR}^{2r}$ for some $0\<\leq r\<\in\ZZ$; almost complex manifolds are the $r=0$ special case. Here, $\2{\IR}^{2r}$ denotes the trivial bundle over $\sM$, which is on a UTM also invariant under its ``half-dimensional'' $(S^1)^n$-transformations.
In particular, the (real) tangent bundle, $\sT_{\!\sM}$, turns out to be a quotient of the direct sum, $\bigoplus_{i=1}^{n+r}L_i$, of $\IC$-line bundles associated with the vertex-ray generators of the multifan, $\pFn{\sM}$---just as in complex-algebraic toric varieties, where this quotient refers to the so-called Euler sequence\cite{rCLS-TV, rBeast2}. 
However---and unlike in complex-algebraic toric varieties, 
({\bf1})~these $L_i$ are $\IC$-valued {\em\/smooth\/} (not a priori holomorphic) line bundles, and
({\bf2})~the quotienting map, $\oplus_{i=1}^{n+r}L_i\onto\sT_{\!\sM}$, is in general not guaranteed to be $(\IC^*\<\supset S^1)^n$-equivariant compatibly with the complex structure of the line bundles. That is, the quotienting map, 
$\oplus_{i=1}^{n+r}L_i\onto\sT_{\!\sM}$, may not have a $\rT$-equivariant inverse, so as to identify $\oplus_{i=1}^{n+r}L_i$ with $\sT_{\!\sM}\<\oplus\2{\IR}^{2r}$ equivariantly.\footnote{I thank Amin Gholampour for explaining to me the $\rT$-equivariance subtlety.}

Nevertheless, the top exterior power of this direct sum {\em\/does\/} define a 
$\IC$-valued ``anticanonical bundle,''
$\cKs{\!\sM}\<\coeq\wedge(\sT_{\!\sM}\<\oplus\2{\IR}^{2r})$, and 
UTMs moreover have the $\rT$-equivariant Chern class $c^\rT(\sM)\<\coeq\prod_i(1{+}\x_i)$
generated by $\x_i\<=c_1^\rT(L_i)\<\in H_\rT^2(\sM)$, so
$c_1^\rT(\cKs{\!\sM})\<=\sum_i\x_i$\cite{rM-MFans, rHM-MFs}.
Furthermore, unless the $\rT$-equivariant Todd class vanishes, $\sM$ is 
{\em\/cohomologically symplectic\/}: there exists an element $\X\<\in H^2(\sM)$ 
such that $\X^n\<\neq0$\cite{rM-MFans} (Cor.\;4.3); this $\X$ is the unitary equivalent of a K\"ahler class, with $\X^n$ {serving as} the K\"ahler volume-form.
The {\em\/orientation\/} (winding or wrapping index), $w(\s)\<={+}1$ (${-}1$), 
of an $n$-cone, $\s\<\in\pFn{\!\sM}$, 
indicates (dis)agreement in the affine chart $\sU_\s\<\subset\sM$ between 
the orientation induced from the complex structure of 
$\sT_{\!\sM}\<\oplus\2{\IR}^{2r}$ 
and that of $\sM$ itself.
The $w(\s)$-signed degree of both a cone and its facet, $\s\<=\sfa(\q)$, 
determines a so-called Duistermaat--Heckman measure over the multifan $\pFn{\!\sM}$\cite{Masuda:2000aa,rHM-MFs} (see also\cite{Berglund:2022zto}) and an {\em\/omniorientation\/} on $\sM$\cite{Buchstaber:2001aa, Masuda:2006aa, Ishida:2013aa, buchstaber2014toric}.\footnote{\label{fn:deg} The degree of a $k$-face, $d(\q^{\sss(k)}):=(k{+}1)!{\cdot}\!\Vol_{k+1}(\q^{\sss(k)})$, is the $(k{+}1)!$-fold volume of the {\em star-pyramid\/} over $\q^{\sss(k)}$\cite{Batyrev:1993oya}. Faces with $|d(\q)|\<>1$ and the cones they span encode $(\IC^n/\ZZ_{d(\q)})$-charts in the toric space; their desingularization is encoded by a subdivision of the cone.
	For simplicity, we tacitly identify the torus manifold-specific orientation-function with the multifan-specific $w$-function\cite{rM-MFans, Masuda:2000aa, rHM-MFs}. Subtleties in ``reading'' a multifan (affecting its fit for our application) are exemplified by the case of $S^4$\cite{rMP-TO+MFs}.} The $w(\s)$-sign of cone and facet degrees turns out to be the {\em\/only\/} extension of standard complex-algebraic toric geometry computations that is necessary and sufficient in the various computations of \mbox{Refs.\cite{rBH-gB, Berglund:2022dgb, Berglund:2024zuz, Hubsch:2025sph, Hubsch:2025teh}}{---wherein $w(\s)$ is required to be continuous over the multifan;
$\pFn{\sM}$: $w(\s)$ changes sign only where the multifan flip-folds, i.e., when a cone folds ``over''/``under'' its adjacent cone specified by the multifan, 
$\pFn{\sM}$. This specification turns out not to have been explored in the mathematical literature, so the precise identification of suitable UTMs remains an open question}.

\paragraph{{Back} to GLSMs:}
Section~\ref{s:TST} started with GLSMs containing $n{+}r$ complex (chiral, $(2,2)$-super)fields, $X_i$, each with its own, a priori independent phase-transformation, $X_i\<\to\l_iX_i$, generating the $\{\l_i\<\neq0,~i=1,\dots (n{+}r)\}=(\IC^*)^{n+r}$ complex-torus actions. The $U(1;\IC)^r\<\approx(\IC^*)^r$-subgroup was gauged, specifying the partitioning of the $X$-field space into separate gauge-orbits (see Section~\ref{s:GO}), leaving the remaining $(\IC^*)^n$-transformation to characterize the fan of the toric $n$-fold. For example, in~\eqref{e:q1-4} the top three rows correspond to this latter, un-gauged $(\IC^*)^3$-transformation on the toric 3-fold, $\FF[3]m$, while any two of the $q_a$-rows specify the gauged $U(1;\IC)^2\<\approx(\IC^*)^2$-transformation.

The $(S^1)^n$-``circle transformation'' specified by the multifan 
$\pFn{\!\sM}$ in UTMs may be identified with the phase-angle transformation implemented by multiplying with $\l\<\in\IC^*$ while restricting to $|\l|=1$ and leaving any accompanying ``radial scaling'' by multiplying with $|\l|\<\in\IR_+$ simply unspecified.\footnote{\label{ft:IC}
This polar-coordinate identification of the ``radial'' and ``circular'' parts of the $\IC^*$-transformation is referred to as the ``locally standard'' $\IC^*$-transformation, identifying, in turn, the chart $\sU_\s$ of every unit-degree cone $\s\in\pFn{\!\sM}$ with an affine copy of $\IC^n$---up to chart-wise conjugation.}
{In turn, this phase-angle transformation,
$X(\x)\to e^{i\vq}X(\x)$ with 
$\vq\<\coeq\arg(\l)\coeq\tfrac1{2i}\log(\l/\l^*)$ the phase-angle of $\l$, 
is exactly the QFT-standard $U(1)$-transformation of complex scalar fields.}
Conversely, complex-algebraic toric varieties may be regarded as UTMs in which 
a {\em\/unique\/} $\IR_+$-``radial scaling'' is associated with each $S^1$-``circle transformation,''
determined by the {\em\/overall\/} complex structure of the variety,
and by combining them into the hallmark complex toric transformation,
$S^1{\times}\IR_+\<\to\IC^*{\<=U(1;\IC)}$.
This comparison provides for the fact that there exists {\em\/a continuum\/} of UTMs corresponding to any given multifan. 
More generally, and starting from multifan $\pFn{}$-specified $(S^1)^n$-transformations, we seek a real $2n$-dimensional UTM, $\sM$, corresponding to 
$\pFn{}$, which admits {\em\/some suitable\/} $S^1\<\iff\IR_+$ association for each of the $n{+}r$ generators to correspondingly extend each $S^1\leadsto\IC^*$-transformation that will:
\begin{enumerate}[labelsep=13pt]
 \item coincide with the canonical association and $\IC^*$-transformation within the algebraically ``rigid'' subclass of complex-algebraic toric varieties;\footnote{\label{fn:frigid} Danilov's characterization as ``frigid toric crystals'' comes to mind\cite{rD-TV} (p.\,100).}
 \item extend to (some of the continuum of) UTMs corresponding to the VEX multifans produced by the transposition mirror construction in Sections~\ref{s:stock} and \ref{s:tGLSM}\cite{rBH-gB, Berglund:2022dgb, Berglund:2024zuz, Hubsch:2025sph, Hubsch:2025teh};
 \item be consistent with all physics-relevant (mostly cohomology ring, but also \mbox{{\em\/metric\/}\cite{Butbaia:2024tje, Constantin:2024yxh, Berglund:2024uqv, Constantin:2024yaz, Berglund:2024psp, Rahman:2026mfy})} computation hallmarks of mirror duality,
\end{enumerate}
and, ultimately,
\begin{enumerate}[resume, labelsep=13pt]
 \item be consistent also with geometric (conifold and other) transitions\cite{rReidK0, Green:1988uw, rGHC, rCGH1, Candelas:1989ug, rAGM01, rAGM06, rAGM04, Avram:1997rs, rB-IS+ST}.
\end{enumerate}

While this may seem like a ``tall order'' and too restrictive to meaningfully extend the framework of complex-algebraic toric geometry (and so leave models such as $\FF{\!\sss\ora{m}}[c_1]$ with $|\ora{m}|\geqslant3$ with no mirror reflexion), the fact that there exists a {\em\/continuum\/} of UTMs corresponding to every multifan gives hope. This does, however, make the task of identifying the ``right'' (possibly non-complex-algebraic) UTMs to serve as the ambient space ``$B$'' in Equations~\eqref{e:B=*A1} and~\eqref{e:B=*A2} akin to seeking a needle in a haystack.

\paragraph{A likely subclass:}
To this end, let us close by calling attention to a very special subclass of UTMs, so-called {\em\/topological torus manifolds} (TTMs)\cite{Ishida:2013ab},
which seem excellent candidates as they 
{have a {\em\/smooth\/} (rather than algebraic, as in toric varieties) 
$(\IC^*)^n$-action with an open dense orbit and}
exhibit several promising features:
\begin{enumerate}[labelsep=13pt]
 \item 
 Each TTM is uniquely specified by a {\em\/topological fan,} defined over 
 the ``ground field'' $\sR\approx\IC{\times}\ZZ$, conveniently parametrized 
 as $\bM{b&0\\c&v}\in\sR$, with $b,c\in\IR$ and $v\in\ZZ$:
  \begin{enumerate}[labelsep=6pt]
   \item
   Each complex-torus $0\<\neq\l\<\in\IC^*$-transformation is $\sR$-parametrized,
\begin{equation}
    \l{\cdot}X \coeq |\l|^b\,e^{i(c\,\log|\l| + v\,\arg(\l))}\,X,
\end{equation}
   \nitem such that

   \item 
   the $b$-projection, $\pFn[b]{\!\sM}$, is a plain, non-self-crossing fan in 
   $\IR^n$ generated by a continuous choice of (not necessarily lattice) 
   $\IR^n$-vectors, and

   \item 
   the $v$-projection, $\pFn[v]{\!\sM}$, is a $\ZZ^n$-lattice multifan in 
   $\IR^n$.
  \end{enumerate}

 \item 
 Restricting to $c\<=0$ and $b\<=v\in\ZZ$ recovers the familiar complex-algebraic 
 toric varieties as a special case of TTMs.
 Omitting the $(b,c)$-{data reverts to the broader class of} UTMs.
 
 \item The cohomology rings of the TTM $\sM$ are completely determined solely by 
 its $v$-projection multifan, $\pFn[v]{\!\sM}$.
\end{enumerate}

This last feature is especially encouraging, {as} it insures that there exists a 
$(b,c)$-continuum of TTMs the cohomology ring of which reproduces all the various computations in Refs.\cite{rBH-gB, Berglund:2022dgb, Berglund:2024zuz, Hubsch:2025sph, Hubsch:2025teh}, and wherein the ``nice'' subclass\cite{Yu:2011aa} further restricts
$\vec{b}_i\!\equiv\!\vec{v}_i\,\textrm{mod}\,2$.\footnote{I thank Amin Gholampour and Zengrui Han for alerting me to these manifolds and discussing their properties.}

It remains to determine which (if any) of these $\sR$-extensions of each VEX multifan $\pFn[v]{\!\sM}$ can guarantee {an} underlying TTM to admit the requisite {complex line-}bundles, with adequate $\IC$-valued sections, {$X_i\<\mapsto\n_i\<\in\pFn[\sss(1)]{\sM}$ and $\G(\pFn{\sM})$,} so as to define as their zero locus the Calabi--Yau hypersurfaces of our ultimate interest (see Refs.\cite{Berglund:2024zuz, Hubsch:2025sph} for a catalog of claims and conjectures with complementary justification)---and this is currently being actively explored; see, e.g.,\cite{Cui:2025Equ, Cui:2025Kly}.

\section{Conclusions and Outlook}
\label{s:CODA}
In this sequel to Ref.\cite{Hubsch:2025teh}, the overall general structure of worldsheet sigma models is reconsidered with the specific focus on constructing pairs of worldsheet models the ground states of which are transposition-mirror Calabi--Yau hypersurfaces~\eqref{e:B=*A2}---even when one of them is a hypersurface in a non-weak-Fano variety, $\sX\<\coeq\{f(X){=}0\}\<\subset A$. Its transposition-mirror is then constructed as a hypersurface $\chX\<\coeq\{g(Y){=}0\}\<\subset B$ in an ambient space $B$ that cannot be a complex-algebraic toric variety but would seem to be a special type of a unitary torus manifold (UTM), the precise type of which remains to be determined---so-called topological torus manifolds (TTMs) providing promising candidates.

\paragraph{{The transpose-ambient space:}}
To help with this determination, Section~\ref{s:tGLSM} details the construction of the transposed-GLSM, thus providing the ``bill of materials'' required in this construction, and which a usable candidate UTM, $\sM$, must be able to provide. In particular, UTMs do admit $\IC$-valued line bundles with $\IC$-valued sections that may serve as local coordinates on the ambient space, and their top exterior power defines the anticanonical $\IC$-line bundle with $\IC$-valued sections the zero-locus of which is to define the Calabi--Yau hypersurfaces of our ultimate interest.

However, these $\IC$-valued line bundles on a UTM, or even on the additionally qualified TTM, are a priori defined as {\em\/smooth\/} rather than holomorphic bundles, and it has been conjectured that flip-folded regions (composed of $w(\s)\<={-}1$ cones) in the multifan encode obstructions to a global (almost) complex structure\cite{Berglund:2024zuz}: a top-dimensional unit-degree cone $\s$ in an $n$-dimensional (multi)fan $\pFn{\!\sM}$ corresponds to a $\IC^n$-like local chart $\sU_\s\subset\sM$, and the sign of $w(\s)$ determines whether its orientation agrees with the unitary structure (the complex structure of $\sT_{\!\sM}\<\oplus\2{\IR}^{2r}$) or is opposite.

This prompts the conjecture\cite{Berglund:2024zuz} that
when two adjacent cones $\s_i,\s_j$ flip-fold ($w(\s_i)\,w(\s_j)\<<0$) across their common facet 
$\vs_{ij}\<=\s_j\cap\s_j$ a chart-wise local complex structure {\em\/need not\/} transfer holomorphically from $\sU_{\s_i}$ to $\sU_{\s_j}$: with respect to the global {\em\/unitary structure\/} of $\sM$ (=\,complex structure of $\sT_{\!\sM}{\oplus}\2\IR^{2r}$),
if $\sU_{\s}\<{\approx_{\sss\IC}}\IC^n$ for $w(\s)\<>0$
then $\sU_{\s'}\<{\approx_{\sss\IC}}\7{\IC^n}$ for $w(\s')\<<0$ if both charts admit a ``locally standard'' $\IC^*$-transformation.
Then, flip-folded parts of a multifan encode chart-wise conjugation 
local obstructions to a complex structure on their glued union,
$\sU_{\s_i}\<{\uplus_{\vs_{ij}}}\sU_{\s_j}$.
However, as seen in Figure~\ref{f:3F5cases},
 both side-panes of $\pDN{\FF[3]5}$,
 the back-pane of $\pDN{\FF[3]{\sss4,1}}$, and
 the front- and back-panes of $\pDN{\FF[3]{\sss3,2}}$
are themselves self-crossing, and so presumably they cannot correspond to ``locally standard'' $\IC^*$-transformation in the corresponding $\sU_\s\<{\approx_{\sss\IR}}\IR^4$-charts, so a more detailed (rigorous) analysis is needed to establish {\em\/if\/} and {\em\/precisely how\/} UTMs with flip-folded multifans fail to be (almost) complex.
\begin{remk}\label{r:acpx}
In lieu of such more precise analysis, and showcased by the multitopes
$\pDN{\FF[3]5}$, $\pDN{\FF[3]{\!\sss4,1}}$ and $\pDN{\FF[3]{\!\sss3,2}}$ that
rather prominently feature flip-folded ``extensions'' (depicted as the frontmost, ``hanging'' panes in Figure~\ref{f:3F5cases}), we may only infer that a (chart-wise glued) almost complex structure {\em\/may\/} have local obstructions in a UTM, in the local charts corresponding to the flip-folded regions in its multifan, $\pFn{\!\sM}$.
\end{remk}

\paragraph{{Localized Obstructions:}}
If a UTM, $\sM$, does have such obstructions to an almost complex structure and the anticanonical hypersurface of our ultimate interest, $\sX\<\in\sM[c_1]$, intersects it then $\sX$ itself ``inherits'' the obstruction and so cannot be almost complex---which would imply that the definitions~\eqref{e:SuSyGeom1}--\eqref{e:p,q} must be modified. How this happens and whether (and how) string theory may accommodate such ``defects''---or if they turn out to break the target space supersymmetry {induced} by this complex structure---is a tantalizing question, which remains open for now.\footnote{This may well be desirable: ``real world'' physics is {\em\/not\/} supersymmetric, and phenomenologically acceptable supersymmetry breaking is still work in progress.} 
In turn, such ``defects'' may also turn out to be innocuous if an ambient UTM (from a {\em\/continuum\/} that corresponds to any given multifan) can be found where $\sX\<\subset\sM$ ``misses'' the obstruction locus.

\begin{figure}[htb]
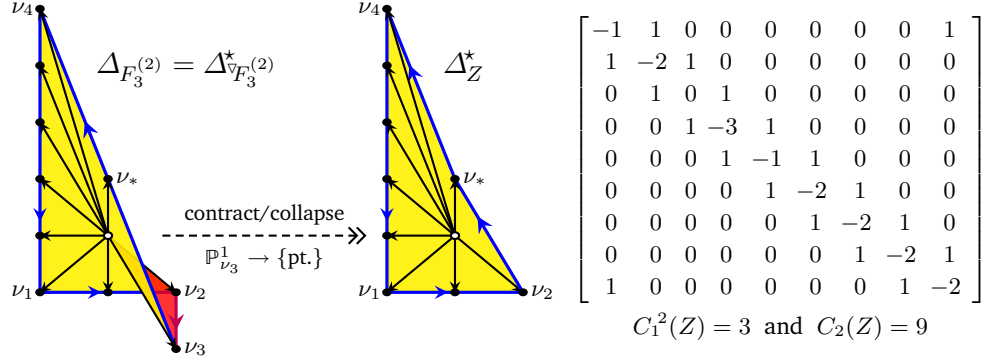

$$
\vC{\TikZ{[thick, xscale=.9, yscale=.75]
     \path[use as bounding box](-1.3,-2)--(6.4,4);
     \fill[yellow, opacity=.9](0,1)--(-1,4)--(-1,-1)--(1,-1)--(0,0)--cycle;
      \draw[blue, very thick, midArrow={pos=.45, end=stealth}](-1,-1)--(1,-1);
     \fill[red, opacity=.8](0,0)--(1,-1)--(1,-2)--cycle;
      \draw[Rouge, very thick, midarrow=stealth](1,-1)--(1,-2);
      \draw[blue, very thick, midArrow={pos=.75, end=stealth}](-1,4)--(-1,-1);
      \draw[-stealth](0,0)--(1,-1)node[right=-1pt]{\fnSz$\n_2$};
     \fill[yellow, opacity=.9](0,0)--(0,1)--(1,-2)--(0,0)--cycle;
      \draw[blue, very thick, midarrow=stealth](1,-2)--(-1,4);
      \draw[-stealth](0,0)--(-1,-1)node[left=-2pt]{\fnSz$\n_1$};
      \draw[-stealth](0,0)--(1,-2)node[right=-1pt]{\fnSz$\n_3$};
      \path(0,1)node[right=-1pt]{\fnSz$\n_*$};
      \draw[-stealth](0,0)--(-1,4)node[left=-2pt]{\fnSz$\n_4$};
      \foreach\y in{0,...,3}\draw[-stealth](0,0)--(-1,\y);
      \draw[stealth-stealth](0,1)--(0,-1);
     \foreach\y in{-1,...,4}\fill(-1,\y)circle(.7mm);
     \fill(0,1)circle(.7mm); \fill(0,-1)circle(.7mm);
     \fill(1,-1)circle(.7mm); \fill(1,-2)circle(.7mm); 
     \filldraw[fill=white](0,0)circle(.6mm);
     \path(-.33,3)node[right]{$\pDN{\FF[2]3}=\pDs{\tP\FF[2]3}$};
     \draw[densely dashed, line width=.8pt, 
           -{Straight Barb[sep=-1pt].Straight Barb[]}](.8,0)--
           node[above]{\scSz contract/collapse}
           node[below]{\scSz $\IP^1_{\n_3}\to\{\text{pt.}\}$}++(3,0);
     \begin{scope}[xshift=5.1cm]
     \fill[yellow, opacity=.9](0,1)--(-1,4)--(-1,-1)--(1,-1)--cycle;
      \draw[blue, very thick, midArrow={pos=.45, end=stealth}](-1,-1)--(1,-1);
      \draw[blue, very thick, midarrow=stealth](0,1)--(-1,4);
      \draw[blue, very thick, midArrow={pos=.75, end=stealth}](-1,4)--(-1,-1);
      \draw[-stealth](0,0)--(1,-1)node[right=-1pt]{\fnSz$\n_2$};
      \draw[blue, very thick, midarrow=stealth](1,-1)--(0,1);
      \draw[-stealth](0,0)--(-1,-1)node[left=-2pt]{\fnSz$\n_1$};
      \path(0,1)node[right=-1pt]{\fnSz$\n_*$};
      \draw[-stealth](0,0)--(-1,4)node[left=-2pt]{\fnSz$\n_4$};
      \foreach\y in{0,...,3}\draw[-stealth](0,0)--(-1,\y);
      \draw[stealth-stealth](0,1)--(0,-1);
     \foreach\y in{-1,...,4}\fill(-1,\y)circle(.7mm);
     \fill(0,1)circle(.7mm); \fill(0,-1)circle(.7mm);
     \fill(1,-1)circle(.7mm); 
     \filldraw[fill=white](0,0)circle(.6mm);
     \path(-.33,3)node[right]{$\pDs{Z}$};
     \end{scope}
            }}
\quad
{\fnSz\MM{\left[
\begin{array}{@{~}c@{~~}c@{~~}c@{~~}c@{~~}c@{~~}c@{~~}c@{~~}c@{~~}c@{~}}
\!-1\,& 1 & 0 & 0 & 0 & 0 & 0 & 0 & 1 \\
 1 &\!-2\,& 1 & 0 & 0 & 0 & 0 & 0 & 0 \\
 0 & 1 & 0 & 1 & 0 & 0 & 0 & 0 & 0 \\
 0 & 0 & 1 &\!-3\,& 1 & 0 & 0 & 0 & 0 \\
 0 & 0 & 0 & 1 &\!-1\,& 1 & 0 & 0 & 0 \\
 0 & 0 & 0 & 0 & 1 &\!-2\,& 1 & 0 & 0 \\
 0 & 0 & 0 & 0 & 0 & 1 &\!-2\,& 1 & 0 \\
 0 & 0 & 0 & 0 & 0 & 0 & 1 &\!-2\,& 1 \\
 1 & 0 & 0 & 0 & 0 & 0 & 0 & 1 &\!-2\,\\
\end{array}
\right]\\[18mm]
 C_1\!^2(Z)=3~~\text{and}~~ C_2(Z)=9
}}
$$
\caption{Collapsing an exceptional $\IP^1$-like cell in a toric space corresponding to a flip-folded multitope}
 \label{f:2F3TTM}
\end{figure}

That such ``defects'' {should be} localized is most easily seen by considering the two-dimensional case of the (extended, complete) Newton multigon of $\FF[2]3$\cite{Berglund:2022dgb, Hubsch:2025sph} depicted below on the far left
in Figure~\ref{f:2F3TTM}.
As specified in~\eqref{e:B=*A1} and~\eqref{e:B=*A2}, the Newton multigon of 
$\FF[2]3$ is equated with the multifan-spanning multigon of the UTM, $B\<=\tP\FF[2]3$, in which the transpose-mirror Calabi--Yau hypersurface (here, a 2-torus) is to be identified.
Depicted {in Figure~\ref{f:2F3TTM},
middle} is the polygon spanning the fan of a complex-algebraic (albeit non-weak-Fano) toric variety $Z$ together with the intersection matrix of its (lattice point corresponding) divisors, starting with $D_{\n_1}$ and {encircling} the origin in (arrow-indicated) CCW order.
The $\pFn{\tP\FF[2]3}$ multifan differs from the $\pFn{Z}$ fan only in the additional $\n_3$-ray that subdivides the $\n_{2*}$-cone, but not within it, as would be required in standard complex-algebraic toric geometry. Consequently, 
$\tP\FF[2]3$ is not a standard blowup of the toric variety $Z$; nevertheless, the difference between the two manifolds is localizable to the $\n_3$-encoded exceptional $\IP^1$-like set---which then (with its normal space) obstructs a complex structure on $\tP\FF[2]3$. This makes $\tP\FF[2]3$ (and all UTMs corresponding to flip-folded self-crossing multitopes) a priori {\em\/pre-complex\/}\cite{Berglund:2024zuz, Hubsch:2025sph, Hubsch:2025teh}---so-called mirroring {\em\/pre-symplectic manifolds,} wherein the symplectic structure may degenerate at real co-dimension-2 locations\cite{rK+T-pSympTM}.

\begin{figure}[htb]
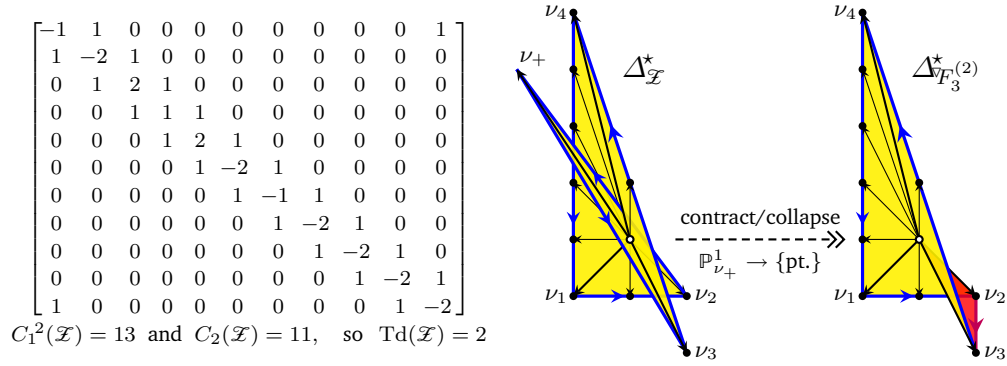

$$
{\scSz\MM{\left[
\begin{array}{@{}c@{~~}c@{~~~}c@{\quad}c@{\quad}c@{~~~}c@{~~}c@{~~}c@{~~}c@{~~}c@{~~}c@{}}
\!-1\,& 1 & 0 & 0 & 0 & 0 & 0 & 0 & 0 & 0 & 1 \\
 1 &\!-2\,& 1 & 0 & 0 & 0 & 0 & 0 & 0 & 0 & 0 \\
 0 & 1 & 2 & 1 & 0 & 0 & 0 & 0 & 0 & 0 & 0 \\
 0 & 0 & 1 & 1 & 1 & 0 & 0 & 0 & 0 & 0 & 0 \\
 0 & 0 & 0 & 1 & 2 & 1 & 0 & 0 & 0 & 0 & 0 \\
 0 & 0 & 0 & 0 & 1 &\!-2\,& 1 & 0 & 0 & 0 & 0 \\
 0 & 0 & 0 & 0 & 0 & 1 &\!-1\,& 1 & 0 & 0 & 0 \\
 0 & 0 & 0 & 0 & 0 & 0 & 1 &\!-2\,& 1 & 0 & 0 \\
 0 & 0 & 0 & 0 & 0 & 0 & 0 & 1 &\!-2\,& 1 & 0 \\
 0 & 0 & 0 & 0 & 0 & 0 & 0 & 0 & 1 &\!-2\,& 1 \\
 1 & 0 & 0 & 0 & 0 & 0 & 0 & 0 & 0 & 1 &\!-2\,\\
\end{array}
\right]\\[18mm]
 C_1\!^2(\sZ)=13~~\text{and}~~ C_2(\sZ)=11,\quad \text{so~~}\Td(\sZ)=2
 }}
\qquad
\vC{\TikZ{[thick, scale=.75]
     \path[use as bounding box](-1.5,-2)--(6.4,4);
     \fill[yellow, opacity=.9](-1,4)--(-1,-1)--(1,-1)--(0,0)--cycle;
      \draw[blue, very thick, midArrow={pos=.45, end=stealth}](-1,-1)--(1,-1);
      \draw[blue, very thick, midArrow={pos=.75, end=stealth}](-1,4)--(-1,-1);
      \draw[-stealth](0,0)--(-1,-1)node[left=-2pt]{\fnSz$\n_1$};
      \draw[-stealth](0,0)--(1,-1)node[right=-1pt]{\fnSz$\n_2$};
      \foreach\y in{0,...,3}\draw[thin, -stealth](0,0)--(-1,\y);
      \draw[thin, -stealth](0,0)--(0,-1);
     \fill[yellow, opacity=.9](0,0)--(1,-1)--(-2,3)--(1,-2)--cycle;
      \draw[blue, very thick, midArrow={pos=.55, end=stealth}](1,-1)--(-2,3);
      \draw[blue, very thick, midArrow={pos=.55, end=stealth}](-2,3)--(1,-2);
     \fill[yellow, opacity=.9](0,0)--(1,-2)--(-1,4)--(0,0)--cycle;
      \draw[blue, very thick, midarrow=stealth](1,-2)--(-1,4);
      \draw[-stealth](0,0)--(1,-2)node[right=-1pt]{\fnSz$\n_3$};
      \draw[-stealth](0,0)--(-1,4)node[left=-2pt]{\fnSz$\n_4$};
      \draw[thin, -stealth](0,0)--(0,1);
      \draw[-stealth](0,0)--(-2,3)node[above right=-3pt]{\fnSz$\n_+$};
     \foreach\y in{-1,...,4}\fill(-1,\y)circle(.7mm);
     \fill(0,1)circle(.7mm); \fill(0,-1)circle(.7mm);
     \fill(1,-1)circle(.7mm); \fill(1,-2)circle(.7mm); 
     \filldraw[fill=white](0,0)circle(.6mm);
     \path(-.33,3)node[right]{$\pDs{\sZ}$};
     \draw[densely dashed, line width=.8pt, 
           -{Straight Barb[sep=-1pt].Straight Barb[]}](.8,0)--
           node[above]{\scSz contract/collapse}
           node[below]{\scSz $\IP^1_{\n_+}\to\{\text{pt.}\}$}++(3,0);
     \begin{scope}[xshift=5.1cm]
     \fill[yellow, opacity=.9](0,1)--(-1,4)--(-1,-1)--(1,-1)--(0,0)--cycle;
      \draw[blue, very thick, midArrow={pos=.45, end=stealth}](-1,-1)--(1,-1);
     \fill[red, opacity=.8](0,0)--(1,-1)--(1,-2)--cycle;
      \draw[Rouge, very thick, midarrow=stealth](1,-1)--(1,-2);
      \draw[blue, very thick, midArrow={pos=.75, end=stealth}](-1,4)--(-1,-1);
      \draw[-stealth](0,0)--(1,-1)node[right=-1pt]{\fnSz$\n_2$};
     \fill[yellow, opacity=.9](0,0)--(0,1)--(1,-2)--(0,0)--cycle;
      \draw[blue, very thick, midarrow=stealth](1,-2)--(-1,4);
      \draw[-stealth](0,0)--(-1,-1)node[left=-2pt]{\fnSz$\n_1$};
      \draw[-stealth](0,0)--(1,-2)node[right=-1pt]{\fnSz$\n_3$};
      \draw[-stealth](0,0)--(-1,4)node[left=-2pt]{\fnSz$\n_4$};
      \foreach\y in{0,...,3}\draw[thin, -stealth](0,0)--(-1,\y);
      \draw[thin, stealth-stealth](0,1)--(0,-1);
     \foreach\y in{-1,...,4}\fill(-1,\y)circle(.7mm);
     \fill(0,1)circle(.7mm); \fill(0,-1)circle(.7mm);
     \fill(1,-1)circle(.7mm); \fill(1,-2)circle(.7mm); 
     \filldraw[fill=white](0,0)circle(.6mm);
     \path(-.33,3)node[right]{$\pDs{\!\tP\FF[2]3}$};
     \end{scope}
            }}
$$
\caption{Collapsing an exceptional $\IP^1$-like cell in a toric space corresponding to a double-winding multitope}
 \label{f:2F3TTM+}
\end{figure}

In turn, the UTM corresponding to the multifan spanned by the multitope 
$\pDN{\FF[2]3}$ in Figure~\ref{f:2F3TTM}
may also be regarded as a ``blowdown'' of a UTM, $\sZ$, which is {\em\/almost complex\/} by\cite{rM-MFans} (Thm.\;5.1), as it corresponds to the double-winding multifan spanned by the star-shaped double-winding/wrapping multitope in Figure~\ref{f:2F3TTM+}.
With the Todd genus\;$=2$, this almost complex UTM, $\sZ$, cannot possibly be a complex-algebraic toric variety, which resonates with one of Danilov's prescient remarks (see fotnote~\ref{fn:frigid}). 
The sequence of blowdown-like local surgeries in Figures~\ref{f:2F3TTM}--\ref{f:2F3TTM+},\footnote{The precise geometry of these surgeries is far from clear, although the relevant cell decomposition would seem to be straightforward.}
$\sZ\<\dto\tP\FF[2]3\<\dto Z$, should aid in determining the geometry of 
$\tP\FF[2]3$ in terms of the almost complex UTM $\sZ$ and the non-weak-Fano toric variety $Z$. 
{With its geometry determined, even partially}, this local surgery sequence {would provide} a template and a general prescription to describe the geometry of all UTMs corresponding to flip-folded multitopes and multifans. However, I defer this analysis to a subsequent endeavor.

In conclusion, this note examines the worldsheet $(2,2)$-supersymmetric GLSM constructions featuring explicitly continuous deformation families of Hirzebruch scrolls with (secondary) deformation families of Calabi--Yau hypersurfaces in each scroll. When those ambient scrolls are not even weak-Fano, the transposition-mirrors of the Calabi--Yau hypersurfaces are by a transposed{-}GLSM construction nevertheless indicated as hypersurfaces in non-algebraic ambient spaces identified as (suitably $\IC$-ringed) unitary torus manifolds, the precise geometry and physics consequences of which remain to be determined. 
{However, it {\em\/is\/} evident that such deformation families connect distinct toric variety ambient spaces and so also the Calabi--Yau hypersurfaces in them via non-algebraic deformation equivalences realized as explicitly continuous deformations. 
Although focused on the $\ora{m}$-web of Hirzebruch scrolls (see Figure~\ref{f:3F5trip}
and Corollary~\ref{C:cases}) and Calabi--Yau hypersurfaces in them, 
the present analysis is framed so that it can be extended to other (abelian) GLSMs. 
The transposed-GLSM in Section~\ref{s:tGLSM} then enables the study of their transposed-mirror reflexions. 
(This extends the 1992 transposition-mirror construction\cite{rBH}, as discussed in the Introduction before Section~\ref{s:QFT+MD}; see footnote~\ref{fn:hBHK}.)
By omitting the superpotential, the GLSM describes the ambient space (Remark~\ref{r:noW}); the deformation construction of anticanonical superpotentials (Section~\ref{s:AC}) reveals their origin-enclosing poset structure (Figure~\ref{f:defPiX} and Remark~\ref{r:IPp}), mirroring the combinatorial framework of toric geometry to their non-algebraic generalizations, an as-yet only incrementally qualified\cite{rBH-gB, Berglund:2022dgb, Berglund:2024zuz, Hubsch:2025sph, Hubsch:2025teh, Hubsch:2026lir} subclass of unitary torus manifolds\cite{rM-MFans, Masuda:2000aa, rHM-MFs, Masuda:2006aa}.
}

\paragraph{Acknowledgments:}
First and foremost, I should like to thank Charlene Cheng for her kind invitation and the opportunity to present this work.
 I am deeply thankful
 to Per Berglund for decades of collaborations, including on many of the topics discussed here, 
 to Mikiya Masuda for guidance through unitary and topological torus manifolds,
to Yong Cui, Amin Gholampour, and Zengui Han for convincing me of the TTM's promising nature,
and 
 to Elijah Sheridan for insightful discussions about vexing.
 I am grateful to
 the Mathematics Department of the University of Maryland and to 
 the Physics Department of the University of Novi Sad, Serbia,
for recurring hospitality and resources.

\paragraph{Abbreviations and Special Terms:}
The following abbreviations and portmanteaus are used in this manu\-script:\\
\noindent 
\begin{tabularx}{\textwidth}{@{}lX}
GLSM & Gauged linear sigma model
  (a class of worldsheet quantum field theories)\cite{rPhases}.\\
MPCP & \2Maximal \2Projective \2Crepant \2Partial 
  (desingularization)\cite{Batyrev:1993oya}.\\
multitope & {A Latin/Greek heterotic portmanteau of ``(possibly) \2{multi}-layered+poly\2{\smash{tope}}''; see footnote~\ref{fn:mT}.}\\
reflexion & This rarely used spelling indicates also the reflexive re-examinations focal herein.\\
transpolar & A portmanteau of ``\2{trans}pose''+``\2{\smash{polar}}''; see footnote~\ref{fn:tP}.\\
TTM  & Topological torus manifold\cite{Ishida:2013aa}.\\
UTM  & Unitary torus manifold\cite{rM-MFans}.\\
VEX  & Mnemonic contraction for ``not necessarily con\2{VEX}
       \2{EX}tension''\cite{Berglund:2024zuz, Hubsch:2025teh}, originally\cite{rBH-gB}.
\end{tabularx}

\begingroup
\footnotesize\raggedright
\def\rasp{\leavevmode\raise.45ex\hbox{$\rhook$}}
\providecommand{\href}[2]{#2}\begingroup\raggedright\endgroup

\endgroup

\end{document}